%% file: desc_white_paper.tex
\documentclass[11pt,headsepline,cleardoubleempty,twoside,openright]{scrbook}

\usepackage{whitepaper}

\begin{document}

\begin{titlepage}
\begin{center}

\vspace*{20mm}

{\Huge\bfseries\scshape Large Synoptic Survey Telescope} 
\linebreak 
\linebreak 
{\Huge\bfseries\scshape Dark Energy Science Collaboration} 

\vspace*{20mm}

\input{authorlist}

\vspace*{5mm}
\begin{center}
{\bf Version 2.0\\

\vspace*{\stretch{0.2}}

October 31, 2012}
\end{center}

Prepared by the LSST Dark Energy Science Collaboration\\

\end{center}
\end{titlepage}

\include{abstract}

\tableofcontents

\include{introduction/introduction}

\include{collaboration/collaboration}

\include{analysis/analysis}

\include{simulations/simulations}

\include{workplan/workplan}

\bibliographystyle{SciBook}
\bibliography{references.bib,introduction/references,simulations/cosim/cosim,analysis/weak-lensing/wl,supernovae/references,analysis/large-scale-structure/lss,analysis/large-scale-structure/extra_lss,analysis/strong-lensing/references,analysis/clusters/clusters,analysis/clusters/clusters_extra,analysis/photoz/photoz,analysis/theory/references,analysis/theory/references_extra,analysis/theory/references_extra_p2.bib,workplan/photoz/references}

\end{document}

%% file: authorlist.tex
Abate, Alexandra$^{32}$, Aldering, Greg$^{18, 33}$, Allen, Steven W.$^{29, 30}$, Ansari, Reza$^{14}$, Antilogus, Pierre$^{16}$, Applegate, Douglas$^{29}$, Astier, Pierre$^{16}$, Aubourg, Eric$^{1}$, Bahcall, Neta A.$^{26}$, Bard, Deborah$^{29}$
% ...

\vspace*{5mm}

\begin{table}[htdp]
\centering
{\renewcommand{\arraystretch}{0.8}
\begin{tabular}{p{10cm}p{10cm}}
$^{1}$Astroparticule et Cosmologie, IN2P3, Paris, France & $^{26}$Princeton University \\
$^{2}$Argonne National Laboratory & $^{27}$Purdue University \\
$^{3}$Brookhaven National Laboratory & $^{28}$Rutgers University \\
$^{4}$Brown University & $^{29}$SLAC National Accelerator Laboratory \\
$^{5}$California Institute of Technology & $^{30}$Stanford University \\
$^{6}$Centre de Calcul, IN2P3, Lyon, France & 						$^{31}$Stony Brook University \\
$^{7}$Centre Physique des Particules de Marseille, IN2P3, Marseille, France & 		$^{32}$University of Arizona \\		
$^{8}$Carnegie Mellon University & 							$^{33}$University of California, Berkeley \\
$^{9}$Cornell University & 								$^{34}$University of California, Davis \\
$^{10}$Fermi National Accelerator Laboratory & 						$^{35}$University of California, Irvine \\
$^{11}$Harvard University & 								$^{36}$University of California, Santa Barbara \\
$^{12}$Jet Propulsion Laboratory & 							$^{37}$University of California, Santa Cruz \\
$^{13}$Johns Hopkins University & 							$^{38}$University of Chicago \\
$^{14}$Laboratoire de l'Acc\'{e}l\'{e}rateur Lin\'{e}aire, IN2P3, Orsay, France & 	$^{39}$University of Cincinnati \\
$^{15}$Laboratoire de Physique Corpusculaire, IN2P3, Clermont-Ferrand, France &        	$^{40}$University of Illinois \\
$^{16}$Laboratoire de Physique Nucl\'{e}aire et de Hautes Energies, IN2P3, Paris, France & $^{41}$University of Miami \\
$^{17}$Laboratoire de Physique Subatomique et de Cosmologies, IN2P3, Grenoble, France & $^{42}$University of North Dakota \\
$^{18}$Lawrence Berkeley National Laboratory & 						$^{43}$University of Oklahoma \\
$^{19}$Lawrence Livermore National Laboratory & 					$^{44}$University of Oxford \\
$^{20}$LSST Corporation & 								$^{45}$University of Pennsylvania \\
$^{21}$National Astronomical Observatories of China & 					$^{46}$University of Pittsburgh \\
$^{22}$National Optical Astronomy Observatory &  					$^{47}$University of Washington \\
$^{23}$Ohio University & 								$^{48}$Wayne State University \\
$^{24}$Ohio State University &								$^{49}$Yale University\\
$^{25}$Paris Diderot University & 							
\end{tabular}
}
\end{table}

%% file: abstract.tex
\begin{center}

\vspace*{30mm}

{\bf Abstract.} 

This white paper describes the LSST Dark Energy Science Collaboration (DESC), whose goal is the study of dark energy and related topics in fundamental physics with data from the Large Synoptic Survey Telescope (LSST).  
It provides an overview of dark energy science and describes the current and anticipated state of the field.  
It makes the case for the DESC by laying out a robust analytical framework for dark energy science that has been defined by its members and the comprehensive three-year work plan they have developed for implementing that framework.  
The analysis working groups cover five key probes of dark energy: weak lensing, large scale structure, galaxy clusters, Type Ia supernovae, and strong lensing. 
The computing working groups span cosmological simulations, galaxy catalogs, photon simulations and a systematic software and computational framework for LSST dark energy data analysis. 
The technical working groups make the connection between dark energy science and the LSST system. 
The working groups have close linkages, especially through the use of the photon simulations to study the impact of instrument design and survey strategy on analysis methodology and cosmological parameter estimation. 
The white paper describes several high priority tasks identified by each of the 16 working groups. 
Over the next three years these tasks will help prepare for LSST analysis, make synergistic connections with ongoing cosmological surveys and provide the dark energy community with state of the art analysis tools. 
Members of the community are invited to join the DESC, according to the membership policies described in the white paper.
Applications to sign up for associate membership may be made by submitting the Web form at
\url{http://www.slac.stanford.edu/exp/lsst/desc/signup.html}
with a short statement of the work they wish to pursue that is relevant to the DESC.

\end{center}

%% file: introduction/introduction.tex
\chapter[LSST as a Dark Energy Experiment]{LSST as a Dark Energy Experiment }
\label{ch:intro}
\section{Introduction and overview}

The Large Synoptic Survey Telescope (LSST) is a wide-field, ground-based telescope, designed to image a substantial fraction of the sky in six optical bands every few nights.  It is planned to operate for a decade allowing the stacked images to detect galaxies to redshifts well beyond unity. 
The LSST and the survey are designed to meet the requirements \citep{SRD} of a broad range of science goals in astronomy, astrophysics and cosmology, including the study of dark energy -- the accelerating expansion of the Universe. 
The LSST was the top-ranked large ground-based initiative in the 2010 National Academy of Sciences decadal survey in astronomy and astrophysics, which noted that the ranking was a result of ``(1) its compelling science case and capacity to address so many of the science goals of this survey and (2) its readiness for submission to the MREFC process as informed by its technical maturity, the survey's assessment of risk, and appraised construction and operations costs.''

The LSST project is a partnership among the National Science Foundation (NSF), the Department of Energy (DOE) Office of Science, and public and private organizations in the United States and abroad.\footnote{
The total construction cost of LSST is estimated to be about \$665M, approximately 70\% from NSF, 24\% from DOE, and 6\% from private donors to the project.}
The NSF is responsible for site and telescope development and the data management system, while the DOE is responsible for development and delivery of the large-format camera. 
Private contributions have already been used for fabrication of the mirrors and site preparation in Chile.
In April 2012, the camera project received ``Critical Decision 1'' approval by the DOE to move into the 
detailed engineering design, schedule, and budget phase.
In July 2012, the National Science Board of the NSF approved the LSST as a
Major Research Equipment and Facilities Construction (MREFC) project, 
allowing the NSF Director to advance the project to the final design stage and include funds for LSST construction in a future budget request. 
If all continues as planned, construction will begin in 2014 and is anticipated to last five years, followed by a two-year commissioning period before the start of the survey in 2021.

The telescope, camera, and data management system are designed and built by the LSST Project Team, 
which is responsible for producing the facility but not for the scientific analysis of the data, which will be made 
public to the US and Chilean communities and some international partners. 
Hence, the Project Team is not a scientific collaboration in the usual sense. 
In 2008, eleven separate quasi-independent science collaborations were formed to focus on a broad range of
topics in astronomy and cosmology that the LSST could address.
Five of the science collaborations have relevance to the study of dark energy (DE) and have provided invaluable guidance to the Project Team;
however, they have not been formally organized or funded to pursue the full range of investigations required to guarantee that the 
most sensitive constraints on the nature of dark energy can be derived from the LSST data. 

In this White Paper, we describe in detail how the various DE analyses that we expect to perform are sensitive to a large number of potential systematic uncertainties that must be identified, quantified, and minimized in order for the DE investigations to achieve their full potential.  
In addition, new algorithms must be developed and tested to enable those analyses, and an efficient computational and software framework must be established to perform the necessary calculations.  
All of these activities require an extensive, coordinated research effort well in advance of the onset of data taking.
To address that need, we have created the LSST {\bf Dark Energy Science Collaboration} (DESC). 

The DESC is organized around five probes of dark energy enabled by the LSST data:\footnote{The keys 
WL, LSS, SN, Cl, SL, Phz, CoSim, CatSim, PhoSim, CWG, DM, SW, CM, and TC will be used to identify tasks 
in Chapter~\ref{sec:workplan}.}
\begin{enumerate}
\item Weak gravitational lensing (WL) -- the deflection of light from distant sources due to the bending of space-time by 
baryonic and dark matter along the line of sight.
\item
Large-scale structure (LSS) -- the large-scale power spectrum for the spatial distribution of matter as a function of redshift. This includes the Baryonic Acoustic Oscillations (BAO) measurement of the distance-redshift relation.
\item 
Type Ia Supernovae (SN) -- luminosity distance as a function of redshift measured with Type Ia SN as standardizable candles.
\item
Galaxy clusters (Cl) -- the spatial density, distribution, and masses of galaxy clusters as a function of redshift.
\item 
Strong gravitational lensing (SL) -- the angular displacement, morphological distortion, and time delay for the multiple images
of a source object due to a massive foreground object.
\end{enumerate}
These include the four techniques (WL, LSS, SN, Cl) described in the 2006 Report of the Dark Energy Task Force (DETF, ~\cite{DETF}).
The DESC will identify and work to minimize the most significant systematic uncertainties (hereafter referred to as ``systematics'') 
that limit the sensitivity of each probe, beginning with those that are most time-urgent.

The DESC will also address high priority tasks that are common to all five probes:
\begin{itemize}
\item  Calibration strategies for photometric redshifts (Phz). 
\item  Cosmological simulations (CoSim), simulated catalogs (CatSim), and photon-level simulations (PhoSim) with the fidelity needed to fully assess and exploit each probe of dark energy with the LSST.
\item  Cross working group tools for data quality assessment and detection of systematics (CWG).
\item  Realistic data model (DM), software framework (SW), and computing model (CM) to fully address DE science. 
\item  Technical coordination tasks related to the instrument model, calibration, and survey operations (TC).
\item  Theory and framework for combining and jointly interpreting dark-energy  probes (TJP).
\end{itemize}

In the remainder of this chapter, we describe the LSST project, the survey strategy, and the data products that will be delivered by the project.
We give an overview of DE science, with a focus on the theoretical challenges to understanding the 
accelerating expansion of the Universe and the types of measurements that are needed to distinguish between
competing hypotheses. 
We summarize precursor imaging, spectroscopic, and time-domain surveys that are relevant to DE science
and are expected to have mapped parts of the sky to various redshift depths by 2020 
-- ``Stage III'' projects in the parlance of the DETF.
We conclude this chapter with an assessment of the gains that the LSST will bring not only in the DE figure of merit 
defined by the DETF, but also in mitigating systematic uncertainties through the statistical power and combination of
probes provided by the LSST.

In Chapter~\ref{ch:collab}, we describe in more detail the need for the LSST Dark Energy Science Collaboration, introduce the 
minimal governance structure that is being used to get the DESC off the ground, and present the general structure
of the work plan for the next three years.

In Chapter~\ref{sec:analysis}, we describe the major analyses and the primary sources of systematic uncertainties for each DE probe.
We describe the technical tools and framework that will be necessary to fully address DE science with the LSST in Chapter~\ref{chp:sims}.

We present a compilation of all time-urgent, high-priority tasks (designated by H) 
and important longer term tasks (designated by LT) in Chapter~\ref{sec:workplan}.
The tasks are arranged according to the DE probes and cross-cutting issues listed above. 
For each task, we describe the motivation, the planned activities, and the expected deliverables. 
We have identified the important systematic uncertainties and have prioritized addressing 
those that could still inform the final design of the project or survey strategy,
or for which the systematic uncertainty is unlikely to be addressed by Stage III surveys,
or for which the systematic is of unknown size or may need new strategies to address.

This White Paper as a whole provides an integrated picture of the DE science analyses we expect to 
tackle with the LSST data set and the extensive preparatory work that must be accomplished to fully
exploit the statistical power of the LSST data. 
Our goal in producing this White Paper is to assist the funding agencies and reviewers 
in assessing proposals from individual principal investigators who wish to contribute to 
the investigation of dark energy through the study of scientific 
opportunities with the LSST, and the investigation of dark energy probes with existing 
astronomical data sets as they pertain to understanding and optimizing the 
potential of the LSST.

\section{Description of the LSST system}

The LSST system is designed to achieve multiple goals in four main science themes: 
inventorying the Solar System, mapping the Milky Way, exploring the transient optical 
sky, and probing dark energy and dark matter. These are just four of the many areas on 
which LSST will have enormous impact, but they span the space of technical challenges 
in the design of the system and the survey and have been used to focus the science requirements.
LSST will be a large, wide-field ground-based telescope, camera and data management system
designed to obtain multi-band images over a substantial fraction of the sky every few nights.
The observatory will be located on Cerro Pach\'on in northern Chile (near the Gemini South 
and SOAR telescopes), with first light expected around 2019. The survey will yield contiguous 
overlapping imaging of over half the sky in six optical bands ($ugrizy$, covering the wavelength 
range 320--1050\,nm). 

The LSST telescope uses a novel three-mirror design (modified Paul-Baker) with a very fast f/1.234 
beam. The optical design has been optimized to yield a large field of view (9.6 deg$^2$), with 
seeing-limited image quality, across a wide wavelength band. Incident light is collected by the 
primary mirror, which is an annulus with an outer diameter of 8.4\,m and inner diameter of 5.0\,m 
(an effective diameter of 6.5\,m), then reflected to a 3.4\,m convex secondary, onto a 5\,m concave 
tertiary, and finally into three refractive lenses in a camera. This is achieved with an innovative 
approach that positions the tertiary mirror inside the annular primary mirror, making it 
possible to fabricate the mirror pair from a single monolithic blank using borosilicate technology. 
The secondary is a thin meniscus mirror, fabricated from an ultra-low expansion material. All three 
mirrors will be actively supported to control wavefront distortions introduced by gravity and 
environmental stresses on the telescope. The telescope sits on a concrete pier within a carousel 
dome that is 30\,m in diameter. The dome has been designed to reduce dome seeing (local air 
turbulence that can distort images) and to maintain a uniform thermal environment over the course 
of the night. 

The LSST camera provides a 3.2 Gigapixel flat focal plane array, tiled by 189 4k$\times$4k CCD science 
sensors with 10\,$\mu$m pixels. This pixel count is a direct consequence of sampling the 9.6 deg$^2$
field-of-view (0.64\,m diameter) with 0.2$\times$0.2 arcsec$^2$ pixels (Nyquist sampling in the best 
expected seeing of $\sim$0.4\,arcsec). The sensors are deep depleted high resistivity silicon 
back-illuminated devices with a highly segmented architecture that enables the entire array to be 
read in 2 seconds. The sensors are grouped into 3$\times$3 rafts; each contains its own dedicated 
front-end and back-end electronics boards. The rafts are mounted on a silicon carbide grid inside a 
vacuum cryostat, with an intricate thermal control system that maintains the CCDs at an operating 
temperature of 180\,K. The entrance window to the cryostat is the third of the three refractive lenses 
in the camera. The other two lenses are mounted in an optics structure at the front of the camera body, 
which also contains a mechanical shutter, and a carousel assembly that holds five large optical filters. 
The sixth optical filter can replace any of the five via a procedure accomplished during daylight hours.

The rapid cadence of the LSST observing program will produce an enormous volume of data ($\sim$15\,TB of 
raw imaging data per night), leading to a total database over the ten years of operations of 100\,PB for 
the imaging data, and 50\,PB for the catalog database. The computing power required to process the data 
grows as the survey progresses, starting at $\sim$100\,TFlops and increasing to $\sim$400\,TFlops by the 
end of the survey. Processing such a large volume of data, automating data quality assessment, and archiving 
the results in a useful form for a broad community of users are major challenges. The data management system 
is configured in three levels: an infrastructure layer consisting of the computing, storage, and networking 
hardware and system software; a middleware layer, which handles distributed processing, data access, user 
interface and system operations services; and an applications layer, which includes the data pipelines and 
products and the science data archives. 

The application layer is organized around the data products being produced. The nightly pipelines are based 
on image subtraction, and are designed to rapidly detect interesting transient events in the image stream 
and send out alerts to the community within 60 seconds of completing the image readout. The data release 
pipelines, in contrast, are intended to produce the most completely analyzed data products of the survey, 
in particular those that measure very faint objects and cover long time scales. A new run will begin each 
year, processing the entire survey data set that is available to date. The data release pipelines consume 
most of the computing power of the data management system. The calibration products pipeline produces the 
wide variety of calibration data required by the other pipelines. All of these pipelines are architected 
to make efficient use of Linux clusters with thousands of nodes. There will be computing facilities at 
the base facility in La Serena, at a central archive facility, and at multiple data access centers. The 
data will be transported over existing high-speed optical fiber links from South America to the USA. 

For a more detailed discussion, including optical design, the filter complement, the focal plane 
layout, and special science programs, please see the LSST overview paper \citep{LSSToverview} 
and the LSST Science Book\footnote{Available from www.lsst.org/lsst/SciBook} \citep{LSSTSciBook}.

\section{Planned survey strategy and delivered data products} 
\label{sec:surveystrategy}

The LSST observing strategy is designed to maximize scientific throughput by minimizing 
slew and other downtime and by making appropriate choices of the filter bands given the 
real-time weather conditions. The fundamental basis of the LSST concept is to scan the sky 
deep, wide, and fast, and to obtain a dataset that simultaneously satisfies the majority of 
the science goals. This concept, the so-called ``universal cadence'', will yield the main 
deep-wide-fast survey and use about 90\% of the observing time. The observing strategy for 
the main survey will be optimized for homogeneity of depth and number of visits. In 
times of good seeing and at low airmass, preference will be given to $r$-band and $i$-band 
observations. As often as possible, each field will be observed twice, with visits separated 
by 15-60 minutes. The ranking criteria also ensure that the visits to each field are widely 
distributed in position on the sky and rotation angle of the camera in order to minimize 
systematic effects in galaxy shape determination.

The current baseline design will allow about 10,000 deg$^2$ of sky to be covered using pairs of 
15-second exposures in two photometric bands every three nights on average, with typical 5$\sigma$ 
depth for point sources of $r\sim24.5$.  These individual visits will be about 
2 mag deeper than the Sloan Digital Sky Survey (SDSS) data, currently the largest existing optical imaging survey. The system 
will yield high image quality as well as superb astrometric and photometric accuracy for a 
ground-based survey. The survey area will include 30,000 deg$^2$ with $\delta<+34.5^\circ$, 
with the 18,000 deg$^2$ main survey footprint visited over 800 times during 10 years. The 
coadded data within the main survey footprint will be 5 mag deeper than SDSS ($r\sim27.5$).
The main survey will result in databases that include 10 billion galaxies and a similar number of 
stars, and will serve the majority of science programs. 
The remaining 10\% of observing time will be used to obtain improved coverage of parameter 
space, such as very deep ($r \sim 26$) observations (e.g., optimized for SNe), observations with 
very short revisit times ($\sim$1 minute), and observations of ``special'' regions such as the 
Ecliptic, the Galactic plane, and the Large and Small Magellanic Clouds. 

The LSST data system is being designed to enable as wide a range of science as possible. Standard 
data products, including calibrated images and catalogs of detected objects and their attributes, 
will be provided both for individual exposures and the deep incremental data coaddition. About 
2 billion objects will be routinely monitored for photometric and astrometric changes, and any 
transient events (non-recurrent sources with statistically significant photometric change -- about 
10,000 per night on average) will be announced and distributed via web portals in less than 60 seconds. For 
the ``static'' sky, there will be yearly database releases that will list many attributes for billions 
of objects and will include other metadata (parameter error estimates, system data, seeing summary, 
etc). 

\section[Dark Energy overview]{Overview of dark energy science}

Observations reveal that the expansion of the Universe is accelerating, a result that poses a direct challenge to current theories of cosmology and fundamental physics. 
The 2011 Nobel Prize in Physics was awarded for the 1998 discovery of the accelerating Universe using distance measurements of Type Ia supernovae (SNe). In the thirteen years since that discovery, astronomical observations ranging from the cosmic microwave background (CMB) to the spatial distribution of galaxies have validated and refined what we know about the expansion history of the Universe.

Acceleration represents a serious dilemma for fundamental physics. The most prosaic explanation is that vacuum energy, which constitutes about three quarters of the energy density in the Universe, is driving the acceleration; 
however, naive predictions based
on the Planck scale would predict vacuum energy densities roughly 100
orders of magnitude larger than observed, implying that some unknown
physics suppresses such predicted vacuum energy, but leaves a very small
residual that is driving the current expansion rate.
Other explanations are even more radical, ranging from a new scalar field akin to the recently discovered Higgs boson but 44 orders of magnitude less massive, to additional dimensions of space, and many others -- all competing models for the phenomenon referred to as dark energy.

Distinguishing competing hypotheses for acceleration will require precise measurements of the {\it cosmic expansion history} and the {\it growth of structure}. The cosmic expansion history is the observational signature most closely connected to acceleration, and future observations must map the expansion history accurately out to higher redshifts. However, expansion history may not offer sufficient discriminating power between different theoretical models and will have to be combined with measurements of the growth of structure. Structure formation involves a balance between gravitational attraction of matter over-densities and the rapid expansion of the background. Thus, quantifying the rate of growth of structures from early times until the present provides additional tests of the energy contents of the Universe and their interactions.

One alternative explanation for cosmic acceleration is that gravity on cosmic scales is not described by Einstein's general relativity (GR). Such alternatives to the conventional smooth dark-energy hypothesis are referred to as modified gravity (MG) scenarios. MG theories are still being developed, but some observational implications have already emerged.  Since MG theories change the gravitational interaction on large scales, one generically expects the growth of structure to be altered compared to a smooth dark energy model with a similar expansion history. Moreover, wide classes of MG models predict a deviation in how massive structures bend light (observed through gravitational lensing) and accelerate other stars and galaxies. Thus, a combination of lensing and dynamical studies can test the MG hypothesis. A second approach is to work on smaller scales and exploit the transition an MG theory must make to GR on laboratory and Solar System scales (where it is well tested).

Testing models of dark energy and determining whether Einstein's theory needs to be modified will therefore be carried out using many different combinations of probes. For smooth dark energy models, the effect on the expansion rate and growth of structure is commonly characterized using two parameters, $w_0$ and $w_a$, that describe the equation of state of dark energy (if dark energy is simply a cosmological constant, then $w_0=-1$ and $w_a=0$). The $w_0-w_a$ parametrization has the drawback of not being physically motivated and also implicitly dis-favoring models in which dark energy played a role at early times. A more general parametrization, which allows the equation of state to vary freely in a succession of redshift bins, is one way to address this drawback. As discussed below in Section~\ref{sec:gains}, LSST is especially powerful for characterizing dark energy by a larger set of parameters that go beyond the $w_0-w_a$ description.

\section[Precursor surveys]{Precursor surveys}

By the time LSST starts taking data, a number of surveys will have mapped parts of the sky to various redshift depths, yielding important datasets and greatly enhancing the expertise of the dark energy science community. Herein, we summarize these precursor surveys, restricting ourselves to projects that have some relevance to LSST and that have secured the bulk of their funding or other project resources. 

{\bf Precursor imaging surveys.} SDSS I-III has mapped nearly 10,000 square degrees in the Northern hemisphere to a limiting magnitude of $g\simeq 21$ with median redshift $z\simeq 0.4$; luminous red galaxies are mapped out to $z\simeq 0.8$ and quasars to higher redshifts. SDSS-Stripe 82 and CFHT Legacy Survey are significantly deeper but only cover about 200 square degrees. Pan-STARRS (PS1) and KIDS have begun their surveys and expect to reach similar depth as SDSS-Stripe 82.  The Dark Energy Survey (DES) and the Hyper Suprime-Cam (HSC) on the Subaru telescope 
have achieved first light. 
DES will map 5000 square degrees in five filters to image galaxies beyond $z=1$ and obtain photometric redshifts.  
HSC aims to map an area of a couple of thousand square degrees.

Precursor surveys will use four dark energy probes: WL, photometric BAO, cluster number counts and SN. It is expected that the cosmological returns from these surveys will be limited by systematic errors of various kinds, not statistical precision.
Weak lensing results so far have mostly focused on the constraints on the amplitude of matter fluctuations; recent measurements have a typical error bar of approximately 10\%. Both DES and HSC expect to improve weak lensing statistical power by about an order of magnitude. 
Photometric BAO were first detected from SDSS II data \citep{Padmanabhan06} and subsequently measured to higher accuracy at $z=0.55$ with SDSS III  with a 4\% error \citep{Ho12,seo12}. Current and upcoming surveys such as PS1, KIDS, DES, HSC will be able to measure BAO at a higher redshift ($z\approx1$) than the concurrent spectroscopic surveys (see below) but with lower accuracy. 
Making use of current optical surveys (SDSS), cluster number counts currently constrain the dark energy equation of state to about 20\% accuracy alone and can improve the overall dark energy constraints modestly when combined with other probes (CMB, BAO, SN). Current and upcoming surveys (DES, HSC, RCS-2, KIDS, PS1) expect to increase the number of clusters by an order of magnitude. The challenge in  cluster cosmology is the calibration of the cluster mass-observable relation. Current and planned SZ surveys such as ACT, SPT, Planck, ACTPol and SPTPol can help calibrate cluster masses to better than 10\%. In parallel, surveys such as DES will use stacked weak lensing to internally calibrate cluster masses.  Successful and well characterized mass-calibration will greatly enhance the constraining power of cluster number counts. 

{\bf Precursor spectroscopic surveys.} The recently-completed WiggleZ survey has mapped $200,000$ galaxies below $z=1$ over $1000$ square degrees. Between 2014 and 2017, the HETDEX survey will map some 800,000 Lyman-alpha emitters at $z\approx2-3.5$ over approximately 500 square degrees. By 2014, Sloan Digital Sky Survey III will have surveyed $10,000$ square degrees, mapping galaxies out to $z=0.8$ and quasars at higher redshift. After Sloan  3 (AS3) by 2018 will have mapped $3100$ square degrees out to greater depth; other planned projects with greater power include the PFS (Prime Focus Spectrograph) on the Subaru telescope, BigBOSS and DESpec. 

The current measurements from SDSS and WiggleZ have already allowed a measurement of the distance at $z=0.35$ at $2\%$ and at $z=0.6$ at 4\% respectively.  SDSS III will improve on these by more than a factor of 2. It will also measure BAO at $z=2.5$ using the Lyman alpha forest, as will  HETDEX using Lyman alpha emitters. AS3 and other upcoming surveys will make equally precise measurements at higher redshifts. 

The combination of lensing measurements from imaging surveys and dynamics from spectroscopic surveys is a powerful probe of modified gravity theories. This test of gravity has been carried out using SDSS data \citep{Reyes:2010tr} and will be performed at higher precision in the coming years. Spectroscopic surveys also provide us with better cluster redshift determination, which is crucial in cluster counts cosmology (since we use $dN/dz$ as the cosmological probe).  The spectroscopic surveys can also provide cluster velocity dispersions to help calibrate cluster mass more accurately. 

{\bf Precursor time domain surveys.} For dark energy cosmology, Type Ia SN are the primary probe that requires time domain information. Having led to the discovery of dark energy in 1998, SN cosmology has advanced significantly in the last decade. Current measurements constrain the dark energy equation of state parameter (if taken to be constant) to better than $10\%$. These constraints are obtained with a combination of ground based data at low redshift and HST at $z\textgreater1$. Upcoming surveys that will impact SN cosmology cover a wide range in redshift, and aim to tackle systematics such as correlation of SN brightness with host galaxy. These include: SDSS, SN Factory, QUEST, PS1, DES and HST based observations. Redshifts of SN for projects such as DES will be obtained through follow up spectroscopy of host galaxies. 

{\bf Implications for LSST.} By the time LSST is commissioned, precursor imaging surveys will have yielded invaluable expertise with theoretical and observational systematics, analysis algorithms, simulations, and data management techniques. The imaging data size at cosmological depths will increase by an order of magnitude in the next five years, so we can expect new challenges and solutions to emerge prior to LSST. There are several likely areas of synergy and collaboration: photometric calibration methods, cosmological and image simulations, pixel level analysis of multiple data sets and a variety of topics in the control of systematics and optimization of dark energy analyses. 

The combination of all precursor surveys still constitute a dataset that is an order of magnitude smaller than the one expected from LSST. In particular LSST will study galaxies at significantly higher redshifts than precursor surveys. LSST will need to incorporate hard-won advances and lessons from these earlier surveys, and prepare to go significantly beyond them in its control and mitigation of systematic errors. 

Precursor Stage III and Stage IV spectroscopic surveys will provide training sets to calibrate LSST's photometric redshifts and mitigate other systematics. Working out the detailed requirements for spectroscopic galaxy samples is an important goal for the near future. It is especially challenging to obtain adequate spectroscopic samples for the faintest galaxies imaged by LSST. A coordinated effort will be needed to make advances in both the techniques for calibrating photometric redshifts and in obtaining the needed spectroscopic samples. 

\section[Gains from LSST]{Gains from LSST}
\label{sec:gains}

Historically, our understanding of the cosmic frontier has progressed in step with the size of our astronomical surveys, and in this respect, LSST promises to be a major advance: its survey coverage will be approximately ten times greater than that of the Stage III Dark Energy Survey. 

\begin{figure}
\centering
\includegraphics[width=3in]{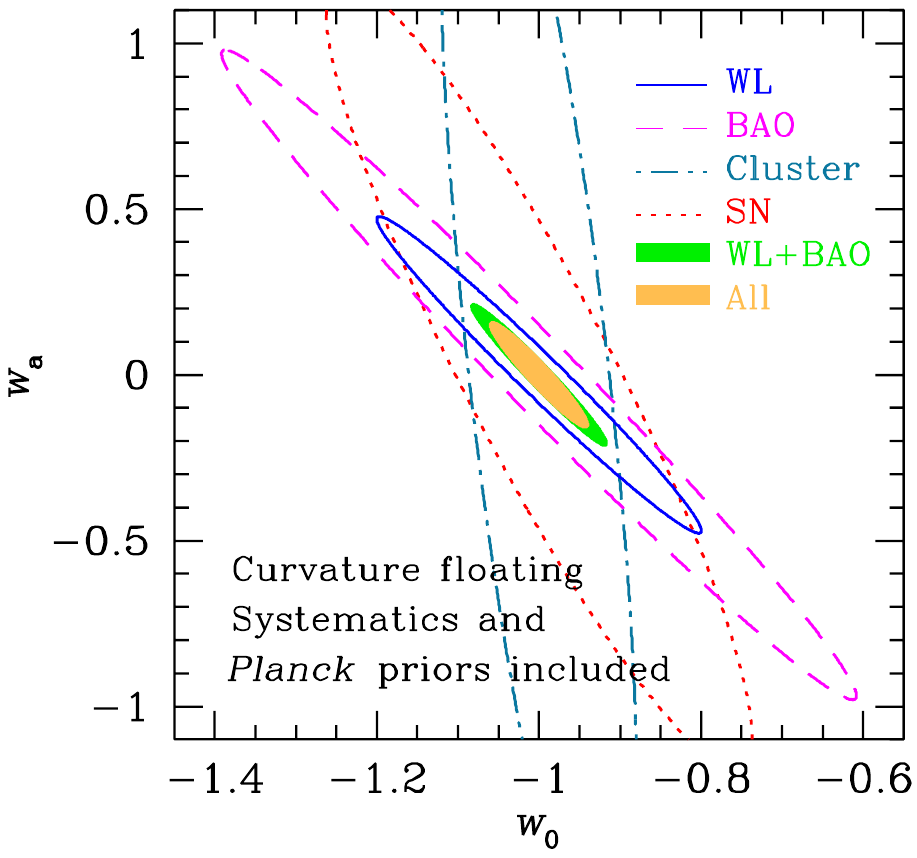}
\caption{Joint $w_0$--$\wa$ constraints from LSST WL (solid line), 
BAO (dashed line), cluster counting (dash-dotted line), 
supernovae (dotted line), joint BAO and WL 
(green shaded area), and all combined (yellow shaded area).
The BAO and WL results are based on galaxy--galaxy, galaxy--shear,
and shear--shear power spectra only. 
Adding other probes such as strong lensing time delay
(Section~\ref{sec:analysis:sl}), and higher-order galaxy and shear 
statistics will further improve the constraints.
For comparison, the areas of the error ellipses of Stage III dark 
energy experiments would be about 10 times larger than that of
LSST \citep[for a comparison of WL+BAO results, see][]{zhan2006}.
\label{fig:cswb}}
\end{figure}

Survey size is a straightforward measure of scientific gain. A less agnostic metric with which to judge dark energy probes is the figure of merit proposed by the Dark Energy Task Force,
which is the reciprocal of  the area of the error ellipse enclosing the 95\% confidence limit in  
the $w_0-w_a$ plane, marginalized over other cosmological parameters. 
Figure~\ref{fig:cswb} shows these projected constraints from four LSST probes: 
WL, BAO, cluster counts, and supernovae. The WL and BAO results are
based on \citet{zhan2006} and \citet{zhan_etal2009}, the cluster 
counting result is from \citet{fang_haiman2007}, and the 
supernova result is based on \citet{zhan_etal2008}. 
While absolute projections are uncertain due to the unknown effects of systematics, the relative gain in the figure of merit of LSST over Stage III surveys consistently comes in at a factor of 5 to 10. 

Future surveys will go beyond the simple $w_0-w_a$ parameterization of the dark-energy 
equation of state; 
these surveys can determine $w(z)$ in bins of redshift, effectively measuring 
{\it modes} -- 
linear combinations of $w(z)$ over the redshift range of interest 
-- with some degree of accuracy. Viewed in this way, the dark energy constraining power of LSST could be several orders of magnitude greater than that of a Stage III survey \citep{albrecht07}.

There is also the tantalizing possibility, discussed earlier, that Einstein was wrong -- that general relativity (GR) is not the true theory of gravity on the large scales probed by cosmology. The true observational test of all modified-gravity (MG) models will be the growth of structure, since MG models generically predict differences in the two scalar potentials (which are equal to one another at late times in general relativity) and modifications to the Poisson equation that relates the gravitational potential to over-densities. These modifications are generally functions of both space and time. Again, no survey will probe these deviations at all redshifts and at all scales, but LSST will go much further than any of its predecessors in its ability to measure growth in tens or even hundreds of spatial-temporal bins and constrain dozens of modes.
Disagreement of even one of these modes with the GR prediction would signal inconsistency with Einstein's theory of gravity. Of course, it is essential to address potential systematics in a wide variety of probes to ensure that any observed disagreement is robust.

While these projections for LSST statistical significance are compelling, they probably do not capture the true nature of the revolution that LSST will enable. The sheer statistical power of the LSST dataset will allow for an all-out attack on systematics, using a combination of null tests and hundreds of nuisance parameters. For example, multiple scans of the same region of sky can be combined to create modes that are sensitive only to atmospheric contamination of the ellipticities. As another example, multiplicative error in measurements of the ellipticity, instead of being ignored, will be detected and studied in as much detail as the signal from cosmic shear. Systematics will be mitigated by combining probes. In the case of BAO and WL, a joint analysis of the shear and galaxy over-densities for the same set of galaxies involves 
galaxy--galaxy, galaxy--shear, and shear--shear correlations, which
enable some calibration of systematics that would otherwise adversely 
impact each probe \citep{zhan2006}. 

Beyond tests of systematics, there is a growing sense in the community that the old, neatly separated categories of dark energy probes will not be appropriate for next generation surveys. Instead of obtaining constraints on dark energy from cluster counts and cosmic shear separately, LSST scientists may use clusters and galaxy-galaxy lensing simultaneously to mitigate the twin systematics of photometric redshift error and mass calibration~\citep{Oguri:2010vi}. Magnification~\citep[e.g.,][]{jain2002} may emerge as just as powerful a tool to measure mass density as galaxy ellipticities; for this, a joint analysis of the density and ellipticity fields is crucial. A homogeneous and carefully calibrated dataset such as LSST's will be essential for such joint analyses. 

The mitigation of systematic errors and joint analysis of multiple probes pose
major challenges in developing the methodology and algorithms for the data analysis. The research and planning required for this undertaking is the primary mission of the DESC. 
The next chapters describe the organization and goals of the DESC, which aim to cover all the key aspects of a long-term plan to tackle the nature of dark energy with the LSST.

%% file: collaboration/collaboration.tex
\chapter[The LSST Dark Energy Science Collaboration]{The LSST Dark Energy Science Collaboration }
\label{ch:collab}

\section[Need]{Need for the Collaboration}

All of the raw data and derived data products from LSST that are described in Section~\ref{sec:surveystrategy} will be made public to the US and Chilean communities, and selected other international partners.  In this sense, the LSST can be more properly viewed as a ``facility'', designed to provide data products, than as an ``experiment'', designed to carry out specific scientific investigations.  The LSST Project Team that has been assembled to design and build the telescope, camera, and data management systems is not a scientific collaboration in the usual sense.  While most of the scientists involved are motivated to work on this project because of their interest in the scientific questions that LSST data can address, they do not ``speak for'' the scientific analyses of those data.  Rather, it is expected that the community will self-organize to perform those analyses and to publish science results once the facility is in operation.

To get things started, a number of quasi-independent scientific collaborations were convened by the Project beginning in 2008.  Their role was to provide advice on technical issues as they relate to specific scientific investigations, and to help articulate the scientific case for LSST as the project made its way through the approval process with the federal funding agencies.  
They were largely responsible for authoring the LSST Science Book  \citep{LSSTSciBook}, which was released in 2009.  At present, there are eleven separate scientific collaborations spanning a broad range of astronomical and cosmological topics.  Five of the eleven (Weak Lensing, Supernovae, Large-Scale Structure/Baryon Oscillations, Strong Lensing, and Informatics and Statistics) have clear relevance to the study of dark energy.  

While these collaborations have been productive, and have provided invaluable service to the Project over the past few years, they have not been formally organized to pursue the full range of investigations required to guarantee that the most sensitive constraints on the nature of dark energy can and will be derived from the LSST data.  As we emphasize below, the various dark energy analyses that we expect to perform are sensitive to a large number of potential systematic uncertainties that must be identified, quantified, and minimized in order for the dark energy investigations to achieve their full potential.  In addition, new algorithms must be developed and tested to enable those analyses, and an efficient computational and software framework must be established to perform the necessary calculations.  All of these activities require an extensive, coordinated research effort well in advance of the onset of data taking.

To address that need, we have created the LSST Dark Energy Science Collaboration (DESC).  The goal of the DESC is to develop a high-level plan for the study of dark energy with LSST data.  This will include the development and optimization of a complete set of dark energy analyses that will be performed with the data, the detailed study of systematic issues that may compromise those analyses, the clarification of the sensitivity of those analyses to various technical aspects of the LSST system design, the generation and refinement of simulations and other tools required to validate the analyses, and the identification and assembly of the computational resources to support such investigations both before and after the onset of data taking.  

An initial ``kick-off'' meeting to establish the DESC was held at the University of Pennsylvania in mid-June 2012.  Since that meeting, over 175 scientists, coming from 49 distinct institutions, have joined the Collaboration.  Their collective expertise covers a wide array of issues relevant to the study of dark energy, ranging from fundamental theoretical cosmology to the detailed characteristics of the LSST hardware and software systems.  Our membership includes not only key scientific leaders of the LSST Project Team, but also key personnel from all of the relevant precursor surveys that will collect data in advance of LSST, and several of the complementary surveys that will be operating in a similar timeframe.  

A large fraction of the members of the five existing collaborations with relevance to dark energy have joined the DESC.  However, we expect that these other collaborations will continue to operate, investigating, for example, issues that are  not related to dark energy.
Maintaining significant overlap in membership between the DESC and these various collaborations will help to ensure that our efforts are coordinated and not redundant.

The DESC will have both informal and formal connections to the LSST Project.  Since many of our members are working directly on the Project, we have direct access to technical information regarding the detailed performance characteristics of the various subsystems.  However, if it becomes necessary for the DESC to communicate specific concerns about the LSST system design that might compromise its effectiveness for dark energy investigations, we will do so officially through the formal channels that the Project has established for interactions with the community. 

\section[Governance]{Governance and organization}

   With the governance structure of a variety of existing collaborations 
as a guide, a governance model for the LSST Dark Energy Science 
Collaboration (LSST DESC) was discussed at the June 2012 meeting 
in Philadelphia. A three-year plan was presented and adopted, beginning with 
a one-year initial phase in which the LSST DESC is established 
with a minimal governance model to get the Collaboration 
off the ground, produce this White Paper, and begin doing science. 
The positions needed for the management team in this minimal model 
were filled by individuals who were ready and able to carry out 
the tasks for the first year.  

  This document describes the first-year minimal governance model, 
which consists of a
management team and an Executive Board. The management team consists 
of the Spokesperson (or co-Spokespersons), Deputy Spokesperson, 
Analysis Coordinator, Computing and Simulation Coordinator, and 
Technical Coordinator.  The Executive Board advises the Spokesperson 
on all scientific, financial, and organizational matters pertaining 
to the Collaboration.

    A Governance Working Group has been 
appointed to continue the deliberations on governance and then 
propose a longer term, more complete governance model for the LSST 
DESC that builds on the one-year initial phase.  In particular, a 
selection process will be proposed for identifying future members 
of the LSST DESC management team.  A Membership Committee will 
also be established to work on more detailed criteria for both individual 
and institutional membership and the procedures for admitting 
new members.
We anticipate that a Collaboration Council will be formed to represent
the full membership of the collaboration.

\begin{figure*}
\centering
\includegraphics[angle=0,width=6.5in]{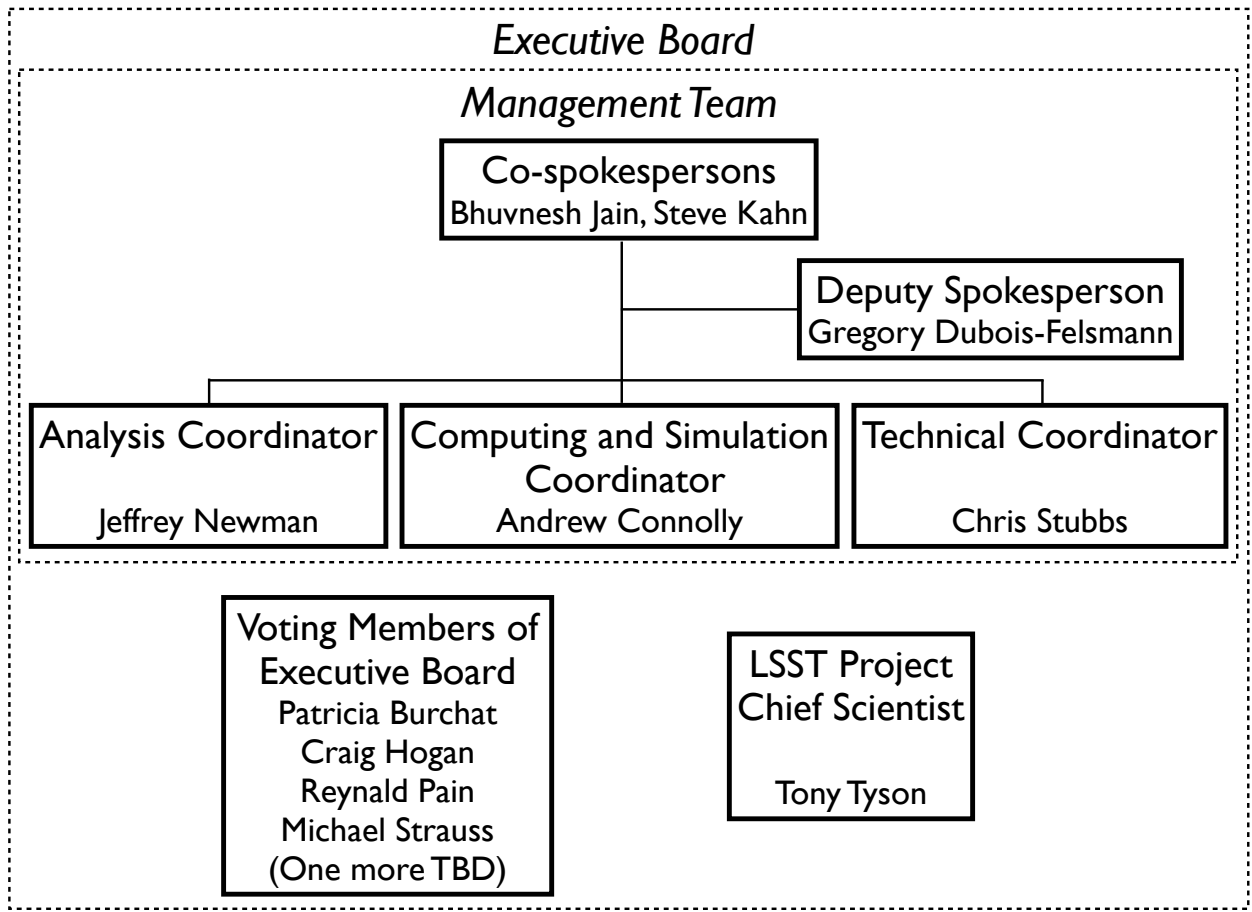}
\caption{
LSST Dark Energy Science Collaboration organizational chart with interim leadership team.
The Executive Board consists of five voting members, the management team, and the 
LSST Project Chief Scientist.
}
\label{fig:OrgChart}
\end{figure*}

     The initial organizational structure is shown in Figure~\ref{fig:OrgChart};
the names correspond to those individuals who will be in their respective 
roles for the first year. After that, members of the leadership 
team will be selected for a term of two-years through a process 
and with roles that will be described in the refined governance model to be 
developed by the Governance Working Group and ratified by the 
Collaboration Council (see Section~\ref{CollabCouncil}).  
It is possible that some terms 
may initially be shorter or longer than two years so that terms
for some positions are appropriately staggered.

The work of the LSST DESC will benefit greatly from interactions with
collaborations carrying out precursor surveys.  
We have appointed four people as liaisons with projects
outside the LSST: 
Josh Frieman, Jason Rhodes, Natalie Roe, and Michael Strauss.

\subsection[ManagementTeam]{The management team}
\label{ManagementTeam}

The details of the management structure, including the specific roles and responsibilities of individual managers,
and the roles and responsibilities of additional Working Groups that may be defined by the management team,
will be established in a Management Plan, proposed by the Spokesperson and ratified by the 
Executive Board.  Here, we outline the general roles of the members of the management team.
\begin{enumerate}
\item {\bf Spokesperson:}
The DESC will be led by a Spokesperson or Co-Spokespersons,
who have overall responsibility for all scientific, technical, organizational, and financial aspects of the Collaboration. 
They are the contact persons for the DOE Office of High Energy Physics and other funding agencies.
The (Co)Spokesperson is the chair of the Executive Board. 

\item {\bf Deputy Spokesperson:}
The Deputy Spokesperson will be responsible for overseeing many of the ongoing operations of the Collaboration.

\item {\bf Analysis Coordinator:}
For the first three years, the Analysis Coordinator is responsible for managing and achieving 
the list of analysis tasks identified in this White Paper as high priority, 
both in specific science areas and in cross-cutting areas. 

\item {\bf Computing and Simulation Coordinator:}
The Computing and Simulation Coordinator is responsible for planning for and coordinating computational resources,
and the development of associated software and simulations necessary to carry out the work of the DESC.

\item {\bf Technical Coordinator:}
The role of the Technical Coordinator is to coordinate the interactions between those with technical knowledge of the components of the LSST system and those on the analysis teams who are improving the understanding of the impacts 
that particular technical issues might have on the science analyses. 

\end{enumerate}

\subsection[Chief Scientist]{LSST Project Chief Scientist}
\label{ChiefScientist}
The LSST Project Chief Scientist is part of the LSST Project team 
and acts as an interface between the LSST Project and the science community.
In the case of the DESC, he/she helps to ensure effective communication of DESC scientific results with the LSST Project. 

\subsection[ExecBoard]{The Executive Board}
\label{ExecBoard}
The Executive Board will consist of five voting members and non-voting {\em ex officio} members 
who represent the management team.
The five voting members will be chosen to bring broad knowledge of the community,
the agencies, and the science goals to the collaboration. 
These members of the Executive Board work with the management team to help define and ratify policy issues as they arise; they do not directly manage the technical and scientific work of the collaboration. 
The (Co)Spokesperson, Deputy Spokesperson, Analysis Coordinator, Computing and Simulation Coordinator,
Technical Coordinator, and LSST Project Chief Scientist are the non-voting {\em ex officio} members of the DESC Executive Board.
The Spokesperson will chair the Executive Board.

\section[Working Groups]{Working Groups}
\label{WorkingGroups}
The areas spanned by the Analysis, Computing and Simulation, and Technical Coordinators are each quite broad.
Therefore, the Management Plan may include the definition of a working group structure for 
efficiently carrying out tasks. 
The initial Management Plan defines a layer of Working Groups under each Coordinator, and an initial set of conveners.
\begin{itemize}

\item
Analysis Working Groups
\begin{enumerate}
\item Weak Lensing --- Michael Jarvis, Rachel Mandelbaum
\item Large Scale Structure --- Eric Gawiser, Shirley Ho
\item  Supernovae --- Alex Kim, Michael Wood-Vasey
\item  Clusters --- Steve Allen, Ian Dell'Antonio 
\item  Strong Lensing --- Phil Marshall
\item Combined Probes, Theory --- Rachel Bean, Hu Zhan
\item Photo-$z$ Calibration --- Jeff Newman (acting)
\item Analysis-Computing Liaison --- Rick Kessler
\end{enumerate}

\item
Computing and Simulation Working Groups
\begin{enumerate}
\item Cosmological Simulations --- Katrin Heitmann
\item Photon Simulator --- John Peterson
\item Computing Infrastructure --- Richard Dubois
\item  Software --- Scott Dodelson
\end{enumerate}

\item
Technical Working Groups
\begin{enumerate}
\item System Throughput --- Andrew Rasmussen
\item Image Processing Algorithms --- Robert Lupton
\item Image Quality --- Chuck Claver
\item Science Operations and Calibration --- Zeljko Ivezic
\end{enumerate}

\end{itemize}
The evolution of these Working Groups and the process for selecting new Working Group conveners will be defined 
by the management team through the Management Plan.

We expect that there will be significant overlap between members of Working Groups under each Coordinator
as DESC collaborators contribute to tools, the project, and science.
For example, the areas covered by the Computing and Simulation Working Groups 
are essential to the work of the Analysis Working Groups. 
The areas covered by the Technical Working Groups parallel the LSST subsystems 
and also interface in essential ways with the Analysis Working Groups.

\section[Policy development]{Development and execution of LSST DESC policies and procedures}
\label{Policies}

\subsection{Governance Working Group}
\label{GovWorkingGroup}

   A Governance Working Group has been established and charged with 
conducting deliberations on and making recommendations for the 
long-term governance, by-laws, and policies of the LSST DESC.  After 
being ratified by the Collaboration Council (see Section~\ref{CollabCouncil} below), 
these will go into effect after the one-year initial phase.

\subsection{Membership Committee}
\label{MembershipCommittee}

   A Membership Committee will be established in the initial 
phase to oversee the membership eligibility requirements for 
the LSST DESC, described below (see Section~\ref{Membership}), 
and to make recommendations
for adjustments to the membership process as suggested by 
experience.

\subsection{The Collaboration Council}
\label{CollabCouncil}

We anticipate that a Collaboration Council
will be formed after the initial phase.  
Representing the full membership of the 
Collaboration, the Collaboration Council will have overall 
responsibility for LSST DESC policies and procedures, including

\begin{enumerate}
\item
         Ratifying the LSST DESC organizational structure, 
         by-laws, and policies proposed by the Governance 
         Working Group.
\item
         Approving proposed modifications or additions to the
         by-laws and policies.  Such modifications or additions
         may be proposed by the Spokesperson or a Collaboration
         Council member.
\item
         Appointing the Committee that proposes the candidate or 
         candidates for Spokesperson, either through nomination 
         or election, depending on the recommendation of the 
         Governance Working Group.
\item
         Reviewing and approving policies on membership and appointing
         a Membership Committee to consider applications for membership.
\item
         Developing a publication policy that is consistent with the 
         LSST Publication Policy 
         and appointing a Publication Board to execute that policy. 
\end{enumerate}

\section[Membership]{Membership}
\label{Membership}

The DESC will have two categories of membership for scientific collaborators: 
full and associate.  

\subsection{Associate membership}

Associate membership is envisioned as providing a path toward full membership;
applicants will be associate members while they develop their detailed 
application for full membership.

\subsubsection{Becoming an associate member}

All those who participated in the creation of the DESC in 2012 and 
signed the membership form are regarded as associate members.  
The signup process defined in the original DESC outreach is still 
available at this time,\footnote{Applications for associate membership may be submitted via Web form at \url{http://www.slac.stanford.edu/exp/lsst/desc/signup.html}.} and a similar online signup process for 
associate membership will be maintained in the future.

The applicant will be asked to 
write a brief proposal (a few paragraphs) for associate membership, 
which will be considered by the DESC Membership Committee.
An individual wishing to join the DESC should use the online process or 
may contact the Spokesperson(s) or the Chair of the Membership 
Committee at any time. 

\subsubsection{Criterion}

Applicants who identify work they wish to pursue that is relevant 
to the DESC will be granted associate membership.

\subsubsection{Rights of associate membership}

Associate membership grants the scientist access to DESC communication 
tools and documents. 
Associate membership does not bring with it access to LSST Corporation (LSSTC) 
communication tools, documentation, or data products.

The associate membership period allows scientists to learn enough 
detail about the activities, needs and coverage of tasks in the DESC 
to write a proposal for full membership.

Initially, no time limit for associate membership will be imposed. 
Beginning one year before the expected start of LSST commissioning,
all persons who have been associate members for at least one year 
will be expected to submit a proposal for full membership or 
become inactive in the DESC.
There will be no minimum time required in the associate member
state before submitting an application for full membership.

\subsection{Full membership}

The transition from associate to full membership is based on a 
written proposal (a few pages) describing the specific contributions the 
scientist is proposing to make; 
these could be tasks outlined in the DESC White Paper, or tasks 
that are argued to be important for achieving the science goals 
of the DESC. 

\subsubsection{Criteria}

The criteria for joining as a full member are: 
(i) the proposed level of commitment, generally expected to be 
at least 30\% of research time over a few years; 
(ii) the importance of the proposed task(s); and
(iii) the need for more effort on the proposed tasks within the DESC.

\subsubsection{Process}

After a transitional period in which the initial setup of the 
collaboration is completed, the process for new applications 
will be as follows.

Faculty and full-time staff scientists must apply for full membership 
and may describe the proposed contributions of students and postdocs 
in their groups. 
Additional graduate students and postdocs may subsequently become 
full members at the request of an existing full member of faculty 
or senior staff rank at the same institution, as long as the 
request specifies an commitment as to the tasks to be performed by 
the new member, consistent with the above criteria.

Proposals may be submitted at any time, and are considered on
a rolling basis throughout the year.

The proposal is considered by the Membership Committee. 
The Membership Committee makes a recommendation to the 
Collaboration Council, which may permit it to take effect
without comment, or may, upon request of Council members
or the applicant, vote on accepting or denying the application.  
If the application is denied, the Membership Committee will
normally
give feedback to the applicant and encourage a resubmission. 
As a general rule we plan to be able to respond to a membership
proposal within two months of its submission.

\paragraph{Transitional process}

During the one-year initial phase, the 12 members of the 
Executive Board (i.e., the Management Team plus the five 
voting members and the LSST Project Chief Scientist) 
plus the 
conveners of the 16 Working Groups will become 
temporary full members of the DESC. 
The Executive Board will constitute the initial Membership 
Committee and may seek advice on membership from the leaders 
of the Working Groups. 

All associate members who responded to the initial call 
for participation in the DESC or subsequently joined 
will be invited to submit applications for full membership. 
After these are processed, a Collaboration Council will be 
formed and a Membership Committee will be appointed by the Council. 
The temporary full members (i.e., the leadership team) 
will then submit their own proposals for full membership 
to be considered by the Membership Committee.

\subsubsection{Rights of full membership}

Full members retain the rights of associate members.
In addition, it is intended that they attain the privileges 
granted to members of LSST Science Collaborations, 
including access to LSSTC communication tools and documentation, 
and internal LSST project data such as simulated datasets 
and the output of Data Management data challenges.

The membership policies of the DESC will be submitted to the
LSSTC Board 
to enable its ratification
of this relationship.
The DESC will regularly inform the LSST Project of the list of
its full members.

Full membership is also a prerequisite for becoming an author of
publications in the name of the DESC.

\subsection{Authorship}

A full publications policy, including criteria for authorship,
will be developed by the Collaboration Council once it has 
been formed.
The guidelines described herein are indicative of how the full
policy may be expected to develop.

\subsubsection{Criteria}

The transition from full membership to authorship status is based on 
contributions to date to the DESC,
demonstrated engagement in the DESC,
and the level of current and future commitment, 
generally expected to be at least 50\% of research time over a few years.

All full members of at least two years' standing
will be eligible to apply for authorship status, beginning one year 
before the start of the LSST commissioning period.

\subsubsection{Process}

The application for authorship will consist of a written 
proposal describing contributions already made and proposed 
future contributions to the DESC effort.

\subsection{Additional topics}

\subsubsection{Members of the Project Team}

The LSST Publication Policy specifies that ``Builders" 
(engineers, scientists, and other individuals whose contributions have been vital to the development
of the LSST infrastructure)
will be invited to add their name to the list of contributing authors 
for key science papers.  

In addition, 
members of the LSST project are encouraged to join the DESC. 
Any member of the project team who wishes 
to become a member of the DESC should apply for associate membership 
for access to the DESC communication tools, etc. 
For further engagement, they should apply for full membership, 
and in their proposal may count their contributions to the LSST project 
in their time commitment to the DESC.
Project members will still be asked to make a specific commitment
to the life of the DESC, expressed by activities such as 
attending meetings, assisting with the development of white papers 
or other reports, and joining one or more of the DESC
working groups.

\subsubsection{International affiliates}

Membership (associate or full) for individuals at institutions
outside the US and Chile (except for IN2P3 in France) cannot be 
granted until the international affiliate institution establishes 
its access to LSST data rights, by MOU with the project.  
Once a named set of individuals at a foreign institution is awarded 
data rights, they are welcome to apply for associate membership 
with the DESC, and we expect that those applications will be accepted.  
Transition to full membership will require engagement and 
an application for full membership, even for individuals from 
international affiliates that have established their data rights.

\subsubsection{Support staff membership}

There will be a category of membership for nonscientific support staff that provides them access to the collaboration databases and communication tools.

\section[Overview of work plan]{General structure of work plan for next three years}
\label{sec:workplan_overview}

As indicated above, key elements of the DESC work plan will involve the investigation of systematic issues that may compromise dark energy analyses, the development and implementation of algorithms in an appropriate software infrastructure to enable those analyses, the development and optimization of large-scale simulation tools necessary for testing such algorithms, and the development of a computing environment capable of supporting all of these calculations.

This is a large and complex effort, and we have found it necessary to prioritize our activities, especially in the early years, when many of our collaboration members are still ramping up their involvement.  Our prioritization is driven principally by considerations of the time urgency for various tasks to be completed.  Time urgency arises from the following considerations:

\begin{enumerate}
\item
While the LSST system design is fairly mature, some elements are undergoing refinement, and there is still opportunity for input based on the implications for dark energy research.  However, since our primary concerns involve subtle systematics issues, these implications are not easily addressed.  In general, they require full end-to-end simulations to evaluate the scientific impacts.  It is very important that this work be performed early, while there is still time to provide feedback on the design.  A coordinated program to carry out the required simulations, focussing on targeted design issues, is required.

\item
For many analyses, the development and testing of an appropriate set of algorithms and associated software tools will involve a multi-year effort with many steps that must be performed serially along the way.  The initial steps in these chains must occur early in order to guarantee completion prior to the onset of data taking. 

\item
LSST will deliver statistical errors that are lower than all precursor surveys by a large factor. This means that systematic errors and subtle physical effects that have not been studied in any detail by the community become relevant for LSST. There is a need for a research effort to scope out and quantify a variety of effects that may be relevant for LSST. Follow up work can then be carried out in subsequent years for the effects that need to be mitigated or studied in more detail. 

\item
It is important, early in our program, to develop an informed sense of the full suite of dark energy analyses that we will pursue. This is crucial for adequately scoping the overall effort and for ensuring that we are budgeting our resources appropriately. The field of dark energy research has been evolving rapidly over the past few years, and a plethora of interesting new dark energy probes have been suggested.  We need to carry out basic feasibility studies to clarify the practicality of some of these new probes, and their relative power for constraining dark energy and modified-gravity parameters.  A mix of theoretical efforts and quick simulations are required to make such assessments.
\end{enumerate}

Other investigations are not only less urgent, but will benefit from the additional experience that the dark energy community will gain from working on precursor projects, several of which are only beginning to take data now.  In general, we have categorized the explicit tasks we have defined as either ``high priority'' (designated H) or ``longer term'' (designated LT) to distinguish between these two classes.  While we expect most of our effort over the first three years to be devoted to the H-tasks, some investment in LT-tasks is also important for balance.

The general structure of our work plan is outlined in the subsequent chapters of this White Paper.  In Chapter~\ref{sec:analysis}, we discuss the different classes of probes, including for each one a brief overview of the technique, a description of the various processing steps that are involved, and a discussion of the major systematic issues that present the most pressing concerns.  In Chapter~\ref{chp:sims}, we review the key simulation and infrastructure tools that will be required for this work, highlighting those areas that are in need of significant additional development.  Finally, in Chapter~\ref{sec:workplan}, we list the tasks that we expect to pursue over the next three years, along with the motivation, explicit description of activities, and expected set of deliverables in each case.

%% file: analysis/analysis.tex
\chapter[The Analysis Framework and Key Systematics for Investigation]
{The Analysis Framework and Key Systematics for Investigation}
\label{sec:analysis}

\section[Weak lensing]{Weak lensing}
\label{sec:analysis:wl}

   \subsection{Overview}

\input{analysis/weak-lensing/overview.tex}

   \subsection{Analysis steps}
   \label{analysis:wl}
   \input{analysis/weak-lensing/analysis.tex}

   \subsection{Systematics}
   \label{systematics:wl}
   \input{analysis/weak-lensing/systematics.tex}

\section[Large-scale structure]{Large-scale structure}
\label{sec:analysis:lss}

   \subsection{Overview}
   \label{overview:lss}

\input{analysis/large-scale-structure/overview.tex}

   \subsection{Analysis steps}
   \label{analysis:lss}
   \input{analysis/large-scale-structure/analysis_requirement.tex}

   \subsection{Systematics}
   \label{challenges:lss}
   \input{analysis/large-scale-structure/systematics.tex}

\section[Supernovae]{Supernovae}
\label{sec:analysis:sne}

\input{supernovae/supernovae_challenges.tex}
\section[Clusters of galaxies]{Clusters of galaxies}
\label{sec:analysis:clusters}

\subsection{Overview}
\input{analysis/clusters/introduction.tex}

\subsection{Analysis steps}
\label{analysis:clusters}
\input{analysis/clusters/analysis.tex}

\subsection{Systematics}
\input{analysis/clusters/challenges.tex}

\section[Strong lensing]{Strong lensing}
\label{sec:analysis:sl}

   \subsection{Overview}

\input{analysis/strong-lensing/overview.tex}

   \subsection{Analysis steps}

\input{analysis/strong-lensing/analysis-steps.tex}

   \subsection{Systematics}
\input{analysis/strong-lensing/major-challenges.tex}

\section[Theory and joint probes]{Theory and joint probes}
\label{sec:theory}

   \subsection{Overview}
   \label{overview:theory}
   \input{analysis/theory/overview.tex}

   \subsection{Analysis steps}
   \label{analysis:theory}

\input{analysis/theory/steps.tex}

   \subsection{Systematics}
   \label{challenges:theory}
\input{analysis/theory/challenges.tex}

\section[Photo-$z$'s and common issues]{Photometric redshifts and other common issues across multiple probes}
\label{sec:photoz}

   \subsection{Overview}
   \input{analysis/photoz/overview.tex}

   \subsection{Analysis steps}
   \label{analysis:photoz}
   \input{analysis/photoz/analysis.tex}

   \subsection{Systematics}
   \input{analysis/photoz/challenges.tex}

%% file: analysis/weak-lensing/overview.tex
Gravitational lensing \citep{2006glsw.book....1S,2010arXiv1010.3829B},
the deflection of light from distant ``sources'' due to the bending of
space-time by masses (``lenses'') along the line of sight, is a useful
cosmological probe because it is sensitive to all matter, whether
baryonic or dark.  Weak lensing (WL) is gravitational lensing in the
limit that the deflections are very small, causing tiny distortions,
or ``shear'', in galaxy shapes
\citep{2001PhR...340..291B,2003ARA&A..41..645R}.  In the absence of
lensing, galaxy orientations are assumed to be random, so they should
not exhibit statistically significant, coherent alignments.  Lensing
induces small ($\lesssim 1$\%) but coherent shears ($\gamma$),
in background galaxy images, which have typical ellipticities that
are far larger (RMS ellipticity $\sim 0.36$).  Thus, WL is detected
statistically by averaging over many lensed galaxies, for which shapes
have been measured in a way that eliminates contributions from
observational effects that also distort galaxy shapes (e.g., the
point-spread function, or PSF).

Measurements of the growth of large-scale structure of the dark
matter field via weak lensing (WL) are sensitive to dark energy, since
the accelerated expansion of the Universe that is caused by dark
energy opposes the gravitational attraction that would otherwise lead
to increased clumping of dark matter structures.  Because of its
sensitivity to both dark matter and dark energy, weak lensing is one
of the four methods advocated by the Dark Energy Task Force, and is a
major driver of survey design for LSST.

%% file: analysis/weak-lensing/analysis.tex
\subsubsection{Processing of LSST data}

Almost every step in the process of extracting shear information from the images of galaxies
has the potential to induce systematic errors, so it is worth briefly reviewing those
steps.  After the basic data processing (bias and overscan subtraction, flat 
fielding, cross-talk correction, interpolation over defects, etc.),
the first step is to detect the stars and galaxies
in the image.  This requires estimating the level of the (spatially varying) sky background,
convolving the image by some target kernel, and then
finding peaks in the convolved image at some level above the background.  
Crowded regions present a challenge;
since the light in any one pixel may come from multiple objects,
some ``deblending'' algorithm is required to determine how much of each pixel's flux
corresponds to each detected object.  

The objects are next classified as stars and galaxies, typically by deriving the probability
of each source being a star.
Once a robust sample of stars is identified, these stars are used to estimate the 
point-spread function (PSF).
We try to obtain as much information as we can about the two-dimensional 
structure of the PSF, since all aspects of the PSF profile will end up affecting the shapes
of the galaxies.  Thus, complicated models are often used, such as a set of 2-d basis functions,
or models of the telescope optics including the relevant Zernicke functions.  %

The PSF varies nontrivially across the image due to both the telescope optics and random 
atmospheric turbulent variation.  Since the galaxies have different locations than the stars 
we use to measure the PSF, we must interpolate the PSF to the location of each  galaxy. 
The simplest way to interpolate the PSF is to fit a function to the measurements in each
image separately.  However, since the PSF patterns are correlated from one image to the next
(being caused by the same optical system), it is often helpful to incorporate information from
multiple images (potentially all LSST images) to better constrain the variation.  
In addition, the PSF varies with wavelength.  Since galaxies have different
spectral energy distributions (SEDs) than stars, the PSF also needs to be interpolated in 
color space, a detail that has generally been ignored in surveys to
date, but which 
will be critical for LSST.

In addition to being convolved by the PSF, galaxy shapes are also distorted by the telescope
optics and atmospheric refraction.  
This distortion is determined from the world-coordinate system (WCS), 
which maps each pixel location to a physical location on the sky. 
The WCS solution comes both from comparisons of the astrometric positions of stars to an
astrometric catalog (e.g., Gaia\footnote{http::/www.rssd.esa.int/gaia/}) and from overlapping
images where the objects are known to be in the same true position (modulo parallax and proper
motion), even if that position is unknown.

We are now able to estimate the size and shape of each galaxy, correcting for the effects of the 
PSF and the distortion.  Since each galaxy will be observed by LSST many times 
($\sim 100$ in each of several filter bands),
we need some way to combine information from the multiple exposures.  This may be done by 
measuring on a stacked image (with the net effective PSF), 
by constraining a single model from all of the original pixels (each with its own PSF and WCS),
or by measuring some parameters on each original image, and then combining them in some way to produce
the final answer.  
Which method will prove to be the best
choice for LSST is still an open question.

Finally, we also care about the brightness of each galaxy and its colors (both for estimating
photometric redshifts and for measuring cosmic magnification).
The optimal algorithm for this step is still being developed, but it may involve using all of 
the images at once, similar to how the shape is estimated.  Furthermore, since a galaxy is 
nearly the same shape in each band pass, the optimal color measurements may use the
images in all the different band passes at once to produce the best estimate of each color.
These fluxes and colors must also be corrected for effects like variable sky transmittance
and extinction by Galactic dust.  

\subsubsection{Additional quantities needed}

In order to properly use the galaxy shape measurements (or sizes/fluxes for 
cosmic magnification), some extra information is required.  First, the predicted gravitational
lensing effect depends on the distance of each galaxy from us; we use
redshift as a proxy for distance.  Since most galaxies observed by LSST will not have spectroscopic
redshifts, we must use photometric redshift (aka photo-$z$) estimates.
For weak lensing, we care less about the specific redshift of each galaxy than we do about having
unbiased estimates of the probability distribution $p(z)$ of each galaxy's redshift.  Thus, our
requirements on the photo-$z$ algorithms tend to be somewhat different
from other probes 
(see Section~\ref{sec:photoz}). 

We also need an estimate of the full
redshift distribution of the galaxies we are using for our statistics (including the rate of
catastrophic outliers).  For some ranges of color/redshift, it can be
challenging to obtain unbiased photo-$z$ estimates, so we may exclude
such galaxies from our 
samples.  This can be an acceptable trade-off between statistical and systematic errors, 
but we need to properly account for such selections by calibrating the final redshift distribution
of the resulting sample to produce the cosmological predictions.

The galaxy shapes require a similar calibration.  The effect of shear on a galaxy's shape
depends on the initial shape.  A given shear cannot make a highly elliptical galaxy 
much more elliptical, so its effect is less than the same shear acting on a round galaxy.
Thus, we need to know the intrinsic shape distribution of the galaxies in our
sample, either to use as a prior during the shape measurement process, or to apply a 
``responsivity'' correction after the fact.  

\subsubsection{Constructing cosmological statistics and covariances}

Working from galaxy catalogs containing positions, photo-$z$ or $p(z)$, and shear estimates for each
galaxy, we can construct cosmological 2-point statistics\footnote{No practical method has yet been
developed to go directly from pixels or catalogs to cosmological parameter constraints without the
intermediate step of constructing such statistics.}.  In a tomographic lensing analysis, there is a
galaxy density field and a shear field in redshift bins (typically between three and six bins), and the most
general analysis that permits us 
to marginalize over systematic errors is to compute all
possible auto- and cross-power spectra, i.e., the galaxy auto-correlation (galaxy clustering), shear
auto-correlation (cosmic shear), and galaxy-shear cross-correlation (galaxy-galaxy lensing), both
within and across redshift bins.  Theoretically, this is best done in Fourier space given that
different modes are uncorrelated;
however, the presence of survey boundaries and masks within
the survey complicates matters, resulting in the frequent use of real-space statistics and the
development of pseudo-Fourier treatments.  In addition to the
cosmologically-motivated statistics, there are also statistics (e.g., $B$-modes, and
star-galaxy cross-correlations) that are mostly sensitive to systematic errors; we must measure
those in order to directly detect and remove systematics in the cosmologically interesting
statistics. 

While two-point statistics are the only ones included in the DETF figure of merit for LSST, they are not the
only cosmologically interesting statistics.  Although they are not explicitly associated with a task
in Chapter~\ref{sec:workplan}, it is important to explore how other statistics besides 2-point functions
can enhance dark energy constraints from LSST.  For example, lensing 3-point functions contain
additional cosmological information and are sensitive to systematic errors
such as intrinsic alignments in different ways than the 2-point functions.  Thus,
3-point lensing statistics can be used both to get additional cosmological information and also to
constrain the impact of intrinsic alignments on the 2-point statistics.
Local maxima in the shear field in lensing maps (called ``lensing peaks'') due to single or
multiple galaxy clusters along the line of sight  provide information similar to that 
provided by galaxy
clusters but are a direct observable of LSST. If systematics can be adequately controlled, lensing peaks will
substantially improve the constraining power of LSST, since they 
include otherwise inaccessible information beyond that in 
2-point statistics. Also, 
because of the great depth of LSST, lensing peaks due to multiple halos along the line of sight will
be prevalent; these have a large share of the
constraining power, and techniques to exploit them will not be required by shallower near-term
surveys.   
Finally, we consider lensing magnification
of fluxes and/or sizes.  As discussed below, the
different sensitivity to systematic errors makes lensing magnification complementary to shear in
several ways.

We must construct not only the cosmological statistics, but also their
covariance matrices (including 
cross-covariances between different statistics).  For this purpose, it is
important to include non-Gaussian effects, which can significantly increase the diagonal terms of
the covariance matrix on the scales used for cosmological parameter constraints.  Furthermore, we must include off-diagonal terms, including those from correlated
shape noise, masking effects, etc.  
This will require a combination of simulation and analytic theory,
including halo-model based covariance estimates; for more detail, see the Theory section (Section~\ref{sec:theory}).

\subsubsection{Cosmological parameter estimation}

Finally, we compare the data with theoretical predictions in order to constrain cosmological
parameters.  To begin the process, we choose a set of cosmological parameters for various CDM, dark
energy, and/or modified gravity models, plus massive neutrinos, primordial non-Gaussianity,
etc. Then we use an emulator (see Section~\ref{sec:theory}) to predict the measured
quantities as a function of the cosmological parameters of interest,
which could involve 
ray-tracing through cosmological N-body simulations generated for a grid of cosmological
parameter sets and interpolating the results to other values of cosmological parameters.

However, the observed statistics include not only cosmological information; there are also 
``nuisance'' effects.  These include observational systematics such as
multiplicative and additive errors in the shear, and photo-$z$ errors; and theoretical uncertainties
such as nonlinear galaxy bias (the relation between observable galaxy clustering and invisible dark matter clustering), 
intrinsic alignments of galaxy shapes with the density field, and
baryonic effects on dark matter halo profiles.  The process of handling these nuisance parameters
includes:
\begin{itemize}
\item Identifying models to describe them.  For example, these could be models based on perturbation
theory for the nonlinear galaxy bias, or atmosphere and/or optics models to describe
how correlations in PSF anisotropies scale with angular separation; or, for some nuisance effects,
we may adopt a flexible but non-physically motivated model with a significant number of free
parameters.  
\item Identifying the range of scales for which we believe our combined cosmological plus
systematics model is valid.
\item Incorporating the nuisance effects into the emulator to calculate how they affect the measured
statistics, as a function of the values of the nuisance parameters.  The procedure for doing so
will depend on the nature of the nuisance effect.
\end{itemize}

We can compare these predictions with the data (given the covariances) to constrain cosmological
parameters while marginalizing over systematic errors.  Given the cosmological parameters and a
flexible model for systematic errors, the parameter space could easily
have hundreds of dimensions.  Parameter constraints in such situations are often done using 
Markov Chain Monte
Carlo (MCMC) to handle the many-dimensional likelihood surface, though this is by no means the only
possible way of doing so.

%% file: analysis/weak-lensing/systematics.tex
\subsubsection{Major systematics and/or ones of unknown magnitude associated with tasks}

For the next few years we plan to focus on systematic uncertainties that satisfy at least one of the
following conditions: 
they are of sufficient importance that we need strategies to ameliorate them, 
but Stage III surveys will not address them at the level required by LSST (or at all);
the systematic is of totally unknown magnitude but could conceivably cause significant issues for LSST; 
or the systematic could motivate a change in LSST survey
strategy (either hardware, observing strategy, or data management).

\paragraph{PSF centroid bias}
\label{sys:wl:psf_centroid}

We are concerned about several potential systematic errors in determinating the PSF.
Since the galaxy images are effectively deconvolved by the PSF to determine their shapes, 
systematic errors in the PSF translate into systematic errors in the shear (or size).
One error that likely will be important for LSST, but which surveys to date have not needed to 
address, is the centroid of the PSF.  Typically, when measuring the shapes of stars, one 
centroids the stellar image before measuring the PSF shape.  This is normally adequate, since we
do not care about any translation of the galaxy positions by the PSF.  However, LSST
will be combining many exposures to measure the galaxy shapes, so the translation component of 
the PSF will cause each galaxy image to be shifted slightly, leading to an effective additional
blurring (convolution).  If this is not accounted for, then the shape estimate will be systematically
rounder than the true shape, leading to a bias.  The solution is to allow the PSF
to include a small centroid shift as part of the convolution, but this requires a substantially different 
algorithm for PSF estimation.  One would have to fix the centroid of each star either using 
an astrometric catalog for the true position, or just using one single fitted position for all 
observations of that star.  However, we are uncertain whether this effect causes a large enough
shear bias that such a change in PSF estimation algorithm is required.  Since we need to inform LSST
Data Management whether the more complicated algorithm 
is required, estimating the magnitude of this effect is one of our short-term priorities.

\paragraph{Small-scale PSF variation and interpolation}
\label{sys:wl:psf_small_scale}

The variation of the PSF on scales greater than about one arc-minute is generally
well constrained by the stars observed in the image.
However, both the telescope optics and  atmospheric turbulence include 
non-negligible smaller-scale PSF variation.  
The atmospheric contribution is particularly 
difficult to estimate, since the effect is random, so existing PSF estimation algorithms provide
no information about the pattern at locations between the observed stars.
Optimal estimators, as well as statistical prior knowledge of atmospheric properties, may be
able to help in this reconstruction.
It is also possible that additional information, obtained from the LSST wavefront and
guidance sensors, may help constrain the PSF pattern on scale smaller than the mean stellar 
separation.

\paragraph{Chromatic PSF effects}
\label{sys:wl:psf_color}

The PSF is wavelength dependent, which means that PSF models constructed 
from stars are strictly speaking correct only for objects with the same SED, and not 
for arbitrary galaxies. The problem is further complicated when the galaxies have 
multiple  components with different SEDs, such as bulge-disk differences, star-bursting regions, 
large HII regions, etc. These chromatic effects may introduce systematic errors in 
shear measurements. 

\paragraph{Star/galaxy separation}
\label{sys:wl:sg_separation}

Weak lensing measurements depend critically on the accurate measurement of the shapes of galaxies. 
As described above, part of this process is PSF deconvolution, which has its own large set of 
associated systematic issues.  
In order to deconvolve the PSF, however, the PSF must be reconstructed using a very pure sample of stars, since interloping galaxies 
in the sample do not have the same shape as the PSF. %
The star selection efficiency is also important. 
The number of stars required to reconstruct the PSF to the level required for LSST WL systematics 
will depend on the algorithm used to model the PSF (see~\cite{2009A&A...500..647P} for an 
approach to this based on the complexity of the chosen PSF model). 
We must evaluate the star sample purity and efficiency that will be required for 
optimal performance of the WL algorithms. 

Galaxy selection is also important here -- if the galaxy sample is contaminated by misidentified 
stars then this will have an impact on the cosmological information we can extract from the data. 
The selection efficiency for galaxies is less important, as long as the effective number of 
galaxies remains high enough to support the planned WL measurements to the required precision. 

\paragraph{WCS}
\label{sys:wl:wcs}

The world coordinate system (WCS) describes the transformation from sky coordinates (R.A., Dec.) to
chip coordinate ($x$, $y$).  Each galaxy has a single position in the sky, but every observation will
be measured at a different position in the image plane.  Thus, errors in the WCS can result
in systematic errors in the shape measurements.  We consider two specific effects.  First, the Jacobian of the transformation is a $2 \times 2$ matrix, which
can be described as a dilation, a rotation, and a distortion.  These affect galaxies differently
than the convolution by the PSF, so errors in the WCS Jacobian will wrongly ascribe some of the
distortion to the PSF, which means the galaxy shapes will be incorrect.
Second, when combining the data from many images of the same galaxy, small errors in the 
centroid on each exposure will lead to an effective blurring of the combined image.  This will
systematically reduce the ellipticity of galaxy shape estimates, leading to a bias if it
is not properly taken into account.  This effect is similar to that of ignoring the centroid shift
of the PSF described above, although the source of the effect and the algorithm 
to properly correct for it is different.

\paragraph{Photometric calibration}
\label{sys:wl:photometry}

Cosmic magnification is measured by computing the cross-correlation
between the density variations of foreground (lens) and distant
background (source) galaxies, selected to be physically unrelated.
There exist a number of systematic/observing errors that can mimic
this signal.  An important example is overlap of the redshift
distributions of the lens and source populations, leading to a much
larger cross-correlation than expected.  This overlap can arise due to
inaccurate photometric redshifts or catastrophic outliers. Other
potential sources of systematic biases include seeing variations,
stellar contamination and dust extinction.  The cosmic signal,
however, does present specific dependence on both the redshift of the
two populations and on the source galaxy counts slope that we can use
to help separate out the impact of these systematic biases.  This
remains an open area of study.  Important areas where magnification
drives analysis requirements are photometric calibration (in
particular spatial variations of residuals), photometric redshift
determination, and star-galaxy separation.

\paragraph{Deblending}
\label{sys:wl:deblending}

LSST will obtain image stacks of unprecedented depth for a WL survey. As survey depth
increases, so too does the fraction of overlapping galaxies. At the faint limit $i \lesssim 26$,
corresponding to $\simeq 94$ galaxies per square arcminute, 99.5\%, 74\%, and 24\% of randomly placed
circular apertures with 4$\arcsec$, 2$\arcsec$,  and 1$\arcsec$ radius are overlapping, respectively. This
high degree of overlap drives the analysis of LSST observations into a new regime where the whole sky is effectively one large
``blend'', and traditional approaches to deblending, or even new methods developed for precursor
projects, most likely will not be sufficient. Any method for measuring shapes in highly blended
images must identify pixels with flux from overlapping objects and either mask or appropriately weight them. With overlaps
occurring predominantly around the edges of galaxies, where the sensitivity to ellipticity
parameters is largest, possible shape-measurement biases must be carefully studied and quantified,
taking into account correlations between the effects of overlaps and galaxy ellipticity, surface
brightness, and size.  The LSST Data Management currently does not have any
WL science-driven requirements on the deblender, and we must address this.

\paragraph{Challenges from multi-epoch data systematics}  
\label{sys:wl:multiepoch}

LSST data will consist of several hundred short exposures of each location in six bandpasses. 
The naive use of such data would be to create a single image stack on a common grid, 
and then to correct galaxy shapes, with the information on systematics extracted from the same 
image stack.
In fact, this method has been widely used in previous ground-based WL surveys, 
for which the number of exposures is small and the pattern of the systematics (e.g., spatial 
variation of PSF) is repeatable to some extent. 
However, stacking is a lossy data compression method, and a significant amount of 
information regarding instrumental systematics (e.g., time- and position-dependent PSF 
variation, image alignment error, residual geometric distortion, flat-fielding error, etc.) 
is irreversibly destroyed during the process. 
To overcome the pitfalls of this conventional approach and obtain both unbiased and 
accurate galaxy shapes from the data, we must apply considerable effort to develop 
optimal multi-epoch data processing algorithms and also efficient dithering 
strategies between telescope pointings.  

\paragraph{Photo-$z$}
\label{sys:wl:photoz}

While the magnitude of the weak lensing shear depends on distances to both the lens and source,
the kernel is quite broad in redshift and thus relatively wide redshift 
bins can be used to study the evolution of the weak lensing signal.  This means the accuracy 
of individual photometric redshifts is not very important, as long as the distribution of 
photometric redshift errors is known to high precision. Therefore, one must accurately 
determine the mean redshift and catastrophic outlier fraction of galaxies in each redshift bin, 
along with the effective bin width.  To achieve this we ideally require an unbiased likelihood 
distribution of possible redshifts for each galaxy.

To inform the photometric-redshift team of appropriate requirements for their algorithms,
we must determine to what precision the photo-$z$ distribution must be known in order to avoid
degrading the constraints on dark energy parameters.  Then, given this precision, 
we need to investigate the 
resulting requirements for spectroscopic observations to calibrate the photometric redshifts.  
The spectroscopic sample must be robust, complete, and accurate so it does not itself introduce bias 
on the cosmological constraints \citep{2012arXiv1207.3347C}.  It must also be drawn from a wide 
enough area of sky to minimize sample variance in the photo-$z$ calibration \citep{2012MNRAS.423..909C}.  \\

\paragraph{Measurement issues associated with galaxy shape measurement and PSF correction}
\label{sys:wl:galaxy_shape}

Several systematic errors are related to measurement of the galaxy shape and correction
for the effects of the PSF, which we recognize here for completeness, but which we 
deem low priority for the LSST DESC in the next three years because progress by Stage III surveys and other
groups is likely to be sufficient:

\begin{itemize}
\item Multiplicative and additive shear errors: in addition to
 progress by Stage III surveys, lensing groups worldwide are
 participating in an ongoing series of blind challenges (GREAT08,
 GREAT10, GREAT3) to identify and minimize these errors.  These
 challenges have not yet assessed systematics in
 magnification analyses, but are likely to soon.
\item Underfitting bias: a bias results when galaxy models attempt to
 use small spatial scale information that has been destroyed by the
 PSF.  This issue has seen rapid progress by groups around the world
 since being identified \citep{Bernstein2010,VoigtBridle2010}.
\item Noise rectification biases: because shear estimation is a
 nonlinear process, pixel noise causes not just scatter but
 also shear biases, which must be controlled \citep{MelchiorViola2012}.
\item Charge-transfer inefficiency (CTI): space-based surveys have led the
 way in mitigating the effects of CTI.  Ground-based surveys are less
 affected because the higher background usually fills the traps in
 the sensor.  This is an area of low risk for LSST.
\end{itemize}

LSST must ultimately reduce these systematics more than Stage III
surveys will, but LSST is similar enough to Stage III surveys that we
will be able to leverage their experience.  In fact, for some of
these systematics (particularly additive shear and CTI), LSST's large
number of independent, dithered, and rotated exposures substantially mitigates the risk. In Chapter~\ref{sec:workplan}, we include a long-term task associated with testing the feasibility of using shape measurement methods from Stage III surveys for LSST, to address any of the above issues.

\paragraph{Lensing peaks}
\label{sys:wl:peaks}

Lensing peaks in shear maps will be an important probe for constraining dark energy significantly beyond the LSST
baseline figure of merit if systematics can be adequately controlled. Near-term surveys are not
likely to address issues for lensing peaks. If lensing peaks are to be a key probe for LSST, we need
to quickly establish this fact and the computational model of how the analysis will be done, since
these decisions might drive computational requirements. To establish lensing peaks as a key probe
for LSST it is necessary to delineate the effect of LSST-specific systematic errors on lensing
peaks: shear errors, photo-$z$ errors, masking, variable depth, baryon effects, and intrinsic alignments,
and to determine the effect of these errors on cosmological inference for LSST when lensing peaks
are included. Important work on these topics is ongoing \citep{2010PhRvD..81d3519K,2011PhRvD..84d3529Y} and preliminary results are positive.

\subsubsection{Systematics without tasks in Chapter~\ref{sec:workplan}}

Below we briefly discuss some weak lensing systematic errors that are not explicitly
associated with tasks in Chapter~\ref{sec:workplan}.

\paragraph{Object detection}
\label{sys:wl:detection}

The detection of galaxies for shape estimation is subject to a number of potential systematics 
that can bias weak lensing measurements.
One bias arises from a
correlation between the shape of the PSF and the shape of the galaxy.  A galaxy whose ellipticity 
is aligned with the ellipticity of the PSF will be detected at a higher S/N than an otherwise 
identical galaxy whose major axis is 90 degrees misaligned with that of the PSF 
\citep{kaiser00, bernstein02}.
This ``PSF selection bias'' can be mitigated by defining a shape-independent galaxy selection 
criterion. 

A different selection bias comes from the fact that detection algorithms preferentially detect round objects over highly elliptical objects of the same flux and size.  As a result, galaxies that have been gravitationally sheared orthogonally to their intrinsic shape are more likely to be detected.  This leads to a ``shear selection bias'' that is proportional to the lensing signal, but has the opposite sign.  The net result is a reduction in the inferred amplitude of the lensing signal \citep{hirata03}.  
In addition, size cuts to eliminate stars may preferentially retain sheared galaxies, increasing the measured signal amplitude. 
Priors on the distributions of galaxy sizes and shapes, which could be constrained from the higher S/N deep-drilling fields,
will be helpful to reduce both selection and shear calibration 
biases \citep{miller07, kitching08, voigt10, melchior10, bernstein10}.

Masks around field boundaries, bright stars, and image artifacts can also introduce biases by 
rejecting elliptical objects that overlap the mask boundaries.
For example, a bad column or stellar saturation bloom will intersect more galaxies whose 
major axes lie parallel to rows in the image than it does galaxies whose major axes are aligned 
with columns \citep[c.f.][]{huff11}.
The large number 
of exposures of each object in LSST will help to mitigate many biases from image artifacts.

Finally, there are subtle issues with the likelihood of detecting an object, correlated with
variable sky levels, proximity to neighbors, or the local slope of the sky background 
(for example in the wings of a bright foreground object).  We do not expect LSST to have a 
more difficult time dealing with any of these biases than Stage III surveys.
However, they merit some attention once 
the results from the Stage III surveys are known and their correction methods can be analyzed
using LSST-like simulations.

\paragraph{Predictions}
\label{sys:wl:predictions}

Theoretical predictions for lensing statistics are challenging for three partially related 
reasons: the nonlinearity of the underlying mass density field on scales of interest, the 
need for predictions of quantities that go beyond the power spectrum, and the inclusion of 
physics that is not well understood such as gas physics in galaxy clusters and intrinsic 
alignments. The common tools used to tackle these challenges and obtain predictions of lensing 
statistics (and their covariances) are simulations and the analytical halo model. N-body 
simulations are a mature tool and computational resources enable large numbers of simulations 
to be performed -- this is essential for obtaining accurate covariances. However spanning a 
multi-dimensional parameter space remains a challenge and emulators are being developed as 
described in Section~\ref{sec:cosim}. Tackling higher order correlations and other statistics and 
modeling baryonic physics is useful for multiple reasons and is part of our ongoing and 
planned work -- see discussion in Section~\ref{sec:theory}.

\paragraph{Constraints from the data}
\label{sys:wl:cosmology_constraints}

LSST lensing data will be an essential component in constraining dark energy's effects on the 
growth of large structure. The clustering of galaxies provides a measure of how nonrelativistic 
species move through the large scale structure gravitational potentials, assuming galaxy
bias is well characterized.
Lensing gives a different perspective, showing how photons 
move through the potentials. While in General Relativity the two probes will be consistent,
evidence of inconsistency 
in these  tracers could well be a signal of a modification to gravity on cosmic scales 
\cite{Zhang:2007nk}. To extract these powerful signatures of dark energy requires careful 
characterization and simultaneous fitting for degenerate astrophysical systematics such as 
intrinsic alignments,
stochasticity in shear-position correlations, nonlinear correlations and relativistic effects. 
Quantifying the dark energy constraints in light of these systematics will be a primary focus 
of the Theory/Joint Probe Working Group, and is discussed in more detail in 
Section~\ref{challenges:theory}.

%% file: analysis/large-scale-structure/overview.tex
Measurements of galaxy clustering constrain the cosmic expansion history, the cosmological distance scale, the growth rate of structure, the mass of the neutrinos and  the abundance of dark matter. 
Baryon acoustic oscillations (BAO) in the early Universe imprint a standard ruler in the power-spectrum (or the two-point correlation function), which measures the angular-diameter distance and the Hubble parameter. 
BAO is a special case of using the full shape of galaxy clustering data to constrain cosmological 
models. 
The BAO feature has now been observed in the SDSS, 2dF, and BOSS surveys, 
using both spectroscopic \citep[e.g.,][]{eisensteinetal05,anderson12,xu12}
and 
photometrically selected galaxy samples
\citep{Padmanabhan06,Ho12,seo12}, and has proven to be a robust probe of dark energy. 
A second probe of dark energy, and a potential discriminator between 
dark energy 
and
modified-gravity models, is the growth of large-scale structure in the Universe. 

The galaxy distributions measured by LSST will reveal
information from both of these classes of probes (see \citealp{SciBook}, Chapter 13 for details). 
LSST's most sensitive measurements of the 
growth of structure will involve 
cross-correlations of the galaxy distribution with the shear field 
measured by lensing (so-called galaxy-galaxy lensing) or with the cosmic microwave background (the integrated Sachs-Wolfe effect).
Cross-correlations with external spectroscopic surveys will also
lead to powerful tests of modified-gravity models. 

Beyond dark energy, the large scale power spectrum is a probe of both neutrino mass and primordial non-gaussianity. 
For all these purposes, identifying and removing systematic effects in the maps of the galaxy density, especially on large scales, will be crucial.

%% file: analysis/large-scale-structure/analysis_requirement.tex
\subsubsection{Basic data processing by LSST DM}
In studies of large-scale structure, galaxies are treated as point sources; 
hence, the relevant LSST Data Management output is a 
catalog of the angular position of the centroid of each galaxy,  the galaxy flux, the background flux,  
the radial distance (via photometric redshift), and the galaxy type determined from the $ugrizy$ colors.  
For all of these quantities, the respective errors must also be accurately estimated. 

In practice, the steps towards an accurate estimate of the intrinsic source density are highly non-trivial. 
In this section, we outline only some of the significant steps that can severely affect the data products 
relevant to studies of LSS, since there is a significant discussion in the WL section. 
After the basic data processing (bias and overscan subtraction, flat
fielding, interpolation over defects, etc.),
in order to detect the stars and galaxies
in the image, we need to estimate the level of the spatially and temporally dependent sky background,
convolve the image by some target kernel, and then
locate peaks in the convolved image at some level above the background.  
This is especially tricky when we have a bright foreground object, 
which can lead to an incorrect estimate of the sky background around the bright source. 
The varying sky and seeing conditions will also make this more challenging, as the kernel with which we should convolve
the image is also changing spatially. 
We also need a proper deblending algorithm as light from any one pixel may come from multiple objects.
This will alter our estimate of not only the brightness of the object, but possibly also 
lead to a misestimation of the colors, and therefore the galaxy type. 
In addition, the estimate of the size of the object can be affected by varying PSFs, possibly leading to misclassification as a 
star or galaxy.
Misclassification of the sources can lead to spurious power over different scales of the survey, reducing our ability to use the angular power to constrain the scale of the 
BAO peak or the shape of the overall power spectrum.

\subsubsection{Additional inputs needed}
Interpreting angular power spectra and correlation functions requires that the radial (redshift) distribution of galaxies in 
each photo-$z$ bin be well-determined.  
Several cosmological parameters that will not be determined directly by LSST but rather by CMB and other astronomical probes are also needed as input, including the primordial power spectrum, 
the BAO scale, and the baryon-to-photon ratio.  

A number of complementary external (Stage III and IV) surveys, some contemporary with LSST, will also overlap with the LSST survey area. Spectroscopic surveys, such as BOSS, HETDEX, eBOSS, BigBOSS, DESpec, Euclid, and WFIRST will observe the same galaxy population as LSST and will provide peculiar velocity measurements, to contrast with the relativistic tracer of the large scale gravitational potential given by the lensing shear.  They will also facilitate improved calibration of LSST's photo-$z$ algorithms. CMB lensing and the integrated Sachs-Wolfe effect will provide measurements of relativistic tracers of the gravitational potentials complementary to LSST's shear survey, to constrain modified gravity models and to constrain shear systematics that contaminate the cosmological lensing signal. 
Additional joint analyses of dark energy and modified gravity will be enabled by combining LSST 
with Stage III-IV spectroscopic-redshift surveys.
By probing redshift space distortions, these spectroscopic surveys enable a direct test of the logarithmic derivative of structure growth as a function of spatial scale \citep[e.g.,][]{acquavivag10, samushiaetal12}.  

\subsubsection{Constructing cosmological statistics}
The standard techniques for measuring the clustering of sources in a photometric survey like 
LSST are the angular two-point correlation function and its Legendre transform, the angular power spectrum~\citep[for a complete discussion see][]{Peebles}. Recently, these techniques have been augmented by using photometric redshifts to construct pseudo-volume-limited samples~\citep{Ross07,Hayes12b} that can provide more precise constraints on dark energy~\citep[e.g.,][]{Eisenstein05}. The two-point correlation function quantifies the excess probability over random of finding one source within a given angular distance of another source and is typically estimated  using the Landy-Szalay estimator, which uses binned pair counts for the source-to-source, source-to-random, and random-to-random distances~\citep{LS93}. Naively this is an O($N^2$) process; however, the performance can be dramatically improved by leveraging tree-data structures and parallel computing~\citep{Moore01,Dolence07,Wang12} or by pixelating the data and calculating the moments of the pixelized counts~\citep[see, e.g.,][]{Ross06}. Likewise, the angular power spectrum can be estimated from a pixelized source distribution~\citep{Tegmark02,Padmanabhan06,Hayes12a,Ho12}. There had been some developments in the optimal estimation of both the angular correlation function or the optimal power spectrum in large-scale structure as implemented in ~\citep{Padmanabhan06, Ho12}.

Source data must be properly masked to the survey window function, 
including deweighting or removing
areas within the survey that are severely
affected by 
significant seeing variations, excessive extinction from Galactic dust, obscuration by bright foreground stars or nearby galaxies, 
or 
general instrumental defects. 
Random catalog data must be masked in the exact same manner as the data in order to prevent false power from being introduced in the estimation process~\citep{Scranton02, Wang12}. Higher order correlation functions (or their power spectra analogues) can be used to constrain 
any scale-dependent bias between the source population and the underlying dark matter distribution.

\subsubsection{Cosmological parameter estimation}
Once the angular correlation function or 
angular power spectrum has been measured, the next step is to 
compute the theoretical expectation for the signals, in order to constrain cosmological models 
~\citep[see, e.g.,][]{Ho12}. 
The standard practice for this 
has been 
to start with a theoretical 3-d matter power spectrum, which can be calculated using the Code for Anisotropies in the Microwave Background~\citep[CAMB][]{Lewis00}.
This 3-d power spectrum must be projected into angular space 
for comparison with 
the observed results, which requires an integration over redshift of the convolution of the 3-d power spectrum with the growth function, the sample bias, the Hubble function, the comoving distance, and the source number 
density~\citep{Hayes12a,Hayes12b,Ho12}. 
A Markov Chain Monte Carlo process~\citep{Lewis02}, for example, can then be applied to the calculation of this theoretical clustering statistic, often in conjunction with cosmological parameter measurements from other projects, 
to produce constraints on the cosmological parameters of interest~\citep{Ho12}.

%% file: analysis/large-scale-structure/systematics.tex
Here we describe the major challenges to obtaining dark energy measurements that the LSS working group must address by 2020, emphasizing those in need of immediate attention.    

\subsubsection{Major astronomical systematics and/or those of unknown magnitude}

\paragraph{Dither pattern.}
Dithering 
refers to utilizing a series of slightly offset telescope pointings to fill in gaps between CCD sensors to create a complete image of nearly uniform depth.  
Large dithers of half the 3.5$^\circ$ field of view could also be used to make the survey coverage more uniform as the sky is tiled with various pointings.   
Because the LSST camera has an instrument rotator, 
rotational dithers 
provide 
an additional mechanism for varying systematics caused by anisotropic PSF, 
filter nonuniformity, and detector asymmetry.
A fundamental question being explored for LSST is whether to utilize all of these dithers to achieve a nearly uniform exposure map upon completion of the 10-year survey, or whether this is outweighed by the advantages for transient searches of nearly fixed pointings.  
Dithering is a significant concern for LSS, as any spurious structure that is introduced on scales on the order of the LSST field of view could alter the BAO feature, which occurs on similar angular scales.  

\paragraph{Atmospheric conditions:  throughput, sky brightness, and seeing.}
The LSST observing cadence will visit 9.6\,square-degree patches of the sky in a dithered way on different days in different wavelength bands and at different zenith angles under different observing conditions. The resulting raw overlapping images will have the discontinuous effects of variations in sky brightness, temporal and spatially varying glints and ghosts from bright stars, total throughput variations due mostly to clouds, and variations in the PSF and its angular correlations.   The LSST pipeline, automated data quality assessment (ADQA) monitoring, and calibration protocol will correct for all these effects down to some residual.  The largest surviving effect will be the patchy {\it variances} in PSF and limiting surface brightness. We need to study the impact on LSS science of the power spectrum of these residuals
and 
their covariances. 
For example, patchy limiting magnitude and color errors will affect photo-$z$ and WL magnification in patches correlated on the sky. 

\paragraph{Stellar contamination and obscuration.}
\label{sec:stars}
LSST is a photometric survey, and thus we depend strongly on the color information of each object for its identification.
The color of stars can mimic the color of other extragalactic objects. 
Since the distribution of stars 
has significant angular correlations on much larger angular scales than galaxies,
when stars contaminate our extragalactic samples, we will observe an excess in the large scale. 
On the other hand, 
imperfect background estimation
around 
bright 
stars 
increases the threshold for detection of 
nearby 
objects. 
This will induce an anticorrelation between stars and galaxies, 
and hence another 
source of anomalous large-scale power of extragalactic objects.
We need to study, estimate, and reduce the impact on LSS science of the power spectrum of these effects.

\paragraph{Photometric calibration.}
The need for photometric calibration in LSS is driven by the precision in large-scale power required for the survey.  
Thus, we need to study, estimate, and reduce the impact of photometric offsets in LSST by utilizing large scale simulations that  include templates for photometric offsets. The templates can be created using the small scale simulations in ImSim to investigate the effects of varying conditions on the throughput of each of the photometric bands in combination with  using OpSim to understand the large scale variations across the sky even after the calibration. 

\paragraph{Galactic extinction.}
One of the major challenges to extragalactic astronomy is correcting for wavelength-dependent dust extinction caused by the foreground screen of the Milky Way.
Several methods have been used to 
correct for the effects of dust, but all have significant residual errors that can create structure on angular scales relevant to LSS.  By causing variations in survey depth, dust extinction in the detection band imprints artificial large-scale structure in the observed distribution of galaxies, which cannot be corrected but must be modelled.  The correlated variations in observed galaxy colors can be corrected, but color-selected galaxy samples will have additional angular correlations imprinted by these.  

\paragraph{Photo-$z$ uncertainties.}
Uncertainties in galaxy photometric redshift estimates will suppress the measured power on radial scales up to, and including, the BAO scale.
Radial modes of the largest scales will not be so strongly affected and will be useful for measuring the scale of the matter power spectrum turnover.  The BAO, on the other hand, will have to be measured in a series of redshift shells.  
tO interpret the galaxy clustering signal correctly it will be important to obtain a sample of galaxies with photometric redshifts accurate enough to allow the sample to be confidently split into these radial bins, out to high redshift.  
The effect of the photo-$z$ uncertainties will be to cause these bins to overlap in redshift, giving rise to 
cross-bin correlations.  
We will use this cross-bin 
clustering signal to calibrate the photo-$z$ error distribution~\citep{SciBook}. In addition we can
 perform a cross correlation with available overlapping spectroscopic survey data~\citep{newman08}.
Further study into the utility of these two methods needs to be undertaken with realistic LSST photo-$z$ errors, beyond the Gaussian model, and also assuming a likely spectroscopic calibration sample.
We need to investigate the required photo-$z$ precision and
the maximum allowable bin contamination rate, and to develop methods to deliver this precision and accuracy.

\subsubsection{Systematics from theory}

\paragraph{Beyond linear models.}
The convolution described at the end of Section~\ref{sec:analysis:lss} highlights the importance of increasing the precision of our measurements of the mildly nonlinear bias between the 
galaxy samples
and the underlying dark matter, and of the photometrically determined redshift distribution of the source population. 

Nonlinearities in the power spectrum of the underlying dark matter are caused by the nonlinear evolution of components of the Universe, especially the late-time evolution of matter and baryons. To capture the full extent of the nonlinearities, with a lack of full-fledged nonlinear evolution theory, 
we
will need to simulate the evolution of most, if not all, of the components of the Universe. Extensive research and discussion have been carried out 
on
 multiple fronts \citep{Sanchezetal2008,Sanchez:2009jq}, whether it is by perturbation theory \citep{Carlson:2009it}, dark matter simulations \citep{Hamaus:2010im,heitmann09}, or fitting functions suggested by dark matter simulations \citep{Smith:2002dz}. 

\subsubsection{Cosmological simulation needs}

Cosmological simulations are necessary inputs on multiple fronts in LSS studies.
On one hand, cosmological simulations provide us with predictions of the galaxy power spectrum.
 We will need to use dark-matter simulations to calibrate the theoretical power spectrum. For example, \cite{heitmann09} has simulated the 3-d galaxy power spectrum for a large range of cosmologies (see more discussions on this in 
 Section~\ref{sec:prediction_sim}). An expanded set of these simulations for different types of galaxies would be very useful to 
help 
calibrate the theoretical power spectra.
On the other hand, cosmological simulations will act as inputs to the mock catalog productions on a much larger scale, which will require larger volumes with comparable mass resolution to that which is currently used in the mock generation.

\subsubsection{Improved software/algorithmic needs}

Achieving the statistical precision 
possible for LSST 
requires 
a reconstitution of software analysis tools and methods that will have been applied to Stage III experiments. Every step of the analysis of the galaxy distribution needs to be tested for precision: two-point function estimations, generation of (non-diagonal) covariance matrices, form of the 
(non-Gaussian) likelihood, 
theoretical predictions, and likelihood samplers. Developing and testing these tools is a long-term goal, one in which members of the LSS Working Group will be actively involved.

On shorter time scales, a software framework is needed that will (i) give scientists easy access to project simulations and (ii) allow for parameter extraction from value-added catalogs. The parameter extraction routines will be developed over the coming decade,
but installing a basic set of tools that can be used across working groups is essential in order to quantify the effects of systematics on dark energy in an ``apples-to-apples'' fashion.

\subsubsection{Ancillary data needs}

The ancillary data required for primary LSS analyses overlaps significantly with other working groups, particularly Weak Lensing.  We need a sufficient sample of spectroscopic redshifts for galaxies covering all redshifts to utilize cross-correlation to determine the true redshift distribution of our adopted bins in photo-$z$. See discussion of photo-$z$ uncertainties above and Section~\ref{sec:photoz} for details.
Additional joint analyses of dark energy and modified gravity will be enabled by combination with (Stage III-IV) spectroscopic redshift surveys such as (Big)BOSS, HETDEX, Euclid, and WFIRST.  By probing redshift space distortions, these spectroscopic surveys enable a direct test of the logarithmic derivative of structure growth as a function of spatial scale \citep[e.g.,][]{acquavivag10, samushiaetal12}.

%% file: supernovae/supernovae_challenges.tex
\subsection{Overview}
Observations of nearby and distant type Ia supernovae (SNe~Ia) led to
the discovery of the accelerating Universe driven by dark energy
\citep{1998AJ....116.1009R,1999ApJ...517..565P}. 
SNe~Ia have played a starring role in the current renaissance of
time-domain astrophysics that will reach a new apex with LSST. SNe~Ia
discovered by LSST will provide important new insight into dark
energy, not only through improvements in the precision of constraints
on dark energy parameters ($w$, $w_a$, etc.), but also through novel
tests of dark energy models (its isotropy, for example).

SNe~Ia are exploding stars defined by the lack of
hydrogen and the presence of silicon in their emission, and are
the product of the thermonuclear explosion of a C/O white dwarf
\citep{2000ARA&A..38..191H}.
The brightening and fading of a SNe~Ia are observed through
its photon flux 
at different dates and wavelengths, i.e., multi-band light curves.
The luminosity at peak brightness can be predicted to $\sim 12$\%
accuracy from light-curve shapes and colors.  The ratio between
the observed flux and predicted luminosity at peak brightness
is a measure of the object's luminosity distance. An accurate
redshift is typically obtained from the host galaxy, although it can be obtained
from the supernova itself.

As ``standardizable candles'', SNe~Ia map the expansion
history of the Universe through measurements of their redshifts
and luminosity distances.  The SN~Ia Hubble diagram is compared
to cosmological models to measure the parameters that determine the
dynamical evolution of the Universe.  The redshift range
of SNe~Ia discovered by LSST, $0.1<z<1.2$, spans eras of
both cosmic acceleration and deceleration, and the transition between 
the two.  SNe~Ia are therefore excellent probes of the physical
properties of the dark energy responsible for the accelerated expansion
of the Universe. 

\subsection{Analysis steps}
The LSST data come in the form of pixelized wide-field images observed
in multiple bands and dates.  Within these images lie the measured supernova
flux where the bulk (but not all) of the information from which
discoveries, typing, redshift, and distances are determined, which in turn are
used to make cosmological inferences.
The survey and analysis steps roughly proceed as follows.

\subsubsection{Survey}
LSST, like the most successful SN cosmology projects today, will
perform a ``rolling'' SN search and followup
\citep{2006A&A...447...31A}.  Cadenced reobservations
of the survey region in multiple bands provide multi-color optical light curves
for the many active supernovae within the imager field of view.
These constitute the fundamental data used in the
analysis.
It is absolutely clear that the
observing plan for the main survey and for the deep drilling fields will have a direct and
major impact on LSST's ability to do SN~Ia cosmology.
Exploration and optimization of the phase space of survey parameters is
of fundamental importance.

\subsubsection{Candidate discovery}
The starting point of supernova analysis is discovery.
There are two distinct types of transient discoveries that are made:

{\bf Live discovery.} Near real-time discovery of supernovae is needed to
trigger supplemental observations while the object is still active.  Generally,
this is done by aligning a set of ``new'' images with ``reference'' images,
subtracting the two, and identifying new point sources in the subtraction.

{\bf A posteriori discovery.} The images contain supernovae that were
not discovered in the real-time analysis but yet can be identified from the full time series.

Well-defined discovery criteria are important to quantify selection bias
in the sample.  Supernova searches are susceptible to Malmquist bias
(preferred discovery of intrinsically brighter objects) and 
a bias toward objects
that are less contaminated by host-galaxy light, i.e., away from the
core of the host.
In practice, some
amount of human intervention is needed to distinguish between real and spurious
transients.  As long as the discovery efficiency is well-quantified,
this potential source of systematic error is controlled.

\subsubsection{Triggered external observations}
Supernova typing, redshift, and distance determinations improve with additional
information beyond the light curves and galaxy photometry provided by LSST.
Type Ia classification is  traditionally done through low-resolution rest-frame
optical spectroscopy. 
If the SN has a clearly identifiable host galaxy, the host
spectroscopic redshift can be obtained at any subsequent time.
The rest-frame near-infrared
\citep{2004ApJ...602L..81K, 2008ApJ...689..377W, 2012arXiv1204.2308B} and spectral features
\citep{2009A&A...500L..17B, 2011ApJ...742...89F, 2012AJ....143..126B, 2012arXiv1202.2130S}
have proven to be a promising
route to make SNe~Ia even more standard.

After identification, a transient is classified as a likely SN~Ia or non-Ia.
An early classification is necessary to send pure target lists for external observations
of active supernovae.  
Follow-up resources are scarce so getting supplemental data on all
LSST SNe is not feasible.
A pure and efficient supernova typer that works on both early
data and complete light curves is required for efficient use of non-LSST resources.
Just as with candidate discovery, the efficiency of the typer must be
quantified to monitor potential bias in the SN~Ia subsample that receives additional
observations.

\subsubsection{Extracting multi-band light curves}  
The ``final'' SN photometry (as opposed to real-time measurements for
detection and quick follow-up) needs to be of the highest fidelity. 
This important
step transforms pixelized data to light curves
in physical units necessary for  supernova
analyses.  
Of
extreme importance is photometric calibration, within and across
fields over the whole survey region. Not just the zeropoints but also
the system response as a function of wavelength, time, and position on
the detector must be provided with sufficient accuracy. The SN
photometry will typically be modeled as a time-varying point source on
a static host-galaxy background surface brightness.
All observations of a field will contribute to
this model (as even observations without the SN provide information
about the static background, i.e., the host galaxy). In addition to the
response function, each image should also come with an accurate PSF
model.

The flux calibration is identified as a major challenge.  We must understand
how the performance of the LSST Project calibration system propagates
into the covariance matrix for all supernova photometry.

\subsubsection{Typing, redshift, and distance modulus} 
Construction of the supernova Hubble diagram depends on
determination of the type, redshift, and distance modulus for each SN, which in turn depends on the available
data.

{\bf Photometric data only.} A large fraction of LSST SNe will have no independent typing or redshift measurements;
these ingredients and distances are determined simultaneously from the complete light curves
and host-galaxy photometry.
Work is underway (but more is needed) to determine to what extent the
photometry of the SN itself can yield the classification,
redshift, and distance simultaneously \citep{2011ApJ...738..162S,2012ApJ...752...79H}.
If a sample of SNe~Ia can be classified
photometrically with sufficient purity and efficiency to yield
competitive dark energy constraints, it is possible to imagine much of
the analysis could be done with the LSST data already taken, without the
need for real-time discovery and ancillary follow-up of objects. A
wide array of different approaches to obtain classification and
redshifts can be explored (SN photo-classification and host spectroscopic redshift,
vs. SN photo-classification and SN photo-$z$, SN classification based on
host color/morphology and host photo-$z$, etc.).

{\bf Photometric and spectroscopic data.}
Spectroscopy of the candidate and/or host galaxy
may be available.  With sufficient signal-to-noise, supernova typing and
redshift are straightforward to do with spectroscopy using existing tools.
With this information,
multi-band light curves
of typed SNe~Ia are analyzed using a model that associates light-curve
shapes and colors with an intrinsic absolute magnitude.
Current examples of light-curve fitters are MLCS2k2
\citep{2007ApJ...659..122J} and SALT2 \citep{2007A&A...466...11G}, though
more sophisticated methods are anticipated by the time of LSST analysis.

Ancillary information about host galaxy properties will also play a
role in the analysis; for example, SN~Ia light curves may be better
standardized with knowledge of global host properties such as age, star
formation rate, and stellar mass \citep{2010ApJ...715..743K,2010MNRAS.406..782S,2010ApJ...722..566L,2011ApJ...740...92G,
2011ApJ...743..172D}. These are typically derived from
multicolor photometry of the host (though sometimes with spectroscopy,
and sometimes locally at the SN site), so it is likely that the host
data from the fully combined stack of LSST (and any other available
surveys, especially coordinated with the deep-drilling fields) will be
a direct input into the SN cosmology analysis.

For both cases, with or without spectroscopy inputs,  it is important to
quantify systematic uncertainties and biases that might be introduced
in these analyses.

\subsubsection{Quantifying systematic uncertainty}
The bulk of LSST data
are regressive, in the sense of being of poorer quality compared to
LSST subsets specifically targeted for detailed study or lower-redshift supernovae external to LSST.
We can quantify the systematic error in inferred absolute magnitude that arises due to our inability
to model the object due to missing or poor-quality data:
for a subset of supernovae with excellent data quality we calculate the bias in distances
determined through analysis of higher-quality
data and the same data degraded to LSST-quality.

We must estimate LSST systematic uncertainties based on current
data and models.  Conversely, we must quantify the ``calibration''
data necessary (from LSST or elsewhere) to constrain systematic uncertainties to our goals.
This will be an ongoing activity as our understanding of SNe~Ia
evolves.

\subsubsection{Fitting cosmologies} 
The combined data set consisting of SN light curves, classifications
(which may be a probability rather than a binary determination: Ia or
not), redshifts, host galaxy properties, and any other ancillary data
will be globally analyzed to determine cosmological constraints
in a blind analysis. 

Work must be done to understand how to propagate or
quantify typing uncertainty
into dark-energy parameter uncertainties.

The analysis chain just described takes input data and propagates them into
dark-energy parameters of interest, incorporating statistical and systematic
uncertainties.  Building elements of this chain quickly is critical for our
interactions with the Project, quantification of systematic uncertainties, and
prioritization of critical path DESC activities.

\subsection{Systematics}
The LSST DESC
seeks to maximize the exploitation
of the LSST supernova survey by minimizing statistical and systematic uncertainties.  In this section, we review the critical issues that must be addressed
from which
tasks to be performed in the three years are identified. 
\subsubsection{Projecting the survey to dark energy parameters}
Most of the activities of the SN working group focus on the optimization of survey strategies and the addressing of
systematic uncertainties.  We have an overarching need for a software tool that projects LSST SN surveys into dark-energy 
measurements with high-fidelity statistical and systematic uncertainty propagation.  No such end-to-end tool exists: a new software
project is needed.  We can leverage the vast experience of DESC members in simulations and data analysis.
We will take advantage of the simulation products provided by the Project; 
indeed we want realistic supernova signals in the Project pixel-level simulations. 
Note that the deliverables of most of the SN tasks will either be a contributed module for this project, or depend on the running of
the software.  

This is a high-priority activity because most of our studies need such a tool to make the connection with survey performance.

\subsubsection{Photometric calibration} 
A key ingredient for projecting survey to cosmology is the Hubble diagram covariance
matrix.  The leading source of uncertainty in current cosmology results is photometric color calibration
\citep{2012ApJ...746...85S}. 
The propagation of  calibration uncertainties into dark-energy uncertainties
is sensitive to the covariance
of the calibrated transmission functions for all bands. Project calibration requirements are specific to the structure of the covariance
matrix produced by a specific calibration procedure.  
The Project requires feedback to tune  the
performance of their calibration system while the DESC needs to properly incorporate the
calibration covariance matrices
in the error budget
to project cosmology reach (e.g., the DETF figure of merit).

This is a high-priority activity because the LSST Project now seeks feedback on the calibration strategy for their planning.
Calibration is of importance as it is the limiting source of uncertainty in current experiments.
While previous experiments have examined zeropoint
uncertainties, none have incorporated in their cosmology analysis the calibrated transmission functions and their covariances. 
We therefore need to develop algorithms and software
that correctly propagate these uncertainties.

\subsubsection{LSST Hubble diagram uncertainties}
LSST is confronted by a novel source of uncertainty in the Hubble diagram covariance matrix.
Construction of a SN~Ia Hubble diagram
requires 1) the classification of objects that enter the analysis; 2) their redshifts; 3) their distance moduli. 
Traditional analyses use spectroscopic data for the typing and redshifts, while new work shows that the determination of distance moduli can depend on UV through NIR data as well as spectral features; 
LSST  provides only
optical  photometry from which the three Hubble-diagram ingredients are inferred,
not the full variety of data that can be used to constrain the SN~Ia model.  Inferred absolute magnitudes will be subject to large statistical
uncertainties due to lacking data, or systematic biases if poorly constrained SN parameters are marginalized in an effort to reduce those 
statistical uncertainties.

This is a high-priority activity because this is a ``new'' source of uncertainty that has yet to be quantified.  We currently have no reliable estimate
of the science reach of an LSST-only (i.e., photometry-only) SN program.
The Dark Energy Survey  faces the same problem and will be developing approaches to minimize and quantify these  uncertainties.
The DESC needs to monitor their progress and incorporate their gained knowledge
in our analysis.

\subsubsection{Survey optimization}
The final number of supernovae identified in the LSST data, and their redshift accuracy and photometric signal-to-noise ratio, will depend on the strategy for both main and deep-drilling surveys.
The SN program's projected figure of merit is  sensitive to choice of exposure times, cadence, and solid angle.
Trade-offs between quantity
and quality of light curves will have implications for both
statistical and systematic uncertainties in dark energy parameters;
these have yet to be fully explored. 
The DESC must have the capability to calculate the figure of merit over the parameter space of survey possibilities.

To highlight the importance of this activity, the general current consensus among
members of the LSST SN Science Collaboration is that  the current plan for
most main survey fields (in terms of cadence and filter in repeat
observations) will not be sufficient for SN~Ia light curves of high
cosmological utility, and that only the deep drilling fields will provide useful
data. 

This is a high-priority activity because we suspect that the baseline main survey is not useful for supernova cosmology.  We must confirm
these findings and provide viable alternative strategies to the Project.
DESC members already have software to simulate LSST light curves.
Improvements in the configurability of the Project-provided OpSim
are necessary to enable these studies.

\subsubsection{Value-added external data}
Non-LSST observations can reduce the statistical uncertainty in absolute magnitude of a single object, while such observations of an
unbiased subset of objects
can calibrate the
absolute magnitude bias when only LSST data are used.  Examples of supplemental data include supernova typing
and subtyping spectra, UV and NIR supernova photometry, and host-galaxy imaging and spectra. 
Euclid NIR light curves and BigBOSS or DESpec spectroscopy to get host redshifts are already
known to have excellent potential to expand the science reach of LSST SNe.

The DESC must identify planned resources (or develop new ones) that can leverage LSST data, and quantify
the resulting benefit.
The same tool used to determine the LSST error budget must allow inclusion of new data and be used to
re-optimize the LSST survey and the use of external resources.
This is a high-priority activity because we anticipate that complementary non-LSST data can leverage the power
of LSST-discovered supernovae.  Planning for use of other resources (such as Euclid) requires advance coordination.

\subsubsection{Refining the distance indicator}
The full range of SN~Ia heterogeneity that is correlated with absolute magnitude is not captured in current models.
Biases in inferred absolute magnitudes are manifest in the identification of
peculiar SNe~Ia, and in correlations of Hubble residuals with light-curve shapes, colors, UV and NIR data, spectral features, and
host-galaxy parameters. Indeed, the imperfect modeling of supernova colors (coupled with calibration) is
a leading source of systematic uncertainty in current SN-cosmology analyses
\citep{2009ApJS..185...32K}.  Improvements
to the SN~Ia model will decrease both their intrinsic dispersion and suppress systematic uncertainties.

We will engage in the study of precursor SN data and theory to improve the model for determining
absolute magnitudes and quantify model uncertainties. There is ongoing work run by distinct groups
who have access to their own data.  Almost all relevant groups are represented in the DESC and we will
work to pool our collective knowledge to formulate SN~Ia models that coherently combine the collective knowledge.
While this is not a time-critical activity that must be done in the next three years, it is very important for
LSST DESC scientists to remain active in the cutting-edge advances in supernova standardization.

%% file: analysis/clusters/introduction.tex
Galaxy clusters provide a powerful toolset with diverse applications
in cosmology and fundamental physics. The observed number density and
clustering of galaxy clusters as a function of mass and redshift are
sensitive to both the expansion history and growth of structure in the
Universe, enabling powerful constraints on dark energy and providing
critical distinguishing power between dark energy and modified gravity
models for cosmic acceleration. Measurements of the baryonic mass
fraction in clusters, and of the tomographic lensing signatures
through clusters, provide additional ways to measure cosmological
parameters.  Galaxy clusters provide sensitive probes of the physics
of inflation and, in combination with CMB data, currently provide our
best constraints on the species-summed neutrino masses.

As with all cosmological probes, the key to extracting robust
cosmological constraints from galaxy clusters is the control of
systematic uncertainties, particularly those associated with finding
clusters and relating the observed properties of clusters to the
underlying matter distribution. This requires a coordinated,
multiwavelength approach, with LSST at its core, and the application
of rigorous statistical frameworks informed by cosmological
simulations. (For a recent review of galaxy cluster cosmology see
\cite{2011ARA&A..49..409A}.)

%% file: analysis/clusters/analysis.tex
The primary way in which galaxy clusters probe cosmology is through
measurements of their spatial distribution, as a function of mass and
redshift. Traditionally, this approach has been split into
measurements of the mass function and its evolution
(e.g., \citealt{2009APJ...692.1060V}), and the clustering of galaxy
clusters (e.g., \citealt{2011MNRAS.414.1545F}). However, significant gains
can be realized by modeling all aspects of the distribution
simultaneously \citep{2012MNRAS.423.2503S}.

The four main steps in extracting robust cosmological constraints from
the observed number density and clustering of galaxy clusters are 1)
predicting the statistical spatial distribution of massive halos as a
function of mass and redshift; 2) constructing the catalogs of
observed clusters used for cosmological work; 3) measuring the
relationships between survey and follow-up observables and mass; and
4) extracting the cosmological information of interest using a
self-consistent statistical framework.

Predicting the number and distribution of massive halos (which host
observed galaxy clusters) as a function of mass and redshift requires
large-scale numerical simulations (see \citealt{2011ASL.....4..204B} for
a recent review). These simulations must span the full range of
cosmologies, and the mass and redshift ranges of interest. Systematic
challenges include identifying and incorporating the necessary dark
matter, dark energy, and baryonic physics (the latter in particular can
be challenging); identifying halos in the simulations and quantifying
deficiencies in the halo finders; simulating sufficiently large
volumes while maintaining resolution; and finding methods to enable
fast model comparisons with the real Universe.

The construction of robust catalogs of observed clusters requires
identifying and employing survey observable(s) that are simple to
measure and have minimal, and well understood, scatter with mass.  For
LSST cluster catalogs these observables will involve accurate
measurements of galaxy colors and spatial concentration or
``richness''. The cluster catalogs produced should be as pure and complete
as possible, and have their residual impurity and incompleteness, as
well as the survey selection function, modeled robustly. Accurate
photometric redshift estimates for the clusters are another key survey
requirement.

Constructing and understanding the relationships between survey
observables and mass is the most difficult and complex stage of the
process.  Generally this requires the use of intermediate scaling
relations, involving multiwavelength mass proxies with minimal and
well understood scatter with mass, gathered from follow-up
observations. The absolute mass calibration of these relations must
also be established accurately and precisely, with this latter step
driven by LSST weak-lensing measurements.
 
All of the above information must then be combined within a robust,
self-consistent framework that utilizes fully the information
available, while accounting for the impacts of survey biases,
systematic uncertainties, and their covariance with all quantities of
interest (e.g., \citealt{2011ARA&A..49..409A}).

Galaxy clusters offer an ensemble of additional constraints on
cosmology and fundamental physics.  Tomographic measurements of the
redshift-dependent lensing of galaxies behind clusters provide a
geometric test of dark energy (e.g., \citealt{2009MNRAS.396..354G}). Measurements of the baryonic mass fraction in
clusters tightly constrains the mean matter density, and provides a
powerful, complementary probe of the expansion history \citep{2008MNRAS.388.1265R}.  The combination of X-ray and mm-wave measurements provides
another redshift-independent distance measurement (e.g., \citealt{2006ApJ...647...25B}).  More generally, the availability of both geometric and
structure growth tests for clusters allows one to separate the effects
of dark energy within the context of General Relativity from
deviations due to alternate descriptions for gravity
\citep{2007AIPC..957...21C,2008ARA&A..46..385F,2012arXiv1205.4679R,2011PhRvD..83f3519L}.

As the most massive collapsed structures, galaxy clusters can
additionally be used to probe non-gaussian signatures of inflation
(see, e.g., \citealt{2010JCAP...04..027C} and \citealt{2010CQGra..27l4010K} for
recent reviews), and the neutrino mass hierarchy, through the effects
of neutrinos in suppressing structure growth on the scale of clusters and
smaller \citep{2006PhR...429..307L,2010MNRAS.406.1805M,2010JCAP...01..003R}.

In all of these aspects, cluster cosmology will exploit the full
potential of LSST, and the rich ensemble of multiwavelength data
available in 2020.

%% file: analysis/clusters/challenges.tex
 
The major, recognized challenges to obtaining robust dark energy
constraints using galaxy clusters observed with LSST can be summarized
as follows.

A first challenge is the accurate theoretical prediction of the
distribution and clustering of galaxy clusters as a function of mass
and redshift. In particular, robust predictions for the mass and bias
functions will be required, spanning the full suite of cosmological
models and mass and redshift ranges of interest.  These predictions
should be formed in a way that facilitates fast model comparisons (see
below), for example, by utilizing ``universal'' forms for mass and bias
functions where possible. Particularly important and challenging will
be understanding the impact of baryonic physics on these
predictions. This task will require cosmological simulations,
building on the
infrastructure developed for Stage III projects such as DES.

A second major challenge involves developing optimized ways to find
clusters and form catalogs, both using LSST data alone and in combination
with auxiliary data. (As discussed in the analysis section, the combination of
optical and X-ray data has been shown to provide better performance
for cluster finding than optical data alone; e.g., \citealt{2012ApJ...746..178R}). 
The challenge will be to develop algorithms that generate
cluster catalogs with maximum purity (i.e.,  no false clusters) and
completeness (i.e., no clusters missing). Equally important, however,
will be the ability to quantify and model residual incompleteness and
impurity precisely, so as to be able to marginalize over these fully
in the cosmological analysis. 
This task will rely heavily upon simulated galaxy catalogs and images that accurately represent the galaxy properties of real clusters, which is a topic of intensive investigation by the DESC Cosmological Simulations Working Group.
This task will also require extensive cosmological simulations, 
building on tools developed for Stage III projects.

While theory can predict the distribution of mass quite robustly as a
function of cosmological parameters, surveys are based on {\it
observables} -- measured quantities that relate to mass with complex
scaling relations and scatter. For LSST cluster cosmology, the primary survey
observables (in addition to photometric redshifts) will relate to optical richness,
likely augmented by information in the X-ray and mm-wave (Sunyaev-Zel'dovich effect). 
The single
most pressing challenge for cluster cosmology is to understand the
statistical relationships between galaxy cluster observables and
mass. There are two main aspects to this work: 1) understanding the
form, scatter and evolution of the {\it scaling relations} linking key
observables and mass; and 2) obtaining robust absolute mass
calibrations for these relations.

For the first aspect, nature fortunately provides several observable
quantities that are straightforward to measure and which correlate
tightly with mass, exhibiting minimal scatter across the mass and
redshift range of interest. These {\it low-scatter mass proxies}
provide a critical intermediate step, allowing one to first
statistically link survey observables to these mass proxies, and then
the mass proxies to mass. Studies have shown
\citep[e.g.,][]{2010MNRAS.406.1759M,2010ApJ...713.1207W} that the addition of robust,
low-scatter mass-proxy measurements for even a small fraction (a few percent) of the clusters in a survey can boost the overall cosmological
constraining power by factors of a few or more.  Major challenges for
this work include identifying the optimal combinations of mass proxies
for LSST cluster science; determining how to distribute targeted follow-up
measurements of mass proxies across the survey flux/redshift range;
identifying where thresholding and binning can be effectively applied; quantifying
and modeling the impacts of systematic uncertainties; and actively
gathering the mass-proxy data. This last challenge will involve
collaborations with other agencies, as many of the best mass proxies are provided by X-ray observations.

Equally important is the absolute mass calibration of these
mass-observable scaling relations -- currently the dominant systematic
uncertainty for cluster cosmology.  Here weak-lensing measurements
will play the driving role. The effects of triaxiality, projected structures along the line of sight, and the finite number of background galaxies with measurable shapes together introduce a scatter of $\geq 20$\% in shear measurements for {\it individual} clusters \citep{2011ApJ...740...25B}, making weak-lensing mass a relatively poor mass proxy.
Critically, for {\it ensembles} of clusters, nearly unbiased results
on the mean mass can be obtained \citep{2009MNRAS.396..315C,2011ApJ...740...25B}. This
requires the use of selected radial fitting ranges and
detailed photometric redshift information for the lensed galaxy
population \citep{WtG1, WtG2, WtG3}. In this context, some of the major
challenges for LSST cluster science are the
development of optimized weak-lensing methods for cluster-mass calibration;
the identification of optimal radial ranges over which measurements should be made 
and the set of shear profile models that should be fit;
and determination of 
how best to bin, stack and threshold data, 
whether higher order lensing effects can be effectively utilized,
and how best to quantify and model the impacts of systematic uncertainties,
allowing these to be marginalized over in the cosmological analysis. 
Correlations between lensing measurements and observables that are used to ``stack'' clusters must be quantified and taken into account to avoid bias in the dark energy constraints \citep{2008MNRAS.387L..28R,2012arXiv1204.1577N}.
Here cosmological and ImSim simulations will play an essential role.

Galaxy clusters are crowded environments hosting the highest densities
of galaxies in the Universe. A challenge for cluster cosmology will be
mitigating and modeling the systematic effects associated with source
confusion, and contamination of lensing and other signatures by the
cluster-member galaxy population.  Being the most massive objects in
the Universe, clusters also exhibit the strongest lensing
signals. Understanding how to utilize higher order lensing effects, as
well as tomographic lensing signals in both the weak and strong
lensing regimes, represent additional challenges for the DESC clusters
team.

As with other LSST probes, certain aspects of cluster cosmology, and
particularly those related to lensing measurements, are sensitive to
photo-$z$ systematics. An important and pressing challenge is to 
quantify the impact of these uncertainties on cluster mass calibration, as a function of cluster mass and redshift.  These studies should explore techniques that minimize the need for exhaustive spectroscopic followup.  Current studies by DESC members that utilize the full photometric redshift information for the lensed galaxies in cluster fields provide some promise in this regard (\citep{WtG1, WtG2, WtG3}.
Conversely,
areas where the unique environment of clusters introduces additional demands for
external calibration data should be identified urgently.

Encompassing the above is the overarching challenge of developing a
self-consistent statistical framework to model the distribution and
properties of clusters, including the mass-observable scaling
relations, accounting for survey biases and covariances, and
marginalizing fully over systematic uncertainties. Members of the DESC
Clusters Working Group have been at the forefront of this work in current
cluster cosmology.  Required extensions include the incorporation of
clustering into the analysis framework, along with the related
nuisance parameters.  These issues will be addressed to some extent by Stage III
projects, but LSST has substantially higher accuracy requirements, so parallel efforts within the DESC are almost certainly desirable.  End-to-end testing, from cosmological simulations to
cosmological constraints, will be required for this effort.

Using this modeling infrastructure, a final major challenge is
understanding the computing needs for cluster cosmology.  Currently,
using X-ray cluster catalogs with $\approx 250$ clusters and 100
multiwavelength follow-up observations, CPU farms with 500 cores
require of order a week of run-time for a complete cosmological
analysis that addresses multiple questions. Scaling simplistically
(linearly) with the number of clusters, and expecting LSST cluster
catalogs to contain hundreds of thousands of clusters, one can
envisage significant computing challenges. Understanding the computing
needs and identifying optimized ways in which to speed up the analysis
while maintaining precision and accuracy, with, for example, improved
algorithms and the judicious use of binning, are important challenges.
Cluster cosmology also requires significant input from cosmological simulations, which are essential for extracting precise predictions from cosmological models,  understanding the interplay between systematic sources of error, and testing the data processing and cosmological analysis pipeline in general.  These require substantial computing resources, and significant work will be required in conjunction with the simulations working group to identify these resources.

\subsection{Urgent Priorities}

Among the most pressing challenges described above are those related
to the absolute calibration of mass-observable scaling relations using
weak-lensing methods. The results from the aforementioned studies of,
for example, the optimal radial ranges over which measurements should be made,
the set of profile models that should be fit to clusters, how to
bin, stack and threshold data, whether higher order lensing effects
can be effectively utilized, and how to model systematic
uncertainties, will significantly impact the long-term strategy for
cluster science with LSST and may influence the development of core
LSST software.

A related urgent issue is to evaluate the performance of current LSST
deblending algorithms in crowded cluster environments, and how
this performance impacts the mass calibration described above.  In
particular, we need to evaluate how limitations in the deblending algorithm impact
the cosmological constraints from clusters, 
the degree to which these limitations can be modeled, 
and the requirements for improvements in the algorithm.  
Here and elsewhere, the analysis will involve data from sources such as Subaru and HST, as well as ray-tracing simulations. 
(DES data are not well-matched to this problem because of their depth and image quality.)
The work will link the development of the deblending algorithm (carried out by the LSST Data Management team) to requirements determined by its effect on cosmological constraints.

Equally important is a detailed, quantitative understanding of the impact of photometric redshift uncertainties on cluster science,  and on cluster mass calibration in particular.  Linking to this, the development of techniques that minimize the requirements for spectroscopic followup may have cross-cutting benefits and expedite the progress of cluster cosmology with respect to other dark energy probes. Conversely, areas where the environments of clusters place additional demands on external calibration data should be
identified urgently, allowing preparations for the gathering of these data to be made. 

The final important challenge 
is the development of a self-consistent
statistical framework to model the observed distribution and
properties of galaxy clusters and exploit fully the cosmological
information contained within these data. This framework should enable
reliable estimates of the computing needs for LSST cluster cosmology
(which may be substantial) and robust predictions of the eventual
cosmological constraints from galaxy clusters. Efforts by the DESC in
this regard, utilizing the expertise within the collaboration, are
also very likely to be of benefit to Stage III projects.

For other challenges, we expect that significant near-term progress
will arise naturally from Stage~III experiments like DES -- for example,
in theoretical predictions of the distribution and clustering of
galaxy clusters across a wide range of interesting dark energy models,
and in the development of methods for cluster finding and the construction of
optimized cluster catalogs. The DESC Clusters Working Group
will stay apprised of these efforts, and will aim to recognize if and when
the need for DESC involvement becomes pressing.

%% file: analysis/strong-lensing/overview.tex
Strong gravitational lensing refers to the multiple imaging of a
background object (``the source'' or ``deflector'') by a massive
foreground object (``the lens'') \citep[e.g.,][]{SKW06,Treu10}. The
resulting angular separation, time delay, and morphological distortion
all depend on the lens geometry, quantified in terms of combinations
of the three angular diameter distances between observer, lens and
source \citep{Ref64,BlandfordNarayan92}.  These observables are also
affected by the mass distribution of the lens, which must be modeled
simultaneously using high resolution follow-up imaging and
spectroscopy.

Strong lensing provides at least two independent ways to measure dark
energy parameters based on LSST data. The first one, gravitational
time delays, has a venerable history and is now, in our opinion, mature
for precision cosmology \citep{Ref64,SuyuEtal10,SuyuEtal12}.  The
second one, the analysis of systems with multiple sets of multiple
images \citep{Gav++08,Jul++10}, is a more recent development and will
require more research and development work in order to quantify
precisely its systematics and assess its potential.  Therefore, we will
spend most of this section addressing the first approach,
giving only a brief description and key references for the
second.

\begin{figure}[htbp!]
\begin{center}
\includegraphics[width=0.46\columnwidth]{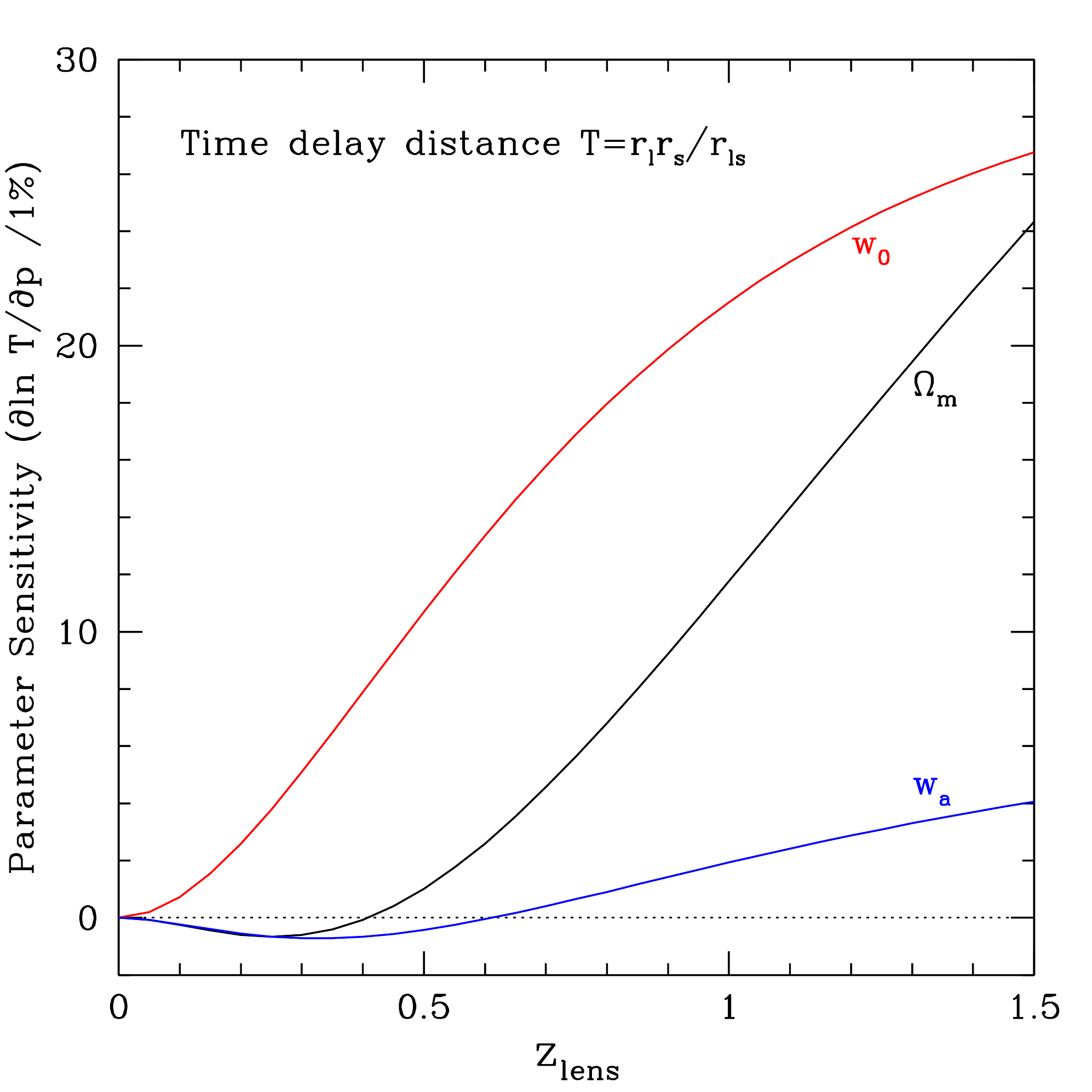}
\includegraphics[width=0.46\columnwidth]{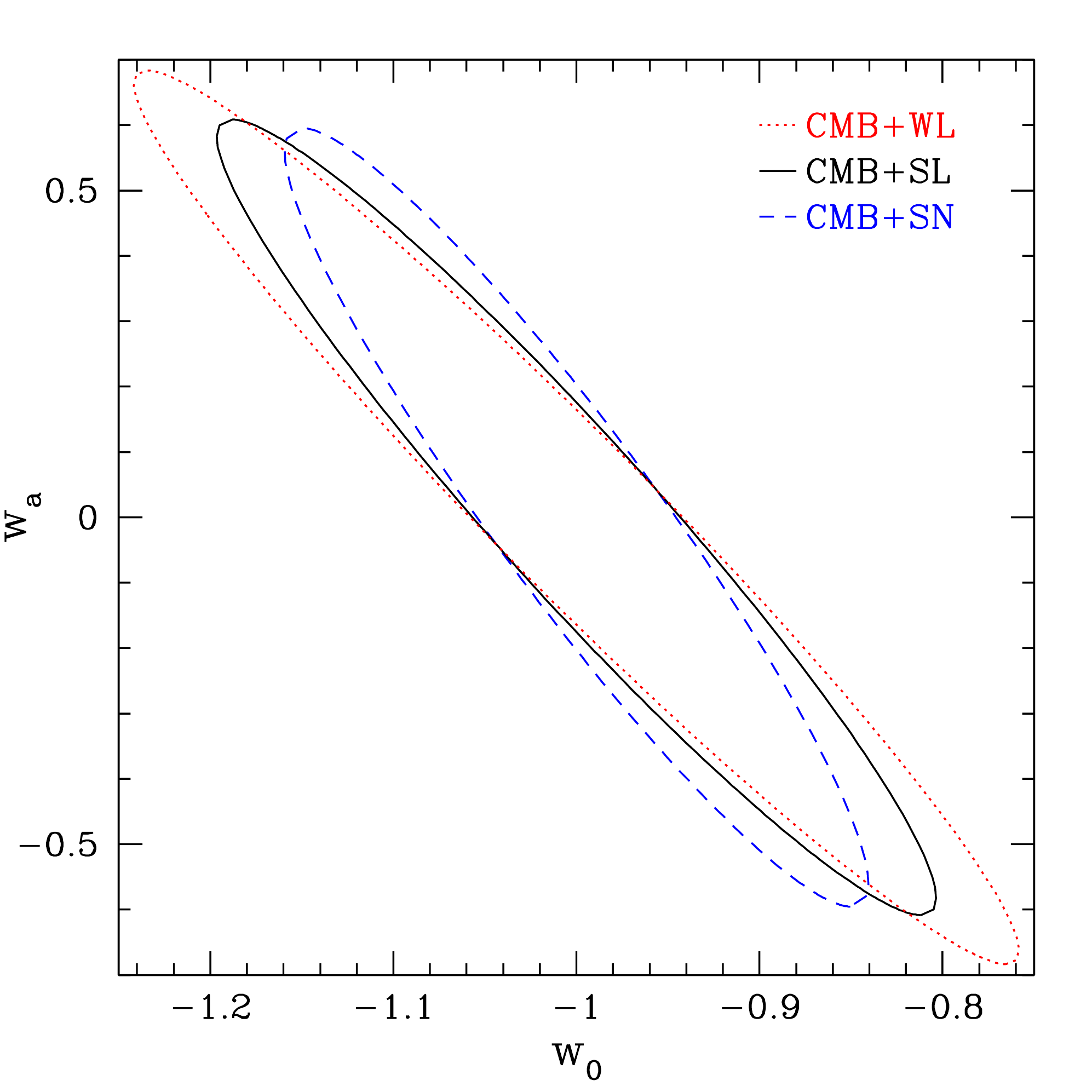}
\end{center}
\caption{Left panel: Sensitivity of the time-delay distance combination 
$T=r_l r_s/r_{ls}$ to the cosmological parameters is plotted vs.~lens 
redshift.  Unlike supernova distances, sensitivities can pass through zero 
(indicating no degeneracy with the other parameters) and have 
opposite signs at the same redshift, giving strong complementarity with 
positive-only supernova distance sensitivities. 
Right panel: Complementarity of time delays (SL) with both supernovae (SN) 
and weak lensing (WL) is shown in the different directions of constraints on 
dark energy parameters, marginalized over neutrino mass and other parameters. 
}
\label{fig:analysis:sl:td-forecast}
\end{figure}

Strong gravitational lensing time delays measure a combination of distances  that
is sensitive primarily to the Hubble constant, but also to the other
parameters of the world model: when many lenses with different lens and source
redshifts are combined, internal degeneracies are broken and dark energy
parameters can be measured \citep{C+M09}. The time-delay distance
provides  significant complementarity with other dark energy probes (see 
\autoref{fig:analysis:sl:td-forecast}) to determine, e.g., the  dark energy equation of state, matter density, Hubble constant, and sum of neutrino masses
\citep[e.g.,][]{Linder11,Col++12,WeinbergEtal12}. 
Furthermore, the main systematics coming from the lens-mass modeling and
line-of-sight dark matter, are interesting astrophysically in
themselves: dark matter substructure and galaxy mass profiles
\citep[e.g.,][and references therein]{Treu10}.  
LSST's wide-field, time-domain survey will be superb for obtaining
large samples of time-delay systems and, with follow up spectroscopy
and high resolution imaging for the lens model, delivering distance-ratio constraints.  
An exciting prospect is the use of time delays to
test gravity on precisely those scales where we expect screening
mechanisms to become active.

We anticipate that 
dark energy measurements from strong lensing with LSST will differ from
previous or concurrent strong-lens surveys in the following key ways:
\begin{enumerate}
\item A high-yield lens search requires all four of  the properties offered by
LSST:  wide area, depth, high image resolution,
and  spectral coverage to distinguish lens and source light. Other surveys
-- current, planned, or concurrent with LSST -- 
will lack at least one of these properties. Only SKA will
surpass LSST in terms of yield; Euclid will perform comparably well,
but is best thought of as a complementary dataset. LSST should be able
to detect $\approx 8000$ time-delay lenses (both AGN and supernovae).
\item Time-domain information will be available 
from within the survey, enabling the measurement
of time delays for thousands of lens systems for the first
time. Precursor surveys such as DES, HSC, and PS1, which should find
lensed AGN in large numbers, will be reliant on monitoring follow-up,
and so will likely achieve only a few hundred measured time delays,
with enormous effort. LSST should provide precision time delays for over
a thousand lenses.
\end{enumerate}
In short, LSST's strengths are the size of its lens sample, which should be an
order of magnitude more in most lens classes (and even more in others), and
its ability to provide measured time delays for the majority of suitable
variable-source systems.

Practical problems to be solved include lens detection (which will be
very sensitive to the image quality and the deblender performance);
image and light-curve modeling (which can be both CPU and skilled-labor
intensive); obtaining and analyzing follow-up data; interpreting the
whole sample of lenses in the context of a well-studied subset; and
finding sufficient experienced researchers to accomplish all these tasks. 
We expand on these topics in the sections below.

%% file: analysis/strong-lensing/analysis-steps.tex
Currently the limiting factor of the time-delay distance method is the
small number of well-measured systems available for this
measurement. Each system, with sufficient ancillary data, can provide
a measurement of the time-delay distance to around 5\%
\cite{SuyuEtal10}.  The dominant known sources of uncertainty to the
error budget are the distribution of mass along the line of sight
(around 4\%), the mass distribution in the main deflector (3\%), and
the time-delays themselves (1-2\%).  The number of known suitable
multiply-imaged quasars is small and those with time delays accurate
to 1-2\% is less than twenty at the time of writing \citep[e.g.,][]{Treu10,Tewes12}. We anticipate that, in the
course of this present decade, only with concerted effort will the
community be able to bring the number of viable systems with measured
time delays to approximately 100.  LSST will revolutionize the field
by increasing the sample size by at least an order of magnitude, with
over 3000 multiply imaged AGN, and several hundred strongly lensed
supernovae expected to be detected well enough to enable time-delay
estimation to be attempted
\citep{OguriMarshall10}.

\subsubsection{Searching for strong lenses}

The first major analysis step required is to {\it find} these time-delay
lenses in the LSST dataset. Such systems are
rare, approximately 0.2 per square degree at LSST single-epoch depth and 
resolution \citep{OguriMarshall10}.  
Finding them in ground-based images (or even shallow space-based images)
is a highly non-trivial task. A number of algorithms have been proposed based
on the morphology, color, and time domain properties of the systems
\citep[e.g.,][]{Kochanek++2006,Oguri++2006}. However,
current algorithms require substantial human intervention and follow-up data
for confirmation. This will be impractical at the LSST scale and therefore
significant effort will need to be spent in the coming years to develop
algorithms with higher purity and efficiency, and test them on precursor
Stage~III survey data. Below we sketch the key features of the time-delay
lens detection pipeline.

{\bf Querying the catalogs.}
The most efficient initial selection will be made by querying the
object and source catalogs, based on photometric, morphological and
time variability criteria together.  In some exposures the lensed
images will be well-resolved, while in others they will not: all
exposures will contain {\it some} information about the variability of
the objects. Interpretation and classification of the candidates based
on physical lens models can begin during this phase.

{\bf Modeling the pixels.}
With such large samples being investigated, we will need to focus on
lens systems that are unambiguously identified using the LSST
data. Moreover, we anticipate the majority of the variability
information coming from the LSST survey data. These desiderata drive
us towards fitting models to the full set of LSST postage stamp
images, in order to disentangle the lens and source brightness
distributions, and optimally extract lightcurves of the lensed
images. Joint analysis with Euclid or SKA at this stage would be very
interesting, with the higher resolution space-based images enhancing
the sensitivity of the LSST measurements. A probabilistic model
comparison between a constrained lens model, and a flexible ``nebula''
model, would provide us with an unsupervised classification (Marshall
et al., in prep.).

{\bf Quality control.}
The basic selection function of the lens 
sample will be understood during the two automated steps above.  Still, human
inspection of the candidates provides two quality control services. First,
rejection of any residual false positives, increasing sample purity. Second,
discovery of particularly interesting systems---such as higher order
catastrophes, multiple source-plane lenses, etc. Beyond this, human
inspection has the potential to discover lens systems that the automated
algorithm missed. The impact of these procedures on the selection function
will need to be quantified and understood.

\subsubsection{Measuring strong lenses}

In order to be able to infer the time-delay distance for each of a thousand
clean lens systems selected from the pool generated in the detection phase, we
will need high precision measurements of the lensed images positions and
varying brightness, and also of the variable source's host galaxy (seen 
distorted into an Einstein Ring). This information will then be combined in
the lens modeling analysis; the time-delay distance is one of the parameters
of such lens models.

{\bf Measuring time delays.}
This will be done using the LSST time
domain data, and in particular the inferred fluxes from the pixel-level
modeling described above. The main sources of systematic error in the
determination of time delays are finite sampling and microlensing
\cite[e.g.,][and Tewes et al.\ 2012, in preparation]{Kochanek2004}. The
extended and well-sampled light curves enabled by the LSST data should
be sufficient for this purpose, but significant work will be needed to
improve and calibrate algorithms to the level of precision and
automation required by the scale of the project.

{\bf Measuring Einstein rings.}
Substantial observational follow-up
with non-LSST facilities will be required for the determination of the
spectroscopic redshifts of the sources and the lens galaxy, and the
stellar kinematics of the lens galaxy (which provides an independent
probe of the lens-mass distribution), and to obtain high resolution
images of the lensed quasar host galaxy. An efficient way to make
these measurements is likely to be with Integral Field Unit (IFU)
spectrographs on either Extremely Large Telescopes (ELTs) on the
ground or with JWST from space. Which facility is used may depend on
the type of lens being studied.

{\bf Modeling strong lenses.}
As well as providing resolved stellar
kinematics of the lens galaxy, IFU observations allow a cleaner
separation of the background and foreground objects, 
and of the source-galaxy stellar distribution and active nucleus, and open up the
possibility of using the resolved velocity field of the source as an
additional source of constraints on the lens model. The width of the
lensed arcs, and the combination of lensing and stellar kinematics,
constrain the density profile of the deflector. One significant source
of systematic uncertainty is the mass distribution external to the
lens itself, both in its physical environment and along the line of
sight. The current approach involves estimating a probability
distribution for the so-called ``external convergence,'' using coarse
measures of projected over-density calibrated with cosmological
simulations \citep{FKW11,SuyuEtal10}. Another source of systematic
error is in the detailed modeling of the lens itself:
overly-simplistic models, or very flexible models with inappropriate
prior PDFs for their parameters, can lead to biased results
\citep{Ogu07}.

\subsubsection{Inferring cosmological parameters from strong lenses}

The final stage of the analysis is the inference of dark energy parameters
from an ensemble of measured time-delay distances. Conceptually, the situation
is very similar to that of the supernova Ia probe. Likewise, one might expect
to be able to improve the accuracy of the constraints on dark energy
parameters by hierarchically inferring the hyperparameters that describe the
population of lenses and their sources, in a joint analysis of all systems. In
principle this analysis could include many more systems that did not have such
high fidelity measurements, but that nevertheless could provide information in
a ``statistical'' analysis. This is a topic that should be explored using mock
catalogs and observations prior to the survey.

\subsubsection{Multiple source plane lens cosmography}

Finally in this sub-section, we turn briefly to our secondary
probe. As far as cosmography with multiple sets of multiple images is
concerned, recent work has shown that in principle it can be
competitive, providing enough systems can be analyzed
\citep{Col++12,Jul++10}.  The main sources of systematic error
are once again the mass modeling of the main deflector, and the
structures along the line of sight \citep{DHB05,D+P11}. From the point
of view of LSST it is useful to distinguish between the cases when the
main deflector is a galaxy or a cluster.

In the case of galaxy-scale systems, the first step is once again finding
them. Compound lens systems are rarer than single multiple imaged galaxies.
\citet{Gav++08} estimate that 1 in 50 galaxy scale lenses should be
identifiable as a multiple source plane system at HST single-orbit
depth and resolution, implying that several hundred galaxy-scale
compound lenses should be contained in the LSST lens sample. The
majority of these systems will have as their first source a faint blue
galaxy rather than an AGN; for these to be detected, the automated
lens discovery algorithms will need to be extended to focus on
extended, non-variable arcs around massive galaxies
\citep[e.g.,][]{Marshall09}. If visible at all in the survey images, the
second arc system could be detectable via image modeling as described 
above---especially 
as the lens perturbation provided by the first source is small,
allowing simple models to be used in the automated classification. Synergy
with Euclid or SKA could be particularly important here.

The time-domain information provided by LSST will not be relevant in all but
the rarest cases: all the cosmographic information comes from the positions
and morphologies of the two arc systems, for which redshifts will need to be
measured. The follow-up strategy is likely to be very similar to that employed
for the time-delay lenses, and the need to model carefully the lens
environment also carries over from our primary probe.

Precise mass models have enabled first constraints on dark energy
parameters. \citet{Jul++10} show that the results from only a single
cluster (albeit a very well studied one) Abell 1689 when combined with
WMAP data are competitive with the other more established methods like
Supernovae and Baryonic Acoustic Oscillations. With many more systems
suitable for such studies that LSST will bring cluster strong lensing
could provide us with a viable complementary technique to
constrain the geometry of the Universe and probe the equation of state
of dark energy.

In the case of cluster-scale systems, the target list will be provided
by cluster catalogs produced from LSST data themselves (e.g., via the
red sequence method), or on other surveys (e.g., X-ray or
Sunyaev-Zeldovich). The challenge will then consist of identifying
many systems of multiple images per cluster (which may well require
space-based follow-up imaging) and determining their redshifts, ideally
spectroscopically, but also possibly photometrically.

%% file: analysis/strong-lensing/major-challenges.tex
\subsubsection{Catalog generation and interpretation \;}
Strong lensing places strong demands on the LSST DM source detection
and characterization software than will other investigations. We are
specifically interested in faint, extended objects, with the complex
morphologies that are typical of gravitational lens systems. The
deflector galaxy will, in the majority of cases, be of greater or
comparable brightness to the system of point images arranged around
it. The distinguishing features of the lensed images will be their
color, and variability: as \citet{Kochanek++2006} pointed out, lens
systems appear as extended sources in difference images, and the
varying quasars and AGN have characteristically blue
colors \citep[e.g.,][]{Richards++2006}.  The challenge is to combine
these algorithms and distill them into a catalog query that will
return a large, high completeness sample of lens candidates that is
orders of magnitude smaller than the parent object catalog. Stage III
surveys have the same problem, of enabling an efficient initial
selection of lens candidates: while they may solve some of the
deblending issues, Stage III image processing pipelines are already
fairly mature.  We have an opportunity with LSST to drive the
development of source detection software, and database schema design,
such that strong lens candidates can be efficiently selected from the
source and object catalogs. In order to do this, we need to be testing
the DM system, {\em as it is developed}, on simulated strong lens
samples.

Multiply-imaged sources will appear as between 1 and 5 sources at each epoch,
depending on the image quality: the LSST source and difference catalogs need
to capture this information and allow lens candidate selection in the presence
of variable image quality. Most of the lensed sources will have magnitudes
close to the detection limit: will difference imaging be sensitive enough to
generate a sufficiently high yield of time-delay lenses? Can a
robust color selection be made, given the blending of lens and source light?
Can the color information in the  difference catalogs be used to robustly
identify strong lenses? How is this variability information best captured and
stored, for the purposes of lens candidate selection? Does it scale to all
objects in the survey, or will this need to be a level 3 calculation on
pre-selected objects? Possible ways of characterizing variability, and its
color dependence, could be explored at Stage III, but only coarse variability
quantification will be possible there.  Making the most of the LSST data will
require dedicated LSST research.

\subsubsection{Automated candidate classification}
Even with optimized catalogs, the false positive rate in the initial lens
candidate sample may be high, as we push down to the single epoch magnitude
limit. The remaining piece of information that we have is the characteristic
pattern of expected lensed image positions (and to a lesser extent, fluxes).
Implementing this constraint means comparing the catalog data for a given
candidate against the predictions of a lens model. Source catalogs that are
faithful to their input images can be used as data in a lens model fit: lens
vs nebula model comparison would then provide an odds ratio utility for
down-selection. Alternative catalog processing and interpretation algorithms
may prove faster and more efficient than such a fit: their exploration is also
desirable.

We expect to be able to reach only a certain degree of purity working from the
pipeline-generated catalogs: we anticipate needing to return to the images to
study samples of around $10^5$ candidates. This could be enabled via the DM
level 3 API. In the first instance,  lens-modeling of cutout images will
refine the automated classification: a ``multi-fit'' to all cutouts from each
visit will preserve all the  information available. Secondly, one of the
by-products of such a fit will be optimally-generated light curves and their
covariance, for the model-deblended sources. 

Again, algorithms and code to do this analysis will be developed for Stage III
experiments, but since the Stage III survey data will not yield sufficiently
long or well-sampled light curves for time-delay estimation, the interest is
mainly in the candidate classification. We can expect the principles of the
solution to be established during this phase. The greater information content
of the LSST data will enable a more exacting classification to be made: can
differential dust extinction be detected and used as a lens signature? How
flexible does the lens galaxy light distribution model need to be for very
accurate lensed source light curves to be derived? Will the lensed source host
galaxy ring need to be included in the model? How useful will this additional
information be? Will the PSF model need to be refined during the fit?

\subsubsection{Checking the lens samples \;} 
At the quality control stage, samples of
several $10^4$ classified  lens candidates and their models will need to be
inspected to check and refine the models, catch unforeseen failure modes, and
rank the systems for follow-up. Who will carry out these tasks? The automated
classification needs to be good, in order to make a pure sample, suggesting
that the time needed to understand a typical candidate, making sure its model
makes sense and potentially re-running it, could be several minutes per
system. A large group of analysts may be needed to do this work. One option is
to use an online crowd-sourcing platform as a ``Lens Zoo,'' to open up access
to the candidates to large numbers of citizen scientists as well as
collaboration members. The Galaxy Zoo user distribution has  a long tail of
highly skilled and motivated volunteers, who could readily assist with the
LSST quality control effort. Such a zoo is under construction in Fall 2012
with the goal of helping find lenses of all kinds, not just time delay or
multiple source plane systems, in Stage III data: if successful, this could
evolve into a useful facility for LSST strong lensing.

The detection of galaxy-scale multiple source plane systems could depend  more
heavily on human inspection: the secondary arc system is likely to be very
faint in most cases, and hence not easily picked up by an automated image
modeler. How well these systems are detected by humans during candidate
inspection will also depend on how the image is presented: our aim is to
sharpen these features, making them easier to spot. Indeed, any visualization
of strong lens candidates seeking to combine many exposures (with varying
resolution) together will need to optimize the final image quality for best
results. Probabilistic deconvolution of the cutout images (that is, fitting a
flexible non-parametric model to all available exposures) can achieve this in
principle, but the challenge is to do it cleanly, without introducing
artifacts, and within a reasonable amount of CPU time. Such an approach would
make the joint analysis of Euclid and LSST images straightforward, and bring
the best out of both datasets. Optimal combination of Euclid and LSST images is a
goal for the strong lensing groups of both missions. How can the different
filters' images be analyzed together to make the best-possible human-viewable
images?

\subsubsection{Inferring time delays in the presence of microlensing}
The light curves of the lensed sources extracted from LSST images will be long,
but not necessarily better sampled than the current state of the art
\citep[e.g.,][]{Courbin++2011}. It has been shown by, for example, 
\cite{Bur++1520,Morgan++2010,Kochanek++2007} 
that
the microlensing effects of the stars in the lens galaxy on the images
of the background point source can be accounted for given good data,
but that the modeling can be very computationally expensive. Can lens
light curve modeling be performed more efficiently without introducing
systematic error in the time delays?

The observing strategy itself could be a source of systematic as well
as statistical uncertainty: short seasons leave gaps in the light curve
that could be problematic \citep[e.g.,][]{PRH2,Kun++97}. Research is
needed to quantify the impact of this on the inferred time delays. The
cadence within a season must be high enough to provide sufficient
sampling of both the intrinsic variability of the source and the
microlensing applied on top of it. In principle, there is a factor of
5 or so in sampling rate available to us by analyzing the light curves
in all filters simultaneously. However, to do this well will require
accurately modeling the color variability of the source, and the
source size and microlensing caustic pattern.

Software developed to implement efficient microlensing correction will be useful
through Stage III, where the follow-up monitoring data are taken in a single
filter. The multi-filter analysis is an LSST-specific project, requiring
development on simulated light curves in preparation for the survey data. The
main motivation for studying this now is to provide feedback on the survey
strategy, and provide estimates of the distribution of precision in time-delays: 
the size of the cosmographic sample will be limited by our ability to measure
time delays.

\subsubsection{Detailed follow-up of 1000 lenses with JWST and Extremely Large Telescopes}
Obtaining and analyzing the deep, high spatial and spectral resolution follow-up
observations that the strong lensing program requires 
represents a significant challenge. 
The calibration and modeling software will need to be made robust and
semi-automated without loss of accuracy, a process that will need to make
extensive use of high fidelity simulated datasets. 
The evolution from Stage
III can be gradual, however, as around $100$ lenses are analyzed in the next
decade. Perhaps the biggest problem we face is one of manpower: at present, 
modeling a single system requires months
of work by an expert analyst. The CPU time required can also run into weeks. 
Scaling up both the hands-on and hands-off parts of the analysis 
will be needed. At the same time, the follow-up data will become more complex.
It is likely that the new integral field unit data will need internal
calibration in terms of flat-fielding and PSF estimation: the deconvolution
part of the procedure may need to be performed semi-blind. With so much
information encoded in, for example, the velocity field of the source, the 
lens models will be able to account for small scale mass structure close to
the critical curves.  In turn, the additional parameters needed to explain
these rich datasets will need to be sampled and marginalized over, indicating
that improved inference engines will be needed.
A major challenge will be developing fast, robust
and accurate software to speed up the modeling process. 
Graphic processing units (GPUs)
offer a solution. The reconstruction methods we use are inherently
parallel and have high arithmetic intensity. As such, they are
particularly suited for high-speed computing with GPUs and in the next
five years we will use this new technology to improve our modeling.

\subsubsection{Accurately characterizing external mass structures} 
A lack of knowledge about the line-of-sight mass distribution are the
dominant source of uncertainty for time-delay
lenses \citep{SuyuEtal10}.  Since surface brightness (or photon
number) is conserved, we expect the additional convergence provided by
mass along the line of sight to average to zero over a large sample of
lenses, provided those lenses are selected in an unbiased
way \citep[e.g.,][]{Seljak1994,Fassnacht++2011}. While the size of the associated
residual systematic error remains to be estimated, understanding the
selection function may become important in correcting for it. Using
more information on each lens' environment should allow the external
convergence to be estimated and the uncertainty in the distances
decreased: however, care must be taken not to introduce large biases
during this process.  Residual systematic errors can arise from the
recipe for assigning mass to light, and from the choice of calibration
cosmological simulation. Modeling the mass in the field on multiple
lens planes will require an improved formalism for describing the
effect of multiple plane lensing on the time delays. A further
complication with modeling the lens environment is how to incorporate
small scale invisible structures: these may cause strong perturbations
to the image fluxes, but relatively weak perturbations on time
delays. Are their populations characterizable as simple mass sheets?
Can their clustering around visible galaxies be used? Clustering on
the group scale is also likely to be important: can we model groups of
galaxies accurately enough to reduce the scatter in the distance
estimates without causing additional bias?

While the analysis techniques for modeling line of sight mass structure will
evolve during stage III, the 1000 lens sample provided by LSST will  place
higher demands. We anticipate spending substantial amounts of time before
first light in investigating environmental effects, by comparing multiple
approaches based on halo models and numerical simulations. We also anticipate
the need for theoretical work to properly model the propagation of light rays
through a inhomogeneous universe to the level of precision required by the
LSST dataset. 

\subsubsection{Establishing the viability of multiple source plane cosmography} 
In both the galaxy-scale and cluster scale multiple source plane lens cases,
significant work will have to be undertaken in the coming years to quantify
systematic uncertainties and how to mitigate them. Detailed forecasts for DE
parameter constraints from plausible samples of LSST lenses need to be
developed. Systematics due to lens modeling (including the first source as a
perturbing lens, in galaxy scale case) are not yet well understood;
nor has the detectability
of multiple image systems using LSST images 
been determined. The small number of available galaxy-scale systems make this a Stage IV,
not III, project. Well-studied clusters of galaxies, each with many sources on
many different planes, are being studied already, with progress on the
systematic uncertainties being made. Galaxy-scale systems should have
different sensitivities to some of the modeling systematics, as their
background sources are large compared to the scales of the caustic features.
Nevertheless, full investigation of the limits of theses systems for dark
energy science is needed.

%% file: analysis/theory/overview.tex
LSST will enable transformative improvements in our understanding 
of dark energy and the nature of gravity on cosmic scales. Realizing 
optimal constraints on the nature of dark energy from LSST will 
involve the collaborative efforts of experts in observation, theory, 
simulation, and data analysis. 
The same LSST survey data can be used for multiple dark energy probes, 
which is a crucial advantage. These probes will form multiple lines 
of attack at the puzzle of dark energy and provide internal cross 
checks that could potentially detect unknown systematics. Moreover, 
when analyzed jointly, these probes will enable self-calibrations of 
systematics, reduce degeneracies between parameters, and strengthen 
the constraints on dark energy properties. Dark energy studies will 
also benefit from the combination and cross-correlation of LSST survey 
data with precursor and contemporary external datasets. It will be a 
major undertaking to coordinate the connections of all these 
multiple efforts into the cosmological constraint analysis pipeline. 
This includes the formation of a software framework to facilitate a 
common platform for testing and integrating data analysis, simulation 
and theoretical prediction codes, and data products.

%% file: analysis/theory/steps.tex
As seen in previous sections, LSST will offer many ways to study 
DE. Here we focus on the joint analysis of the main probes:
WL and LSS two-point statistics, cluster mass function, type Ia SN 
distances, and SL time delay. Potentially powerful probes such 
as higher order WL and LSS statistics will be included as the method 
of joint analysis with other probes becomes mature. 

The general steps of analyzing LSST data for DE investigations 
are as follows. 
\begin{enumerate}
\item Reduce the images to collections of objects and auxiliary data
(such as masks, observing patterns, etc.) with relevant properties 
determined. 

This is primarily performed by the LSST data management (DM) team with 
requirements set by relevant working groups (WGs). Crucial contributions 
from these WGs should be integrated into the DM pipeline well in advance.

\item Extract statistics or quantities of interest from the objects. 

For WL and LSS, these would include not only the statistics (such as 
two-point correlation functions) measured by their 
respective WG but also cross statistics between the shear and galaxy 
fluctuations. For clusters, the main statistic is the mass function 
at different redshifts. For SN and SL, by their very nature, individual 
objects are studied in great detail to derive distances. 

\item Turning statistics into DE constraints

In the usual Bayesian approach, one turns the likelihood of the data 
given the model into the posterior probability distribution of the 
model by sampling the model space with tools like Markov Chain Monte Carlos.

Predicting the likelihood is critical to the analyses. For WL and LSS, 
the likelihood is determined by theory on linear scales, and
simulations and modeling on nonlinear scales. Observational effects
such as the photo-$z$ errors also fold into the likelihood. Since 
clusters are highly nonlinear objects, simulations are needed to 
accurately predict the likelihood of their mass function (and certainly
the mass-observable relation). In the case of SNe and SL, the 
likelihood is driven by intrinsic uncertainties of object properties 
and random measurement errors.

\end{enumerate}

External data may be used in all the steps above. For example, infrared
image data will have to be processed together with LSST image data to 
reduce the photo-$z$ systematics. Cross correlations between the CMB 
and the LSST galaxy distribution will be calculated in step 
(2) and analyzed with other data in step (3) to include the ISW effect
in constraining DE.

%% file: analysis/theory/challenges.tex
\subsubsection{Dark energy forecasting and analysis pipeline} 
\label{challenges:theory:int}

LSST's survey data should reveal far more than whether the equation of 
state parameter for dark energy is $-1$, as it is for $\Lambda$CDM, or not. In looking to address the theoretical fine-tuning and coincidence problems associated with $\Lambda$CDM,  a wealth of alternative explanations for acceleration have been considered. These could be distinguished from $\Lambda$CDM by looking for evidence of a dynamical equation of state parameter, such as arises for a scalar field; or evidence that dark energy
interacts with matter, that it clusters, that it was important in the early 
universe, or that it derives from a modification to GR on cosmic 
scales. To fully utilize LSST's data will require careful determination 
of theoretical predictions for survey observables for the spectrum of 
models that LSST can test. This includes model-independent 
parameterizations, effective field theory approaches to DE and dark 
energy-dark matter interactions, and models of modified gravity, 
such as those derived from massive and higher dimensional theories of 
gravity.

LSST's large survey area and depth will provide an unprecedented 
opportunity to test dark energy's properties at many different spatial 
scales and redshifts. This opens up the possibility of studying the 
transition from linear to nonlinear scales for evidence of screening 
mechanisms present in a variety of modified gravity theories. These 
analyses will require coordinated, detailed simulations to model 
structure growth that can be quite distinct from that in LCDM models. 
Combining, and contrasting, LSST's lensing and galaxy clustering data 
could provide a distinctive test of GR. We will need to develop a pipeline that allows combination with 
complementary external CMB intergrated Sachs-Wolfe (ISW) and lensing data, 
galaxy clustering measurements from spectroscopic datasets, and HI data, amongst others. 
This will enhance LSST data's application as a highly sensitive test of gravity, by 
contrasting relativistic and non-relativistic tracers of the same 
gravitational potentials \citep{Zhang:2007nk,Reyes:2010tr}.

Even neglecting systematic uncertainties, galaxy clustering and lensing 
are not pristine measurements of dark energy and modified gravity. In 
comparing observations from halo to horizon scales, and multiple 
tomographic redshift slices, we will have to include relativistic 
effects and primordial non-Gaussianity \citep{dalal08}, running in 
the primordial power spectrum \citep{Yoo2009,Yoo2010,Bonvin&Durrer2011, 
Challinor&Lewis2011}, and the impact of neutrino mass and 
additional relativistic species, amongst other possible degenerate 
cosmological parameters, to unambiguously disentangle them from the dark 
energy signatures we are looking for.

One of the most important activities is the joint analysis of probes, 
incorporating best estimates of the LSST data model from measurements on 
actual LSST subsystems and other data (at the telescope site, etc.).   This
DESC effort will couple strongly with activities in the LSST  project, both
ways, to determine a realistic data model. This will  decrease the sensitivity
to systematics, and enable the DESC to  prioritize which residual systematics
need most attention.
Likewise, new DE probes and their systematics will be able to be tested for
their statistical competitiveness, and their power to break  parameter
degeneracies, as they are suggested. Multiple source plane strong lens
cosmography is one example of an LSST DE  probe with as-yet-unexplored
potential, that would benefit from study within the joint analysis pipeline.
This pipeline will provide  feedback into the optimal observing strategy and
system design  (including design and quality assurance metrics, the data
management  pipeline, database, which site-specific data to keep, etc.).  This 
critical tie between the LSST DESC and project activities will start  early
in the DESC's activities, with staged updates as LSST progresses  through
R\&D, construction, and into commissioning.

\subsubsection{Simulation challenge}
\label{challenges:theory:sim}

Simulations will be crucial to LSST DE analyses. The challenge
is to accurately predict or quantify (1) the observables such as statistics 
of WL shear and galaxy number density fluctuations, (2) observational 
effects such as those caused by instrumental effects, observing patterns, 
complex masks, etc., and (3) the likelihood function of the observables in 
a given model, so that their uncertainties do not become a dominant source 
of error in the LSST DE analyses. This will require a major effort in
the Simulations WG (see Chapter \ref{chp:sims}). The Theory/Joint Probes 
WG will contribute significantly by analyzing this wide range of effects, 
determining practical requirements on the simulations, exploring ways to
reduce the sensitivity to potentially detrimental effects, and iterating
with the Simulations WG to make sure that the DESC will meet its goal. 

As an example, present N-body simulations (see Section \ref{sec:cosim}) can 
predict the matter power spectrum with an accuracy approaching 1\% on 
scales $k \lesssim 1 \hmpci$ \citep{lawrence_etal2010}. In order to keep 
the uncertainties in the matter power spectrum a negligible portion 
of the errors in the inferred DE equation of state parameter, 
it will be necessary to predict the power spectrum with an accuracy of 
$\sim 0.5\%$ on scales $k \lesssim 5 \hmpci$ 
\citep{huterer_etal2005,hearin_etal2012}. Similarly, precise predictions are 
required for other statistics. Moreover, these levels of precision 
must be achieved for models that span the relevant cosmological 
parameter space. 

Separately, the effects of baryons on the matter power spectrum 
(and hence lensing and galaxy power spectra), abundances of galaxy 
clusters, and density profiles of halos can be significant
\citep{white_2004,zhan_etal2004,jing_etal2006,rudd_etal2008,zentner_etal2008,stanek_etal2009,vandaalen_etal2011,semboloni_etal2011} and can 
induce systematic biases in the DE equation of state parameters
\citep{zentner_etal2008,hearin_etal2009,semboloni_etal2011,hearin_etal2012}. 
It has been demonstrated that a parametrized model can capture the 
feature(s) of nonlinear and baryonic effects \citep[e.g.,][]{rudd_etal2008}. 
This method will be further developed along with simulation tasks
in Task~\ref{task:cosmosim}.

\subsubsection{Galaxy modeling challenge}

A number of DE probes derive DE properties from the 
matter distribution and its evolution over cosmic time, which are 
inferred from measurements of galaxy positions (e.g., BAO and clusters) 
and shapes (e.g., weak lensing). Because of the highly 
complex and nonlinear nature of galaxy formation and evolution, it is 
currently not possible to accurately connect galaxy properties to the 
underlying matter distribution. Significant progress must be made to
reduce the uncertainties to the levels acceptable to LSST.

The relative spatial clustering amplitude for galaxies and dark 
matter is referred to as the galaxy clustering ``bias,'' which is 
generally expected to be scale-independent and deterministic on large 
scales ($\gtrsim 100 \hmpc$). But on very large scales, general 
relativistic corrections and primordial non-Gaussianity could introduce 
a scale-dependence to the bias \citep{baldauf11, bruni12}. These 
effects should be taken into account in analyses. On smaller 
scales, the gravitational evolution of structures creates a 
scale-dependent and partially stochastic bias. If not properly modeled, 
the evolution of bias with cosmic time and scale could be confused with 
the linear growth rate of structures, shifts in the scale of the BAO 
peak, and ``tilts'' in the shapes of the $n$-point correlation functions. 
The precision required on small scales is comparable to those in 
the last item.

Weak lensing measures the correlated ellipticities of galaxies, which 
are determined by a sum of intrinsic and lensing-induced ellipticities. 
Galaxy intrinsic alignments (IAs) are created by the formation of 
galaxies in large-scale gravitational potentials \citep{catelan01}. 
The IA signal can be a significant systematic
contamination to the lensing signal \citep{hirata04}. DE 
parameters and modified gravity parameters can be biased by more than 
100\% if no modeling or mitigation of IA contamination is included in 
the LSST weak-lensing analysis \citep{kirk12, laszlo12}. It is thus
imperative to establish IA models that can accurately predict the 
alignments between orientations of galaxies and their surrounding 
matter distribution. At the same time, one should also explore and
demonstrate methods to mitigate the IA effects, e.g., 
``self-calibration'' with correlations between the weak-lensing shear 
and the galaxy distribution \citep{zhang10}.

\subsubsection{Measurement on very large scales}
\label{challenges:theory:largescale}

Analyzing the clustering on scales comparable to the Hubble scale can 
allow the best constraints on dark energy and modified gravity models 
and test general relativity itself.

We need to improve the theoretical modeling of the clustering on large 
scales. The Kaiser approximation needs to be modified \citep{Szalay1998, 
Matsubara:1999du,Szapudi2004,Papai_Szapudi2008, 
Raccanelli2010,Samushia2012}. General Relativistic 
corrections \citep{Yoo2009,Yoo2010,Bonvin&Durrer2011, 
Challinor&Lewis2011,Yoo:2012se,Jeong2012,Bertacca:2012tp}
and degeneracies with non-Gaussian effects \citep{Bruni2012, 
Maartens:2012rh} must be included. 
To extract the dark energy signatures 
from degenerate cosmological effects, the theoretical 
predictions that simultaneously account for primordial non-Gaussianity, and
relativistic and modified-gravity effects must be refined.

Large numerical simulations are required to understand the large-scale 
structure at high precision or the statistics of rare objects. 
Typically, simulations covering a substantial fraction of the Hubble 
volume trace only the evolution of the dark matter in N-body 
calculations; however, the importance of modeling baryonic effects in 
large cosmological volumes is well recognized \citep{dimatteo12}, including 
changes in the shape of the power spectrum measured by weak lensing. The 
simulation must have enough mass resolution to resolve the 
gravitationally bound dark matter halos that create the potential wells 
for galaxy formation.

Current state-of-the-art dark-matter-only simulations have used several 
hundred billion mass tracer particles in volumes many Gpc on a side, 
requiring thousands of parallel processors and of order 10 million 
cpu-hours per run \citep{springel12, alimi12}. Analysis of the LSST survey 
data will require many simulations of comparable volumes to explore the 
different physical models discussed in the previous paragraph, include 
gas physics in Gpc volumes, and the complicated selection functions 
\citep{overzier12}. This will yield the required precision for numerical and 
``semi-analytic'' models for galaxy formation to accurately compare with 
observations.

\subsubsection{Photo-$z$ modeling}
\label{challenges:theory:photos}

Photo-$z$ errors have a critical impact on all LSST DE probes. 
The effects are twofold: they 
randomize galaxy positions in the line-of-sight direction, causing a 
loss of information; and the uncertainty in the error distribution leads 
to uncertain predictions of the observables. In addition to weakening 
DE constraints, systematic errors in the estimated photo-$z$ error 
distribution can bias the results.

It has been shown that WL constraints on the DE equation-of-state 
parameters are sensitive to uncertainties in the photo-$z$ error 
distribution \citep[e.g.,][]{ma_etal2006}. BAO constraints are 
sensitive to the width of the photo-$z$ error distribution
\citep[e.g.,][]{seo_eisenstein2003}. 
Constraints from a joint analysis of 
the two probes are much less prone to photo-$z$ issues 
\citep{2006JCAP...08..008Z}. 
These results were obtained with relatively simple photo-$z$ error 
models. To assess LSST DE science capability with high fidelity 
and to properly analyze the data in the future, we must include a 
realistic model of the photo-$z$ error distribution. The challenge is to 
determine the photo-$z$ error distribution and its variation accurately
and to model it with as few parameters as possible. With this model,
one can then assess LSST DE capability more realistically and explore 
methods to mitigate the adverse impact of photo-$z$ errors, which should
be demonstrated with precursor data and incorporated in LSST DE analyses.

\subsubsection{Combining with external datasets: theory and systematics mitigation}
\label{challenges:theory:ext}
Galaxy distributions, peculiar motions and lensing shear and 
magnification measurements trace multiple, independent fields within the 
same cosmic volume.  As such these complementary measurements can go 
well beyond the naive cosmic variance limit.
Cross correlating multiple probes, constrasting 
velocity, density and gravitational lensing fields, can provide a 
distinctive test of whether dark energy derives from a modification to 
gravity on cosmic scales \citep{Zhang:2007nk, Reyes:2010tr}. At the heart 
of this technique is the use of the lensing and galaxy clustering data 
in tandem to both test gravity and constrain the galaxy bias (a key 
systematic uncertainty).  When taken together multiple large-scale 
structure probes, or relativistic tracers, such as galaxy lensing and 
the CMB ISW effect, and non-relativistic tracers, peculiar motions, 
and potential kinetic SZ measurements both break cosmological 
degeneracies and calibrate astrophysical uncertainties such as galaxy 
bias, intrinsic alignments, and shear calibration offsets to which their 
responses differ \cite{Laszlo:2011sv}. A number of complementary 
surveys will have notable overlaps with the LSST field, including 
BOSS/eBOSS, BigBOSS, DES/DESpec, Euclid, and the ACT, Planck and WMAP CMB surveys, 
amongst others. Where current datasets have been correlated, the results 
have had a major impact \citep[e.g.,][]{Reyes:2010tr,Hand:2012ui}. The 
cross-correlation of datasets is still quite a nascent area of study, 
offering tantalizing opportunities that have not yet been fully 
developed in a practical setting, with real data 
\citep[e.g.,][]{Baldauf:2009vj,Hikage:2011ut}. LSST will provide one of the 
central datasets with which the full power of cross-correlation will be 
able to be explored.

{\bf LSST and other optical/near-IR surveys.} At optical and near-infrared wavelengths, a suite of  new ground-based surveys are about to come on-line that will provide crucial pre-cursor datasets in advance of LSST. These include the
Panoramic Survey Telescope and Rapid Response System (Pan-STARRS); the
Dark Energy Survey (DES); the KIlo-Degree Survey (KIDS) and the
complementary VISTA Kilo-degree INfrared Galaxy survey (VIKING); and the
Subaru Hyper Suprime-Cam survey (HSC). These experiments offer significant potential
for a broad range of science including most of the science goals
spanned by LSST. 

Planned, space-based survey missions such as the Wide
Field Infrared Survey Telescope (WFIRST) and the European Space 
Agency's Euclid also offer
outstanding potential and synergies with LSST, complementing the
ground-based data in particular by extending search volumes to
higher redshifts.  Euclid  will map the same sky with a broad optical 
band with lower sensitivity, but higher resolution than LSST.  Euclid 
will provide an independent measurement of cosmic shear.  Its 
measurements will be more sensitive to the shear in the inner regions of 
a typical galaxy ($\theta < 0.5''$), while LSST will be more sensitive to 
shear measured in the galaxy's outer isophotes.  By comparing and 
cross-correlating these two measurements, we will be able to better 
characterize systematics in each experiment and obtain a shear power 
spectrum that will be nearly independent of the systematics in each 
experiment. Euclid will also complement LSST photometric measurements by making 
near-infrared measurements with its three broad infrared bands.  These 
measurements should significantly improve LSST photo-$z$ estimates and 
enable higher precision tomographic measurements \citep{Abdalla}.

{\bf LSST and spectroscopic surveys.} 
{\it Breaking cosmological degeneracy.} Spectroscopic redshift surveys 
like BOSS, eBOSS, BigBOSS, DESpec, or Euclid will complement LSST dark energy 
constraints by adding independent cosmological measurements, which break 
degeneracies. The complementarity is enhanced when the surveys overlap 
on the sky so that their different tracers measure exactly the same 
density fluctuations (e.g., LSST lensing and galaxy densities, and redshift-survey 
galaxy density) and permit cross-correlation between the redshift 
survey and photometric survey data to calibrate photo-$z$'s. The redshift 
survey helps measure geometry through 3-d BAO and more general 
Alcock-Paczynski-like uses of the broadband redshift-space power 
spectrum. It also constrains the growth of structure through 
redshift-space distortions and higher order statistics (e.g., 
bispectrum) that should break degeneracies in the lensing survey dark 
energy / modified gravity parameters by providing constraints on 
nonrelativistic tracers of the gravitational potentials (e.g., \cite{Zhang:2007nk}) 
and by helping constrain the effects of neutrino masses and 
inflation parameters on the shape of the power spectrum.

{\bf LSST and CMB.} 
Large-scale CMB fluctuations are mostly 
sensitive to physical conditions at $z =1100$ and low-redshift physics, 
including the integrated ISW effect, for which 
cross-correlation with LSST tomographic slices will provide a measurement  
of the growth of structure during accelerative expansion. The small-scale 
CMB statistics contain a wealth of information about physical 
effects, all of which can be cross-correlated with the LSST data in a 
number of interesting ways.

{\it CMB deflections and cosmic shear.} Just like galaxy-emitted 
photons, the CMB photons on their way from the surface of last 
scattering are gravitationally lensed by the intervening matter 
distribution. This effect was detected for the first time in 
cross-correlation techniques with WMAP data \citep{Smith2007, Hirata_2008}
or more recently though the lensing power spectrum by ACT 
\citep{Das:2011ak} and SPT \citep{vanEngelen:2012va}.  Planck will soon 
release its temperature maps and its polarization maps in 2014.  These 
maps will provide S/N$\sim1$ measurements of the cosmic shear field with 
2$^\circ$ resolution. By 2015, ACTPOL and SPTPOL's wide observations will 
provide a $S/N > 1$ lensing map with $20'$ resolution.  By 2020, we 
anticipate that ground-based CMB polarization maps will provide lensing 
maps with $< 10'$ resolution.  We plan to cross-correlate these CMB 
lensing maps with LSST lensing maps to trace the evolution of matter 
fluctuations over a wider range of redshifts. \cite{Song&Knox2004} 
anticipated that the cross-correlations between Planck lensing and LSST 
will significantly improve the determination of cosmological parameters. 
Since small-scale CMB experiments have much higher-resolution lensing 
maps, the combination of these measurements with LSST will yield even 
more powerful constraints on the growth rate of structure. This will 
increase our sensitivity to the early evolution of dark energy and to 
neutrino mass.

{\it CMB deflection and galaxy counts.} While LSST survey data focuses 
on structure at $z\approx0-2$, CMB lensing surveys provided by the 
Planck satellite and higher-resolution ground-based CMB polarization 
experiments will probe the matter power spectrum at  
$z\approx$1--5.  Cross-correlation determines the bias of the tracer 
galaxies and normalizes the amplitude of the galaxy power spectrum (see \cite{ 
Bleem:2012gm} and \cite{Sherwin:2012mr} for recent measurements of CMB 
lensing cross-correlations with $z\sim $1--2  samples). In conjunction the 
dark energy figure of merit can be improved by a factor of 2 relative to 
neglecting CMB lensing.

{\it Thermal Sunyaev-Zeldovich (tSZ) effect -- measuring the integrated 
pressure.} Over the next 2--3 years, the completion of the SPT, ACT, and Planck
tSZ catalogs will extend our detailed, statistical knowledge of galaxy
clusters out to $z>1$, with a well-defined, nearly mass-selected, cluster sample. Together, these projects expect to find $\sim 1000$ new clusters, mostly at intermediate to high redshifts.
Used in combination with
existing low-redshift X-ray and optical catalogs, they should provide
significant near-term improvements in our knowledge of cluster growth,
and corresponding improvements in the constraints on dark energy and
gravity models. In the LSST era, the development of experiments with
improved sensitivity, frequency coverage and spatial resolution will,
like X-ray observations, provide an excellent complement to LSST both
in cluster finding and additional mass-proxy information.

{\it Kinematic Sunyaev-Zeldovich (kSZ) effect - measuring the integrated 
momentum.} The kSZ effect traces the large-scale velocity fields and has 
the potential of tracing large-scale velocity flows. The first detection 
of this effect was recently made by cross-correlating CMB and 
spectroscopic galaxy data \citep{Handetal2012}.  By using external 
spectroscopic samples (e.g., BOSS, BigBOSS, DESpec, PFS) to determine the 
large-scale momentum field \citep{Hoetal2009} and the LSST data to 
determine the galaxy positions, we can measure the cross-correlation 
between the galaxy momentum field and the kSZ signal.  This 
cross-correlation measures the large-scale distribution of electrons 
around galaxies.  By measuring these correlations as a function of 
redshift, in the context of dark energy, this would provide us with a 
complementary measure of the gravitational fields and their evolution as 
a function of redshift; more generally, it will tell us about how galaxy 
feedback drives gas into the intergalactic medium.

{\bf LSST and X-ray surveys.} X-ray observations currently offer
the most mature and precise technique for constructing cluster
catalogs, the primary advantages being excellent purity and
completeness and the tight correlations between X-ray observables and
mass.

Over the next 2-3 years, X-ray cluster samples constructed from ROSAT,
Chandra and XMM observations will continue to offer important
gains in cosmological constraints. The next major advance at X-ray
wavelengths, however, will be provided by the eROSITA telescope on the
Spektrum-Roentgen-Gamma Mission.  Scheduled for launch in 2014,
eROSITA will perform a four-year, all-sky survey that should detect an
estimated 50,000--100,000 clusters.  The cross-correlation of LSST and
eROSITA-detected clusters in particular offers the potential for the
construction of large, landmark cluster catalogs with exquisite purity
and completeness, and systematic uncertainties controlled to levels
far exceeding those possible with either experiment alone.

X-ray observations provide the best mass proxies for galaxy clusters,
with observed scatters of $<15\%$ over the full redshift range
of interest. Follow-up observations with high throughput X-ray
telescopes will therefore be essential for LSST cluster science. X-ray
observatories like Chandra, XMM, Suzaku,  ASTRO-H, and
eROSITA will be the cornerstones of this work.

{\bf LSST and hydrogen line (HI) surveys.} 
LSST by itself will produce deep imaging 
of a large fraction of the sky and will produce photometric, not 
spectroscopic, redshifts for the galaxies detected.  A complementary 
approach to large-scale structure, under development, is mapping the 
21-cm emission from galaxies (HI surveys).  
HI surveys contemporary with LSST will map the large-scale features in 
HI emission with a technique called intensity mapping, where individual 
galaxies are not resolved. With sufficient improvement in angular resolution (say $10'$ 
or better) and low enough noise in the HI map (say smaller than 
50\,$\mu$K) one will be able to resolve nonlinear concentrations of HI 
corresponding to groups of galaxies at redshifts 1--2.
 In  combination, a Southern Hemisphere HI  survey plus  LSST could provide a  complementary source of redshift information and tell us about the evolution of the HI 
content by galaxy type, which will be of great use for constraining models of galaxy 
formation and evolution.

%% file: analysis/photoz/overview.tex
All LSST probes of dark energy and related physics rely on determining the behavior of some quantity as a function of redshift $z$.  Distances, the growth rate of dark matter fluctuations, and the expansion rate of the Universe are all functions of redshift that can be readily calculated given a cosmological model; dark energy experiments then constrain cosmological parameters by measuring observables dependent upon these functions.  However, it is completely infeasible with either current or near-future instruments to obtain redshifts via spectroscopy for so large a number of galaxies, so widely distributed, and extending to such faint magnitudes, as those studied by LSST.  

Hence, LSST will primarily rely on {\it photometric redshifts} --  i.e., estimates of the redshift (or the probability distribution of possible redshifts, $p(z)$) for an object based only on imaging information, rather than spectroscopy \citep{1985PhDT........17S,1999ASPC..191....3K}.  Effectively, multiband (e.g., $ugrizy$) imaging provides a very low-resolution spectrum of an object, which can be used to constrain its redshift.  Because flux from a relatively wide wavelength range is being combined for each filter, imaging provides a higher signal-to-noise ratio than spectroscopy; 
however, broader filters provide cruder information on the spectrum, and hence on $z$ or $p(z)$.  

For most LSST probes of dark energy (BAO is the primary exception), what will matter most is not the precision with which individual photometric redshifts are measured, but rather the degree to which we understand the actual redshift distributions of LSST samples; if photo-$z$'s are systematically biased or their errors are poorly understood, dark energy inference will be biased as well (e.g., because we are effectively measuring the distance to a different redshift than is assumed in calculations).   For LSST, it is estimated that the mean redshift and $z$ dispersion for samples of objects in a single photo-$z$ bin must be known to $\sim 2\times 10^{-3} (1+z)$ (\citealt{2006ApJ...644..663Z,2006JCAP...08..008Z,2006JCAP...08..008Z,tysonconf}) for dark energy inference not to be systematically degraded.  As a result, this working group's efforts will be primarily focused on making sure that the photometric redshifts we obtain are accurate and well-understood.

%% file: analysis/photoz/analysis.tex
{\bf Survey Photometry:} The determination of a photometric redshift for an LSST galaxy will begin with measurements of the flux from an object in all LSST bands, provided by data management.  
Most efficiently, these may be the aggregate fluxes from combining information from all images in a given band.  However, the maximum information would be obtained by utilizing all single-visit measurements of an object to estimate the photometric redshift (see also Section~\ref{photoz:challenges}).

{\bf Photometric Redshift Determination:} A wide variety of techniques for estimating photometric redshifts (commonly referred to as ``photo-$z$'s'') have been developed, ranging from the most straightforward -- determining the redshift that minimizes the $\chi^2$ or maximizes the likelihood when some set of empirical or synthetic template galaxy spectra are fit to the observed multiband photometry of a galaxy -- to Bayesian methods, to neural-network or 
other machine-learning based estimators that rely on having a training set that matches the galaxies to which the method will be applied \citep[e.g.,][]{2009ApJ...695..747B}.  In current samples, a wide variety of techniques offer very similar performance.  In many cases, for dark energy inference we will want to determine a probability distribution for the redshift of each object, $p(z)$, rather than describing the photometric redshift as a single value with uncertainty.  

{\bf Training Photometric Redshift Algorithms:} Although it is not part of the workflow for measuring photometric redshifts for individual objects, a key ingredient for all LSST photometric redshift measurements is a training set of galaxies with highly reliable redshifts.  For template-based methods, this training set is generally used to refine template galaxy spectral energy distributions and effective photometric calibrations that are then used to determine the posterior probability of a given redshift.  A wide variety of other, training-set--based techniques rely on having a set of objects that can be used to directly map relationships between photometric properties and redshift.  Those methods take advantage of advances made in the field of machine learning in recent decades; however, they in general require a training set with uniform or well-understood completeness that spans the full range in properties of the galaxies whose photo-$z$'s will be measured, as they extrapolate extremely poorly.  

Most estimates are that training sets of 20,000 objects or more will be required so that LSST dark energy inference is not dominated by systematics \citep[e.g.,][]{2009arXiv0902.2782B,2012JCAP...04..034H}. However, as described in Section~\ref{photoz:challenges}, obtaining a complete dataset of this size to LSST depth will pose major challenges.

{\bf Calibrating systematic biases due to incomplete training sets:} Fortunately, even if our photometric redshift measurements have systematic biases (e.g., due to incompleteness in the training set), so long as we can determine the level of those biases,  dark energy inference will generally remain unscathed.  A variety of cross-correlation techniques may be used to determine the actual redshift distribution of a sample, allowing us to detect biases that the likely incomplete spectroscopic training sets cannot.

For instance, cross-correlating the locations of galaxies of known spectroscopic redshift with a photo-$z$-selected sample on the sky, as a function of spectroscopic $z$, provides sufficient information to reconstruct the true, underlying redshift distribution of the photometric sample.   This method can achieve high accuracy even if it is only possible to obtain spectroscopic redshifts for  a bright, biased subset of the galaxies at a given redshift \citep{2008ApJ...684...88N}.  Cross-correlation techniques rely on the fact that all galaxies -- from the faintest to the brightest -- cluster together on large scales, tracing the underlying web of dark matter.  Hence,  galaxies in a spectroscopic sample that are at a given redshift will cluster on the sky solely with the subset of galaxies in other samples that are at or near that redshift.  We can exploit this fact to accurately determine the true redshift distribution of objects in a photometrically selected sample.

If we only measure cross-correlations, redshift distribution estimates are degenerate with the strength of the intrinsic, three-dimensional clustering between the two samples; however, the autocorrelations (i.e., degree of clustering with itself) of the photometric and spectroscopic samples -- some of the most basic measurements that are generally made from galaxy surveys -- can break that degeneracy.   Other cross-correlation techniques for testing photometric redshifts have been developed \citep{2006ApJ...644..663Z, 2006ApJ...651...14S} and are likely to be very useful for LSST, but they do not break that degeneracy.  Further advances on all these current methods (e.g., utilizing lensing tomography as a consistency check, by applying the techniques from Clowe et al. 2012 [in prep.]) could yield even better results.

%% file: analysis/photoz/challenges.tex
\label{photoz:challenges}

In this section, we describe the major issues we will need to address in order to utilize photometric redshift measurements for dark energy science.

{\bf Obtaining large, deep training sets.} Measuring a statistically complete set of $>20,000$ redshifts to full LSST depth will be a major challenge.  The  LSST `gold sample' for dark energy measurements extends to $i=25.3$, more than 10 times fainter than those objects for which redshifts may be obtained (with $\sim$(60--70)\% completeness) at 8--10\,m telescopes in one hour of observing time \citep{2009ApJS..184..218L,2012arXiv1203.3192N}.  At least 100 hours of observation time would thus be required to achieve that success rate for LSST calibration redshifts; with existing instruments, at most 100-300 redshifts can be obtained at a time, while the proposed PFS spectrograph \citep{2012arXiv1206.0737E} should be able to target $>2000$ objects at once.  Even a telescope with a 30\,m diameter mirror would require $>10$ hours to obtain redshifts to LSST depth, and would do so for only a very limited number of objects in a small field of view.  We will require a major international effort, which must be scoped and coordinated.  This will require a long lead-time to spread telescope-time allocations over many cycles, so must begin now.

{\bf Incompleteness in training sets.} Faint galaxy surveys have systematically failed to obtain redshifts for a substantial fraction of their targets. For instance, the DEEP2 Galaxy Redshift Survey, for redshift quality classes with $>99.5$\% (vs.~$>95$\%) reproducibility, obtained ``secure'' redshifts for only $60\%$ (vs.~$75\%$) of the $>50,000$ galaxies targeted \citep{2012arXiv1203.3192N}.  
Other surveys have done worse, with high-confidence redshift measurements for  (21--59)\% of targeted galaxies \citep{2005A&A...439..845L,2008A&A...486..683G,2009ApJS..184..218L}. 
Obtaining $>90\%$ redshift completeness for a red galaxy at the LSST magnitude limit would take roughly one thousand hours of observation time (more than 100 nights) on an 8--10\,m telescope.  Deep infrared spectroscopy from space could lessen these problems, but will not solve them, due to the small field of view (and hence high shot noise and sample variance) of JWST and the limited depth and wavelength range of WFIRST or Euclid spectroscopy.  

 If the sorts of galaxies that fail to yield redshifts are localized in color space, problem galaxies could be excluded from analyses at the cost of reduced sample size; however, this is unlikely to always be possible, and could weaken dark energy constraints.  For LSST, we will need to be prepared for the likelihood that we will  have only incomplete training sets, both in considering photo-$z$ algorithms and in establishing calibration plans.  Stage III surveys will need to make progress on this problem, as it is already a major issue at DES depth, but LSST photo-$z$ calibration requirements are $\sim 2\times$ more stringent, so additional efforts are necessary.

{\bf Erroneous calibration redshifts.}  Even when considering only objects with `robust',  $>95\%$ secure, redshift measurements, existing deep samples have incorrect-redshift rates ranging from (0.3--0.4)\% (for DEEP2, based on tests with repeated independent measurements) to $\gtsim 3$\% (for VVDS).  Since objects with incorrect redshifts are generally at higher redshift ($z \gtsim 0.5$), even DEEP2-like error rates may bias photo-$z$ calibration beyond LSST tolerances.  New, robust calibration algorithms will likely be needed for LSST.  This is also a serious issue for Stage~III surveys; we plan to proceed in cooperation with Stage~III projects on this challenge.

{\bf Impact of cosmic variance in training sets.}
Both training and testing of photo-$z$'s usually utilizes sets of galaxies with known spectroscopic redshifts.  However, in samples covering small sky area, including all current deep spectroscopic samples (e.g., DEEP2 and zCOSMOS), the number of objects observed at a given redshift will have substantially greater variance than expected for shot noise (generally referred to as ``sample'' or ``cosmic'' variance) due to the large-scale structure of the Universe; this is compounded by the fact that galaxy properties, including luminosity and color, correlate with the local matter density.    Sample variance can affect tests of overall redshift distributions and the training of priors for template methods, while training-set--based photo-$z$ methods will imprint any sample variance in the spectroscopic sample onto photo-$z$'s applied across the sky \citep[]{Cun:12}.   
If we can develop improved methods of taking out sample variance in training sets (e.g., correcting for the observed fluctuations in galaxy density with $z$ in the spectroscopic samples), photo-$z$ training and calibration will be more effective.

{\bf Optimizing training sets.}
Developing optimal photo-$z$ algorithms depends upon complete and spanning 
spectroscopic-redshift datasets, especially for training-set methods that cannot extrapolate.  It will be important to develop methods for calibrating and testing the effectiveness of photometric redshifts for faint galaxies, since it is observed both locally and at $z\sim 1$ that the spectral energy distributions (SEDs), or colors, of bright galaxies (whose redshifts are more easily obtained) differ systematically from those for fainter objects.  
Targeted spectroscopic follow-up can focus resources on  gaps in training sets from redshift surveys and on characterizing the tails of the photo-$z$ error distribution, which have outsize impact on dark energy inference \citep{2009arXiv0902.2782B,2010ApJ...720.1351H}.  
Active-learning algorithms may be used to identify problematic regions of parameter space and determine spectroscopic follow-up strategies (Budavari et al. 2013, in prep).  Given the large amounts of telescope time required for photo-$z$ calibration, anything we can do to increase the efficiency of training sets will improve our results.

{\bf Testing cross-correlation techniques.} Cross-correlation methods provide an extremely promising response to incompleteness in spectroscopic training sets.  However, tests of the technique to date have relied on simulated galaxy catalogs covering limited volume; those mock catalogs do not allow tests down to the fidelity required for LSST.    Better tests are needed.

{\bf Excluding potential outliers.} Even with perfect training sets, the limited wavelength coverage and broad bands of LSST result in some regions of color/magnitude space for which the observed colors are consistent with multiple redshifts.  For example, in many cases the Balmer break can be mistaken for the Lyman break, leading to ambiguities between $z\,\sim\,$0.2--0.3 galaxies and $z\,\sim\,$2--3  galaxies.  Rather than use every galaxy for dark energy measurements, those with known problems could be discarded \citep[e.g.,][]{Jai:07, Nis:10}.  
Removal of problematic galaxies will reduce outlier rates, but will also reduce the overall number of galaxies useable for science; we need to explore these tradeoffs to be able to  predict LSST system performance accurately.

{\bf Use of many-band photo-$z$'s for bootstrap calibration.}
Spectroscopic redshift measurements at LSST depth are very time-expensive.  Alternatively, given deep, many-band imaging information over a limited area of the survey, we can use the resulting high-precision photo-$z$'s as secondary calibrators for $ugrizy$ photometric redshifts.   Such efforts have yielded very promising results \citep[e.g.,][]{Ilb:09,Whit:11}, achieving $\sigma_{z}<0.01$ and catastrophic outlier rates of $<10\%$, though primarily for objects  at $z<1$, which are much brighter than the LSST gold-sample depth. 
Substantial investments of telescope time will be required to obtain deep many-band imaging over a significant total area of sky ($\gtsim 10$\,deg$^2$ spread over multiple fields), but this nonetheless may represent a more efficient strategy than relying only on spectroscopic calibrators.  Because of the large time requirements, we must scope our needs and begin observations soon.

{\bf Leveraging space-based, limited-area photometry.}
Upcoming missions such as EUCLID or  WFIRST are likely to obtain near-IR imaging over a large fraction, but not all, of the LSST survey area.  
Simulations of EUCLID-depth $J$, $H$, and/or $K_{s}$-band data combined with LSST demonstrate greatly reduced photo-$z$ scatter at $z>1.4$, where the Balmer break has shifted beyond the $y$-band.  Higher-resolution optical imaging from space will be helpful for testing star-galaxy separation, methods of determining sizes and shapes from LSST data (which can be fed back into photo-$z$ algorithms), and the effects of blending.  For some dark energy analyses, it may be better to utilize improved optical plus NIR photo-$z$'s where available, while in others that rely on uniform selection we will wish to use the space-based data to improve calibrations over the full survey area; we need to explore these tradeoffs.   Partnerships between LSST and future space-based dark energy missions will be mutually beneficial, and collaboration and planning must begin soon to maximize the benefits for LSST studies.  

{\bf Template mismatch and incompleteness.}
In template-based photometric redshift measurements, adding additional spectral templates increases computational requirements and the number of possible degeneracies, so the set is often limited to a small number (6--30) of templates that approximately span the range of expected SEDs.  Galaxies that fall between these templates will have slightly biased predicted redshifts, resulting in a contribution to photo-$z$ errors from this ``template mismatch variance''.  In addition, incomplete sampling of the full galaxy SED distribution will introduce systematic biases in photo-$z$ measurements.  Determination of both the size and contents of an optimal template set is an open question that will affect LSST data-processing requirements.  

{\bf Impact of blended objects.} In DEEP2 data, the rate of objects that appear to be a  single object from the ground but are resolved into multiple galaxies from space rises above 1\% at $z>1$ \citep{2012arXiv1203.3192N}.  This is likely to set fundamental limits on LSST photo-$z$ performance, as no single redshift and SED will describe the blended object.  Since the LSST sample will be fainter and at higher redshifts, this will be a more severe problem for LSST than Stage~III surveys.  

{\bf Synergies with star-galaxy separation.}
Given typical seeing of $\sim0.7$ arcseconds, LSST will be unable to distinguish compact galaxies from point sources (e.g., stars) with size measurements alone.  
Color and magnitude information can aid in classification.  \citet[]{Fad:12} explore both template-based and training-set techniques for star-galaxy separation in COSMOS data, finding promising ($\sim80\%$ completeness) but not perfect performance.  The similarity of methods suggests that probabilistic object classification should likely occur jointly with photo-$z$ estimation.  
However, techniques for doing this and the impact on LSST computing needs have not yet been explored. 

{\bf Whether and how to use single-visit information.}
Each LSST object will be imaged many times, often in different observing conditions.  For filters that are more sensitive to atmospheric conditions 
we will be measuring flux through different effective bandpasses in each visit; other differences in effective wavelength will occur due to variations amongst the QE curves of the LSST detectors.  It should be possible to use these variations to improve photo-$z$ estimates, as we effectively have measurements in more than six bands.  
However, variations in atmospheric conditions will also affect other aspects of photometric measurements (seeing, sky brightness, etc.) that may be difficult to disentangle.  LSST Data Management currently has no provisions for storing filter information for individual visits; if this information would yield significant improvements in photo-$z$'s, we must know that soon to scope any data-storage requirements.

{\bf Determining optimal methods of storing redshift probability distributions.}
Photo-$z$ probability density functions (PDFs) are often asymmetric and multimodal.  A single point estimate (i.e., one number) is not sufficient for most purposes.  
From a data-management perspective, we need an efficient way to store and use this complex information in as few columns in the database as possible, without degrading dark energy measurements.  
An optimal solution has not yet been found, and further work is necessary.  This is very likely to be investigated for Stage~III surveys.

{\bf Algorithm development.} 
To fully exploit the potential of LSST, we will wish to achieve superior performance in photo-$z$ estimation (not only in accuracy, but also in speed and efficiency) compared to the current state of the art.
We will need to develop sophisticated methods both to analyze the photometric data and  to output the results in the most economical yet informative manner.  
Issues not described above include the following.
\begin{enumerate}
\item As shown by Gorecki et al. 2012 (in prep), there can be gains from estimating photo-$z$'s with multiple, different algorithms, so we may wish to store a number of PDFs for each object, or else we will need to determine how to optimally combine them.

\item  Given the large size of LSST samples, we will need to perform all per-galaxy calculations rapidly; this may be challenging if single-visit measurements are incorporated.

\item It will be necessary to develop optimized techniques and store information required to allow automated rejection of galaxies likely to have poor photo-$z$ estimates (\citealt{benitez:00}; Gorecki et al. 2012, in prep.).  

\end{enumerate}

Other issues we will need to address by the time LSST begins operations include developing methods to minimize the effects of spatially varying systematics -- such as seeing, sky brightness, or extinction -- on training data and photo-$z$ estimation, and providing photometric redshifts for time-variable objects such as AGN and supernovae.  There also remain significant issues that will be pursued together with the theory/joint probes working group, including determining the impact of photo-$z$ systematics that are covariant between different dark energy probes and exploring the impact of catastrophic photo-$z$ outliers on probes beyond WL.  Ultimately, we wish to be able to turn LSST dark energy requirements into requirements for photo-$z$ systematic errors, which should feed back to the LSST System Requirements Document; 
frameworks to be developed by the theory working group should enable that work.

%% file: simulations/simulations.tex
\chapter[Simulation Tools and Technical Infrastructure]{Simulation Tools and Technical Infrastructure}
\label{chp:sims}

\input{simulations/chapterintro.tex}

\section[Simulations]{Simulations}
\label{sec:simintro}

\input{simulations/intro/intro.tex}

\subsection[Simulation tools]{Simulation tools}
\subsubsection[Cosmological simulations]{Cosmological simulations}
\label{sec:cosim}
\input{simulations/cosim/cosim.tex}

\subsubsection[Sky catalogs]{Sky catalogs}
\label{sec:mock}

\input{simulations/skysim/skysim.tex}

\subsubsection[Operations simulators]{Operations simulators}
\label{sec:opsim}
\input{simulations/opsim/opsim.tex}

\subsubsection[Photon simulator]{Photon simulator \label{phosimdescription}         }

\input{simulations/phosim/phosim.tex}

\section[Computing infrastructure]{Computing infrastructure}
\subsection[Software framework]{Development of a software framework}
\label{sec:software-framework}

\input{workplan/framework.tex}

\subsection[Computational model]{Development of a computational model}

\input{workplan/ComputingModel}

\section[Technical coordination]{Technical coordination}

In any experiment which pushes the envelope technically and
scientifically, it is necessary to get close to the experiment in
order to carry out the science. This is particularly true of LSST
probes of dark energy, each of which depend critically on understanding
the details of the LSST system, operations, and the properties of
the resulting data. Until now it has proven convenient to use the
LSST baseline system definition and performance requirements to
estimate residual systematic errors, and assume that they are
Gaussian distributed. However, it is often the case that the
non-gaussian and non-stochastic tails of distributions in an
experiment, coupled with subtle sample selection effects, ultimately
determine the achievable precision. Low level systematics which do
not average down with the number of exposures will be particularly
worthy of investigation.

Using end-to-end simulations based on the existing LSST design,
together with performance measurements on existing LSST subsystems
(hardware and software) and measurements at the LSST site, it will
be possible over the next few years to obtain a deeper understanding
of the LSST systems, the statistics of relevant parameters, and
their impact on each of the dark energy probes. One important early
goal will be to incorporate this more realistic information on the
expected LSST data in the end-to-end LSST image simulator ImSim,
enabling joint analysis of multiple probes and a realistic assessment
of LSST's ultimate capability of constraining models of dark energy
physics.

For this to happen there needs to be a two-way interaction between
members of the DESC and the LSST project. Within the DESC, Chris
Stubbs (Harvard) and Tony Tyson (UC Davis, LSST) are charged with
supporting and coordinating this interaction bridging the DESC and
Project. This will involve inquiries that arise in both directions,
and proposed measurements and investigations. The LSST project
technical teams will increasingly require input from the science
collaborations on specific technical design issues that arise as
the designs mature. And the DESC working groups will need deep
understanding of the LSST system. This will require system-wide
involvement to validate algorithms, develop new algorithms, ensure
simulator fidelity, test the simulator and system components, assess
the calibration plans, and explore operations cadence scenarios and
residual systematics. 

One example is the measurement of the shapes of astronomical objects,
which is the key observable for weak lensing dark energy studies. 
The intrinsic shape of a source is degraded/modified by the combination of 
1) atmospheric effects, 
2) optical aberrations, and other effects from the telescope and optics
3) the detector's ``shape transfer function'', and
4) the code and algorithms that are used for shape processing . 
These are not independent, and attaining a detailed understanding of the 
entire shape measurement chain, from the light hitting the silicon 
in the detectors to the bits hitting the silicon on the disk, is imperative
in order to truly understand the shape measurements and the associated
systematic error budget. This endeavor straddles the hardware, the
software, and the science analysis. The DESC will achieve the 
end-to-end understanding that is required in order to make the fullest
(and fully informed) use of the LSST data set. 

Another example is the formulation of dark energy specific system monitoring
metrics for data quality assessment. To facilitate this two-way
interaction, Stubbs and Tyson in collaboration with DESC members
will develop specific joint investigations and will involve key
members of the LSST Project, such as the system scientist, system
engineer, data management scientist, camera scientist, and image
simulation coordinator. Several of these scientists are members of
both the LSST project and the LSST DESC. While parts of some of
these investigations will occur within the LSST construction activity,
the dark energy specific issues will benefit immensely from a
coordinated effort.

%% file: simulations/chapterintro.tex
As emphasized in Chapter~\ref{sec:analysis}, we face a number of challenges in the implementation of the various dark energy probes, at the level of precision we hope to achieve with LSST.  The detailed assessment of those challenges, and the development of new algorithms and techniques to mitigate them, requires that we have available a suite of simulation tools and a technical and computing infrastructure sufficient to support the anticipated analysis effort.  In this Chapter, we review the current state of our capabilities in these areas, and the developments needed over the next few years.   We begin with a broad overview of simulations, and the array of simulation tools we will use for our investigations.  We then briefly cover our computing infrastructure needs, and then conclude with a discussion of how we expect to integrate technically with the LSST Project to ensure efficient transfer of information in both directions.

%% file: simulations/intro/intro.tex
Simulating and modeling the LSST system as a whole (from the
underlying cosmology to the performance of the LSST telescope and
camera) is integral to our understanding of the capabilities of the
LSST in characterizing dark energy. For example, as a Stage IV dark
energy experiment, the LSST must achieve three orders of magnitude
improvement in PSF interpolation (as a function of wavelength and
position on the focal plane), and two orders of magnitude improvement
in shear calibration over published Stage II dark energy surveys. Our
ability to achieve these advances will be dependent on the level of
systematics present within the LSST data and our ability to correct
for these effects. Developing the strategies and algorithms necessary
to attain these improvements requires high-fidelity simulations that
capture the properties of the LSST and how they couple to the
cosmological measurements (including uncertainties due to
perturbations in the optical surfaces, modulation of the PSF by the
atmosphere, variations in photometricity, and complex astrophysical
foregrounds).

\begin{figure}[th]
\includegraphics[angle=0,width=\textwidth]{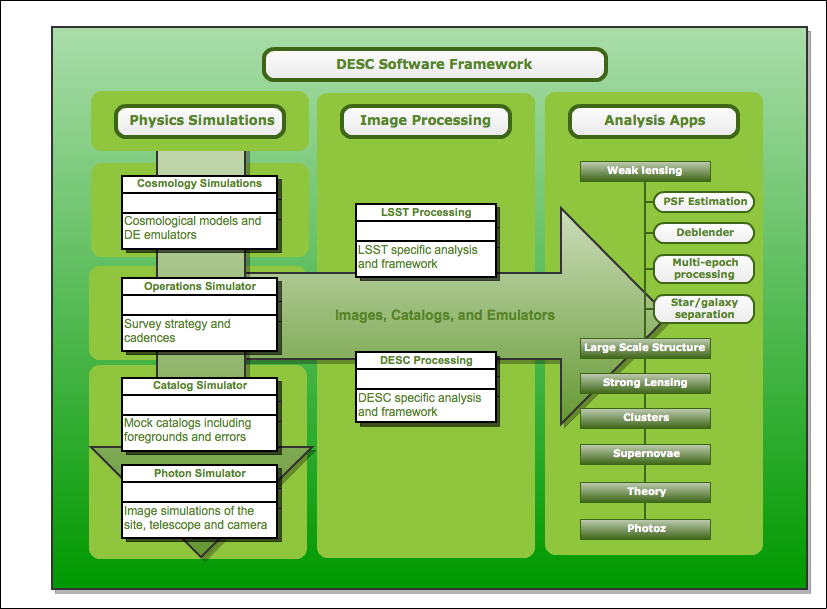}
\label{fig:intro:flow}
\caption{The flow of information through the DESC simulation
framework. Cosmological n-body and hydrodynamic simulations feed the
generation of mock catalogs (with appropriate source properties and
densities, and measurement uncertainties). Coupled with the Operations
Simulator, this parameterized view of the sky can be used either
directly for DESC science analyses or input into the photon simulator
to generate images that are representative of the LSST. The software
and analysis framework will provide a unified interface to all aspects
of the simulation, catalog and image generation, image processing and
analyses.}
\end{figure}

Figure~\ref{fig:intro:flow} illustrates the simulation tools and their
dependencies that will be required as part the DESC software
framework. Cosmological simulations, based on N-body codes~\citep{2009JPhCS.180a2019H,2005MNRAS.364.1105S} and hydrodynamic
codes~\citep{2000ApJS..131..273F,2006NewA...11..273T}, feed into the generation of mock catalogs, which
reproduce the properties of astrophysical sources including redshift,
color, and magnitude distributions. Coupling these catalogs to
sequences of simulated LSST observations (using the Operations
Simulator) enables the generation of data with appropriate densities
and uncertainties. These data can then be used to evaluate the
performance of the LSST system (e.g., how does Galactic extinction or
photometric calibration modulate the densities of galaxies and how can
that imprint systematic errors in the measures of dark energy) or, for
higher fidelity questions, be fed directly into a photon based image
simulator (PhoSim).

The vertical arrow in Figure~\ref{fig:intro:flow} demonstrates the
connection between these tools with the flow of information from the
cosmological models to the generation of catalogs with realistic
uncertainties to the production of simulated LSST images. The
horizontal arrow illustrates the flow of these simulated data sets
into the DESC analysis framework. This incorporates the analysis of
images using the LSST and DESC pipelines and the task-specific
algorithms that will be generated by the individual working groups
as discussed in Chapter~\ref{sec:analysis} (e.g., see the weak lensing high priority tasks broken out in
Figure~\ref{fig:intro:flow}). All of these tasks will need to be
developed within a software framework that combines the simulations
and data analyses, enabling common development tools and providing a
scalable platform for working with the volume of LSST data.

Throughout the following sections we describe the simulation tools
available to the DESC, their capabilities and limitations, and how we
expect to combine these tools within a coherent simulation and
analysis framework that will enable our dark energy studies to be
accomplished.

%% file: simulations/cosim/cosim.tex
Cosmological simulations occupy key roles in achieving the goals of
the DESC. They are required for
theoretical predictions (e.g., the effects of different dark energy
models), for generating mock data sets to understand observational
systematics and selection effects, for data analysis (error estimates,
testing and development of methods), and for optimizing observational
strategies.  No simulation starts from first principles, includes all
known physics, and produces quantities which are directly observed to
perfect precision. Different approximations are appropriate for
different simulations: the appropriateness must be validated for
self-consistency and for consistency with known observations.

In the field as a whole, there are gravity-only simulations (``N-body'')
and`` hydro'' simulations that include gas dynamics and additional
processes (star formation, feedback, AGN feedback, etc.).  Dark matter
simulations have converged between methods
\citep{2005ApJS..160...28H,2008CS&D....1a5003H} for predictions of
key inputs to cosmological probes such as the power spectrum, halo
mass functions, etc., and developments continue in the assignment of
observables (gas, galaxies and their properties) to this skeleton.
For simulations including gas, which can be more directly compared to
observations, development is needed both in simulation techniques and
in further steps to obtain
observables~\citep{2007MNRAS.380..963A}.  Either of these simulation
methods can be used, depending upon the science of
interest. Additionally, subgrid models based on hydro simulations can
be incorporated into large-volume N-body simulations.

{\bf Available codes and tools.}
The collaboration owns a large suite
of simulation and analysis tools that will be readily available to
carry out the tasks identified for the LSST Development Phase, described in later
sections. During the Construction Phase, these codes and tools will be
extended and improved as required.  

\begin{figure}
\center\includegraphics[width=3.5in]{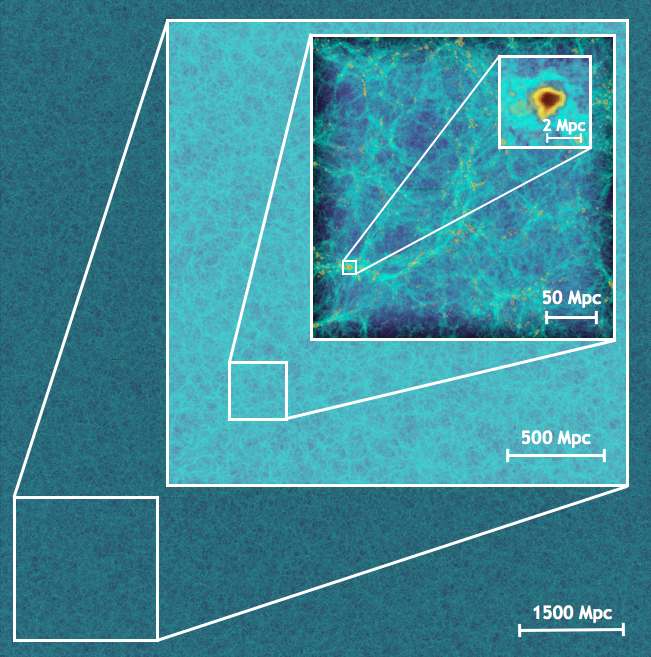}                                                                                       
\caption{\label{sim} Visualization of the density field from a 1.07
trillion particle simulation carried out with HACC on Mira, the new
BG/Q supercomputer at Argonne National Laboratory.}
\end{figure}

For the next set of simulations, up to multi-trillion particle
high-resolution N-body runs will be required. To achieve this scale,
one of the major codes to be used is the Hardware/Hybrid Accelerated
Cosmology Code, HACC~\citep{2009JPhCS.180a2019H}. HACC has been shown
to work on a variety of high-performance computing architectures and
to scale up to the largest systems currently available ($\sim$750,000
cores on Argonne's new BlueGene system Mira), enabling simulations
with more than one trillion particles (Habib et al. to appear in the
proceedings of SC12, see Figure~\ref{sim}). The code has been chosen as
one of the Gordon Bell Award finalists this year (the award will be
given in November 2012). We will also make extensive use of the P$^3$M
and Particle-Multi-Mesh (PMM) codes~\citep{2006NewA...11..273T}, which
have generated N-body simulations with up to 30 billion
particles. All codes are equipped with halo finders and merger trees
are constructed on the fly. In addition, we will use the publicly
available TreePM code {\sc GADGET-2}~\citep{2005MNRAS.364.1105S}.
 
For hydrodynamic simulations, we will use a range of codes, including
the Adaptive Mesh Refinement (AMR) Tree Code, ART, the AMR codes
FLASH~\citep{2000ApJS..131..273F},
RadHydro~\citep{2006NewA...11..273T} and Nyx (Almgren et al.,
submitted), as well as the publicly available Smoothed-Particle
Hydrodynamics (SPH) code {\sc GADGET-2} and the AMR code
Enzo~\footnote{http://lca.ucsd.edu/portal/software/enzo}. Both ART and
FLASH are well-established codes, which include various additional
physics models and built-in analysis tools. The two codes have been
extensively used to investigate the physics of clusters of galaxies and
galaxy formation. RadHydro combines a cosmological hydro code (moving
frame hydro plus particle-mesh N-body) with an adaptive ray-tracing
radiative transfer algorithm~\citep{2007ApJ...671....1T} to
simultaneously solve the coupled evolution of dark matter,
baryons, and radiation. Nyx has been developed very recently with a
first science focus on Lyman-$\alpha$ simulations. A suite of analysis
and visualization tools
exist~\citep{2011ApJS..195...11W,2011ApJS..192....9T,2011arXiv1110.4372B}
and several of the codes have built-in analysis capabilities.

With these codes and tools at hand, we will address the major
cosmological simulation challenges for the LSST DESC. These challenges
cover four key areas.

{\bf Mock catalogs.}  
Cosmological simulations are the backbone of
realistic mock catalogs, which in turn feed into the image simulation
pipeline. The mock catalogs and the outputs from ImSim
(Section~\ref{simulations:skysim}) will be important for all the analysis
working groups for testing analysis and data management pipelines, as
well as for calibrating systematics inherent in the survey strategy or
methods, selection functions, etc.- that is, for survey issues in
general. (Details on the mock catalog generation will be given in
Section~\ref{simulations:skysim}.)

For the generation of mock data sets (as distinguished from the image
pipeline), the physics that LSST will be able to measure must be
included, along with its correlations.  For instance, galactic
properties that will be used (e.g., shapes, luminosities, colors for
photometric redshifts) should be included, which can be done either in
a dark matter or a hydro simulation. The procedures will necessarily
require phenomenological assumptions at some level. Several
simulations exist as starting points (e.g., the data from the
Millennium simulation~\citep{2005Natur.435..629S} has a 'start to
finish' implementation, but is small and still being improved; an
impressive effort under development is the upcoming release of the
Millennium Run Observatory~\citep{2012arXiv1206.6923O}). There are dark
matter simulations with large volume and high resolution, but in which
the identification with observable properties is still under
development.  These efforts will require extensive calibration from
observational data as well.  The higher the dependence upon
simulations for complementing the data, the more accurate they need to
be.  In particular, they need to accurately include the selection
effects of the survey.

A major goal during the LSST Developoment Phase will be the generation
of a detailed mock catalog that provides a significantly improved
alternative to the currently-used catalog based on the Millennium
simulation. The new simulation will cover a significantly larger
volume at comparable mass and force resolution. We will use different
prescriptions to generate the mock catalogs, such as semi-analytic
methods (provided by e.g., Galacticus \citep{2012NewA...17..175B}) and
other methods. The outputs of mock catalogs must be tied to the
requirements of the science analysis groups and the image simulations,
and a major task will therefore be established to ensure that the catalogs will
include all the necessary information to generate the image
simulations reliably. During the LSST Construction Phase, careful
validation of these mock catalogs will be a key task, as will their
extension to a large suite of cosmological models.

{\bf Prediction tools.}  
\label{sec:prediction_sim}
LSST measurements that will further our
understanding of dark energy will probe deep into the nonlinear regime
of structure formation. Precision predictions in this regime will
therefore be essential and can only be obtained from detailed
simulations over a wide range of cosmological models, augmented by
observational inputs. These predictions will be important for all
analysis teams and their associated tasks; these will be discussed in
each of the sub-sections in Chapter~\ref{sec:analysis} in the context of each probe.

For theoretical predictions, optimal strategies for generating a
simulation campaign in order to explore a range of different proposed
models exist \citep{heitmann09}.  It has been shown that
accurate prediction tools, so-called emulators, can be built from a
(relatively) limited set of high-quality
simulations~\citep{2006ApJ...646L...1H,2007PhRvD..76h3503H,heitmann09,2010ApJ...713.1322L}. A
full set of simulations covering the model space of interest for LSST
DESC is not yet in hand.  Many proposed modifications of gravity and
models for dark energy have not yet been simulated, and new proposals
are still appearing regularly.

In addition, different probes require different analysis tools for the
simulations. Prominent examples are ray-tracing methods for weak and
strong lensing predictions. Some of these tools are available in the
collaboration already \citep[e.g.,][]{2009A&A...499...31H,
  2011MNRAS.414.2235K}. However, further work will be needed to
develop analysis tools for all relevant probes and types of
simulations, including simulations with non-standard physics, and to
ensure that these tools meet the accuracy requirements for LSST. The
simulation group will coordinate the development and use of these
different tools and make a detailed plan for needed additions.

In summary, the major challenges to be overcome for building the
required prediction tools are: (i) achieving the high accuracy needed
for deriving reliable dark energy constraints, (ii) accounting for and
modeling possible astrophysical systematics as described below, (iii)
generating a large simulation suite that encompasses a wide class of
dark energy models and theories of modified gravity, (iv) ensuring that
the available analysis tools are adequate and develop new ones as
required.

During the LSST Development Phase, the simulation group will work
closely together with the analysis groups to map out the required set
of simulations and generate the first set of these simulations. The
requirements will include determination of volumes, resolution,
required accuracy, and the set of cosmological models to be
covered. Some work has already been carried out in this direction. The
longer-term goal during the LSST Construction Phase will be to
generate precision prediction tools from a large suite of simulations
for the various probes. Again, close collaboration with the analysis
working groups will be required. While one might hope that some of
these prediction tools will be developed as part of some of the Stage
III missions, such as the Dark Energy Survey, the accuracy
requirements for LSST DESC will be much more stringent
\citep{2009arXiv0912.0201L}. Although some of this work will build
upon progress for Stage III experiments, it is unlikely that those
tools will meet LSST DESC requirements.

{\bf Data analysis.} 
For data analysis, many large-volume simulations (possibly in the
thousands) are needed to understand the statistical scatter possible
within the survey volume. For any observable, the scatters and their
correlations must be sufficiently accurate - extracting the required
covariance matrices needs a large number of simulations. Some of these
requirements may be ameliorated by new techniques that require smaller
simulation volumes; these should be implemented and extended (see,
e.g., \citealt{2011ApJ...737...11S}). Having some simulations early on,
to train and develop analysis methods, especially those particular to
the new possibilities provided by LSST, will be extremely useful.

During the LSST Development Phase we will map out the requirements for
these simulations and produce a set of simulations that can be used as
a testbed for new methods and approaches.  During the LSST Construction
Phase, these methods have to be refined and the simulations further
improved.

{\bf Astrophysical systematics.}  
At the small length scales probed by
LSST, baryonic effects become important and have to be quantified. Two
prominent examples are baryonic effects on the power spectrum for weak
lensing measurements and for estimates of cluster masses. In general,
one may distinguish between two classes of problems: (i) models for
assigning observable quantities to halos of known mass, shape,
environment, etc. and (ii) models for modifying statistics like the
power spectrum to account for baryonic effects (this second class can
use results from (i) as inputs). Again, specific problems will be
discussed in detail in the sub-sections of the different probes. As for
the analysis tools, some progress will be made in this area as part of
Stage III dark energy experiments.

%% file: simulations/skysim/skysim.tex
\label{simulations:skysim}

An understanding of how measurements of dark energy depend on statistical
and systematic uncertainties relies on a detailed knowledge of how the
signatures of dark energy are imprinted within the properties of
astrophysical sources. To address these questions for the LSST prior
to operations, and thereby optimize the returns of the survey,
requires that we simulate the small- and large-scale cosmological
signatures expected within the LSST data stream (e.g., from galaxy
clusters to Baryon Acoustic Oscillations) as well as the astrophysical
foregrounds that might systematically bias the cosmological signal.

Many of the dark energy tasks described in Chapter~\ref{sec:analysis} will
be undertaken either using catalog data (with the appropriate
uncertainties assign to the measured attributes) or will use catalogs
as inputs to the image generation.  Development of strategies and
algorithms to optimize the performance of the of the LSST will,
therefore, depend on how well we can characterize and model the
properties of the observable Universe.

The fidelity required for mock galaxy catalogs is dependent on the
science task at hand. Dark matter simulations incorporate the effects
of standard gravity, including the collapse of halos and the
persistence of subhalos (usually identified with galaxies). They do
not, by definition, include baryonic effects, which must either be
included directly using hydrodynamic simulations (see e.g., the code
comparison paper by \cite{2012MNRAS.423.1726S}), or by the use of ad
hoc prescriptions. For large-scale structure, positions of galaxies
along with luminosities need to be produced. These can be generated by
using dark matter positions (e.g., taking galaxies of a given density
and assigning galaxy properties to individual dark matter particles,
e.g., \cite{1998MNRAS.300..945C,addgals};
or, through halo and sub-halo models, using observed correlation functions and/or luminosity, to constrain
the number of galaxies per halo of a given luminosity and then
generating this number of galaxies within each dark matter halo, or
using this information to match subhalos to luminosity
(e.g., \citealt{2000MNRAS.318..203S,2000MNRAS.318.1144P,2003MNRAS.339.1057Y,
2006ApJ...647..201C}). In some cases luminosity is
assigned first, then color, e.g., \cite{2009MNRAS.392.1080S}. In other
cases, the assignment is done in one step by matching to a “similar”
galaxy in observations, e.g., \cite{1998MNRAS.300..945C,addgals,2012ApJ...747...58S}, or is derived by following the subhalo’s history
(semi-analytic models such as reviewed in \cite{2006RPPh...69.3101B}
and \cite{2012NewA...17..175B}).

As galaxy properties are all tuned to observational data, there is
heavy reliance upon the availability of relevant observations for
calibration and validation of the catalogs (e.g., correlation
functions, and luminosity functions for galaxies in different
wavebands). Observations must be compared to the mock catalogs; in
particular, any property that is heavily relied upon within dark
energy analyses (e.g., colors of galaxies for cluster finding or for
estimating the galaxy selection function) needs to be accurately
reflected in the mocks to make any conclusions or analysis based upon
the mocks valid. This also requires a detailed understanding of the
biases present within all of the observational data sets
(\ref{sec:workplan:catalog}).

\begin{figure}[ht]
\begin{center}
\includegraphics[angle=0,width=0.4\textwidth]{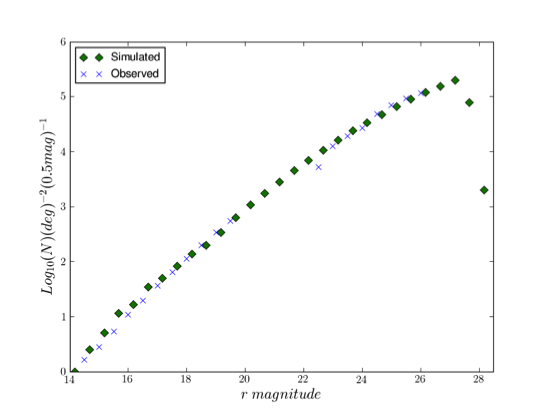}\hfil\includegraphics[angle=0,width=0.4\textwidth]{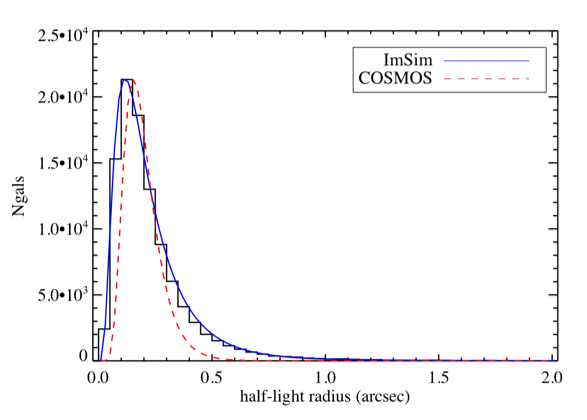}
\end{center}
\caption{Models of the observable Universe include the clustering
properties of galaxies, distributions of stars that follow the
Galactic structure, and solar system sources. These are designed to
reproduce the key observable properties that drive the survey. For
example, the left panel compares the simulated density of galaxies
(green points) with observed galaxy number counts (crosses) and the
right panel compares the predicted size distributions of galaxies
(blue line) with that observed from the COSMOS HST survey.}
\label{fig:numcounts}
\end{figure}

Mock catalogs currently used within the LSST collaboration have
focused on reproducing the distributions and properties of
astrophysical sources that significantly impact the technical
performance of the project \cite{2010SPIE.7738E..53C}. This
fidelity is designed to match the gross properties of the survey: the
distributions of galaxy and stellar magnitudes, the redshift
distributions, and the size distributions. To accomplish this,
extragalactic catalogs have been derived from the semi-analytic models
of \cite{2006MNRAS.366..499D} (based on the Millenium survey N-body
simulations). Figure~\ref{fig:numcounts} shows a comparison of the
distribution of galaxy number counts and sizes within the LSST
simulations with observational data sets. Initial comparisons showed
deficits of faint galaxies within the simulations, which led to the
development of a cloning procedure designed to reproduce the observed
densities of galaxies to $r>28$. This result highlights the need for detailed
validation of the simulated data.

In the context of the DESC tasks outlined in Chapter~\ref{sec:workplan}, the
limitations of the current implementation of the cosmological catalogs
arises, principally, from the limited cosmological information present
within the data.  The catalogs only extend over a single LSST pointing
(i.e., large-scale BAO signatures are not present). Larger catalogs
are generated by tiling the individual pointings across the
sky. Cosmologically variable sources, such as Type Ia supernovae, are
not simulated as part of the catalog generation. Weak lensing
signatures are also not incorporated within the data; they must be
coupled to the underlying dark matter mass distributions.
Baryonic modifications to the dark matter distribution need to be
either modeled in the analysis or somehow grafted onto the dark matter
distribution.  Again, the accuracy of these corrections needs to be
high enough to not undermine the desired precision of the
observations.

Beyond the cosmological properties of the data, astrophysical
foregrounds will limit the information on dark energy that can be
extracted from the LSST. Biases in the cosmological signals will arise
because of the distribution of stars within our Galaxy (where the
halos of bright stars can modulate the efficiency of detecting
galaxies); because of differences between the colors of stars used to
derive a PSF and the colors of galaxies to which that PSF is applied;
because of the modulation of the density and colors of galaxies as a
function of position on the sky due to Galactic extinction and
reddening; and, because identification of candidate Type Ia supernovae
can be contaminated by other variable sources.

To model these effects, stellar distributions must incorporate
positions, velocities and parallaxes for the 9 billion sources that
reside within the LSST footprint \cite{2008ApJ...673..864J}. Densities of stars
(which impact the interpolation of the PSF, how well the telescope can
guide, and the depth of the galaxy catalogs in the presence of bright
stars) must be modeled as a combination of thin-disk, thick-disk, and
halo stellar populations.  Current foreground implementations used in
the LSST match the densities of stars observed by the Sloan Digital
Sky Survey (SDSS) but do not replicate regions of high stellar density
close to the Galactic plane. Variability is assigned in an ad hoc
manner to approximately 10\% of the stars using light curves that model
transient and periodic behavior, but not to supernovae. 

In Section~\ref{sec:workplan:catalog}, we will
describe the enhancements necessary to make the current simulated
catalogs applicable to the proposed DESC trade studies and
analyses. This includes: making the data accessible to all of the DESC
collaborations, coupling the catalogs to the cadence of the
observations, incorporating realistic errors within the measured
properties, and generating magnitudes and errors that map to the
conditions under which the simulated data were ``observed''.

%% file: simulations/opsim/opsim.tex
The efficiency and performance of the LSST depends not only on its optical system. The observing strategy, weather, scheduled down time, properties of the site and the mechanical and electronic properties of the telescope and camera will all impact the area and depth of the survey over its ten-year lifespan. To characterize these dependencies and to enable studies that can optimize survey performance the project has developed an operations simulator (OpSim). OpSim has been used extensively throughout the LSST project to compare the performance of different sites, to study the impact of readout time and slew time on survey efficiency, and to undertake trade studies such as the trade-off between field of view and mirror size.

OpSim simulates sequences of observations based on a parameterized description of the LSST survey. It takes as input historical weather records and measures of the atmospheric conditions for the Cerro Pachon site (taken over a 10-year time period). From this, a sky model is generated that predicts sky 
brightness \citep{1991PASP..103.1033K}, atmospheric seeing, and cloud coverage as a function of time. Coupled with a mechanical model for the telescope (e.g., one that predicts slew and settle times) and a model for the camera shutter, readout and filter exchange times, estimates of the efficiency of the survey can be made as a function of integration time.

\begin{figure}
\begin{center}
\includegraphics[angle=0,width=0.4\textwidth]{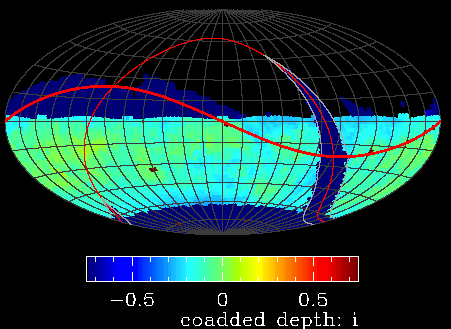}\hfil\includegraphics[angle=0,width=0.4\textwidth]{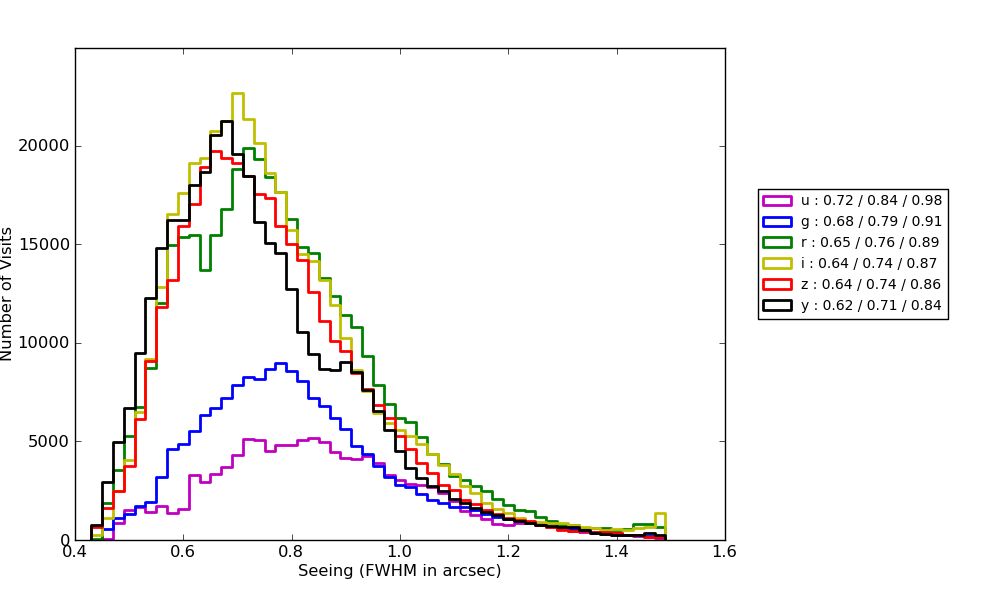}
\end{center}
\caption{The left panel shows the depth after 10 years in the $i$-band relative to the nominal depth of the survey. Areas outside of the main survey reach a shallower depth due to less sampling. The right panel shows the distribution of seeing values for all observations within a single 10-year simulation from OpSim.}
\label{fig:opsim}
\end{figure}

Observing sequences are simulated in a Monte Carlo fashion. Given specified sets of cadences (e.g., how often an exposure must be repeated), OpSim ranks all potential telescope pointings based on their ability to satisfy the LSST science requirements and the cost of slewing to that location. This process is undertaken dynamically, with the scheduler deciding which patch of the sky represents the best target for each subsequent observation. Figure~\ref{fig:opsim} demonstrates the outputs of OpSim. The left panel shows the depth, relative to the nominal survey, as a function of position on the sky in the $i$ band and after 10 years of observations (note the main survey area of 18,000 degree$^2$ goes deeper than the extension of the survey along the North Galactic Spur). The right panel shows the resulting distributions of seeing values (including the atmosphere, telescope, and camera contributions) that accompany this 10-year simulation.

The current version of Opsim separates the science cadences from the scheduling component of the simulator. This enables science cases to be input and evaluated (e.g., by changing the required cadence for how often a field must be observed and with what time separation between observations).  Outputs from the simulated runs (pointings as a function of time, and observing conditions for each pointing) are made available through a database and through a series of summary statistics.

The DESC will use OpSim results as input data for simulations of representative regions of the sky (either as sequences of catalogs or images). These data will be used in evaluating the calibration of the LSST data, its photometric performance, the impact of different cadences on our ability to extract variability from sequences of observations (i.e., to generate SNe light curves and to measure lens induced time delays).  To accomplish this, a number of enhancements will be needed to address the tasks outlined in Chapter~\ref{sec:workplan}. These include: the integration of OpSim with the catalog generation and calibration procedures to provide estimates of the uncertainties on the measured properties of sources (e.g., to test the impact of photometric redshifts and measures of large-scale structure); the inclusion of dithering  (currently OpSim uses a fixed pattern of fields with only rotational dithering) to enable studies of the impact of dithering on the PSF and large scale structure; the generation of a lookahead model for the scheduler that will enable optimization of survey strategy by predicting the expected survey conditions as a function of time of night and position on the sky.

%% file: simulations/phosim/phosim.tex
A significant fraction of all dark energy measurement systematics with LSST
will come from the complex physical effects of the atmosphere, telescope, and camera
that distort the light from astrophysical objects.  This distortion is
imprinted in the images, and therefore the most direct and robust way of mitigating and
understanding these systematics is through high-fidelity image simulations.
We may use codes for simple image simulations through a parameterized
point-spread-function (e.g., GalSim), but for detailed simulations in the analysis tasks outlined in
Chapter~\ref{sec:workplan}, we will require significantly more detailed simulations. 

To transform the catalogs of astrophysical objects into images, the DESC will use a
high fidelity simulator developed in cooperation with the LSST Project called the photon
simulator (phoSim).  The photon simulator does this using a photon Monte Carlo
approach.  A Monte Carlo  approach produces simulated images efficiently and encodes an arbitrary
complexity of atmosphere and instrument physics in terms of photon
manipulations.  The basic approach of the simulator is described below.

To simulate images, photons are first sampled from the catalogs of astrophysical objects
described in the previous section.  The direction of the photons is chosen from the source positions and spatial
models, and the wavelengths of the photons are sampled probabilistically from
the spectral energy distributions (SEDs) assigned to each object.  Then, the
photon simulator simulates the atmosphere by propagating the photon through a
series of layers placed vertically between the ground and 20 km.  Each layer
has some probability of eliminating the photon, either through cloud opacity,
molecular absorption, or Rayleigh scattering according to atomic physics and
atmospheric structure models.  Each layer also has some probability of
altering the trajectories of the photons by refraction of atmospheric
turbulence.  The turbulence is modelled by phase screen perturbations having a
Kolmogorov spectrum up to an outer scale where the turbulence is driven, and
the turbulence intensity as a function of height is modeled for the LSST
observing site.  The phase screens are drifted during the simulated exposure, so each
photon hits a different part of the screen depending on its intial position and
arrival time.  

\begin{figure}[b!]
\epsfig{file=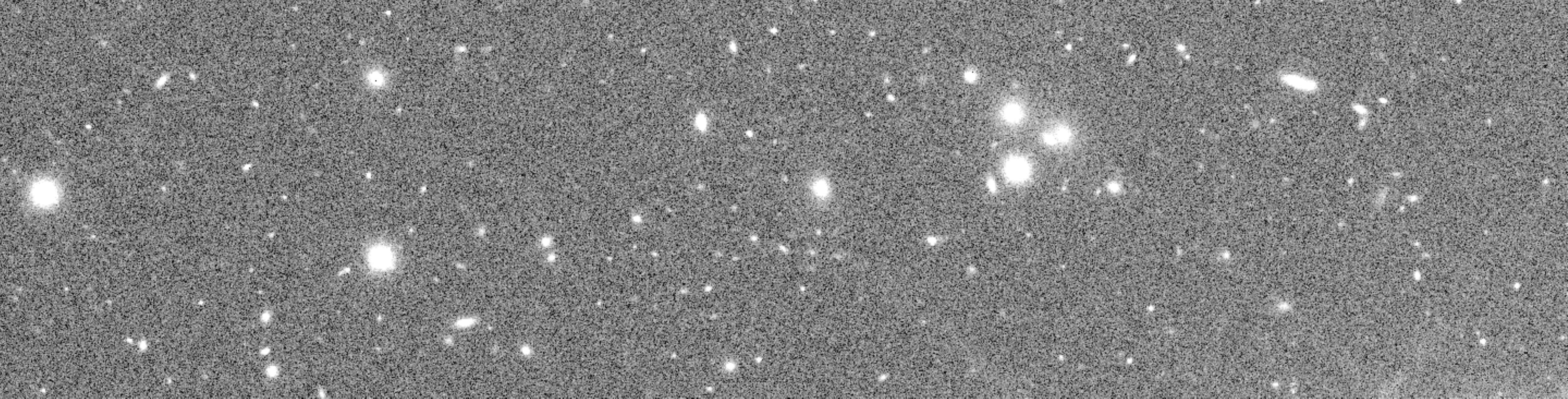,width=6.5in}
\caption{\label{ampimage} An example of a single amplifier image (2k by 512
  pixels).  We have previously simulated millions
  of these images.  Every photon used to construct these star and galaxy images has been
  simulated through detailed atmosphere, telescope, and camera physics models.}
\end{figure}

\begin{figure}[h!]
\begin{center}
\epsfig{file=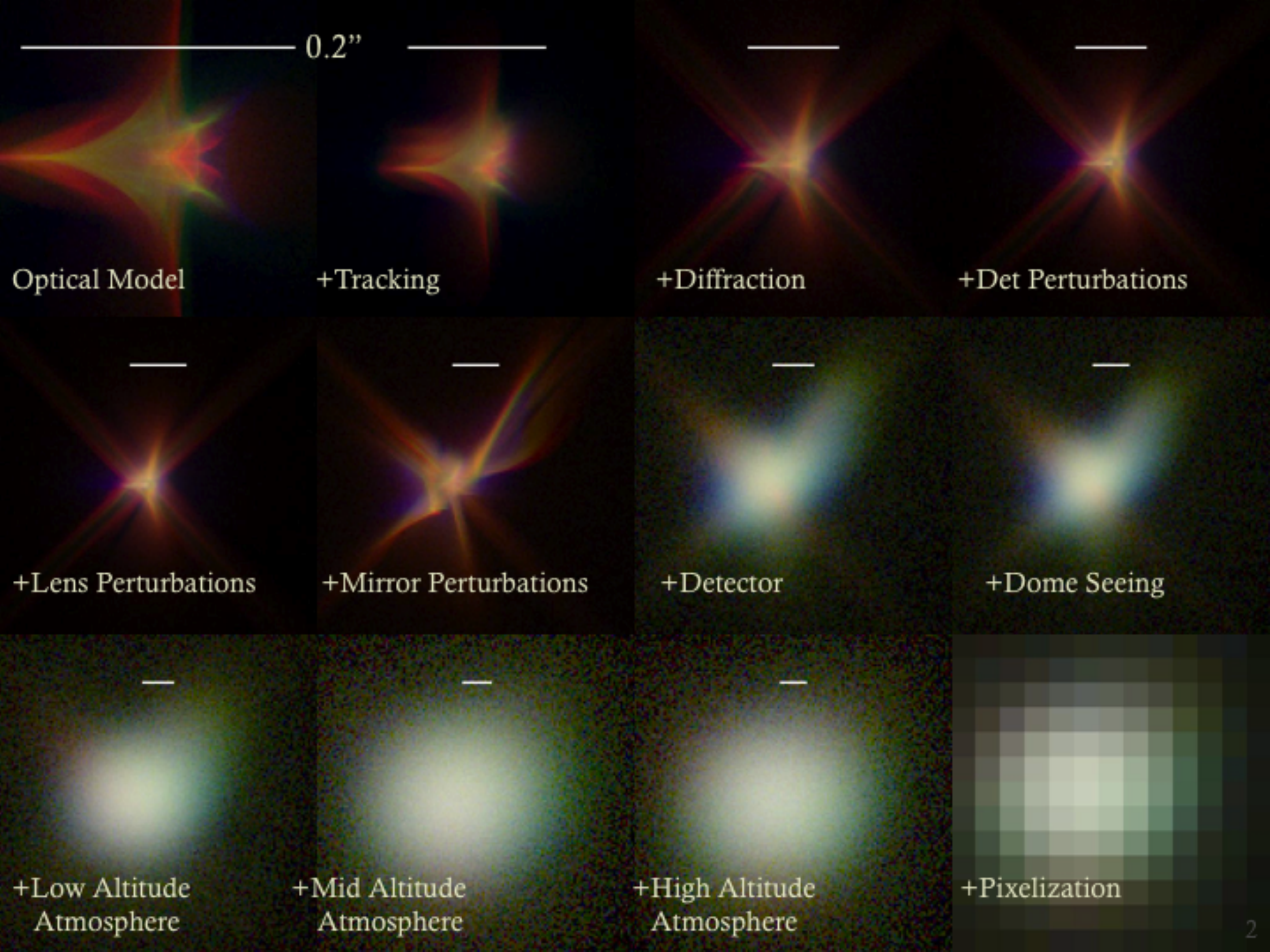,width=5.0in}
\caption{\label{twelvepsf} Simulations of a single star with successively more
  physics turned on in the photon simulator.  The images overlay simulations
  in the u,r, and y filters in their respective RGB colors, which highlights
  the wavelength-dependence of the simulations.  The images demonstrate the
  multiple physics models that contribute to the size, shape, and centroid of the PSF as
  well as the photometric response.
}
\end{center}
\end{figure}

After the atmosphere, the photons are then raytraced through the telescope and
camera.  To do this, every photon is either reflected and refracted through the series of mirrors and lenses
correposnding to the LSST optical design.  During each photon interaction, misalignments and surface
perturbations are modeled for every degree of freedom from the ideal mirrors
and lenses.  This contributes significantly to the image quality.  Photons are removed
as they interact with the mirror coatings, lens anti-reflective coatings, and
filter multilayer coatings.  When the photons enter the Silicon of the CCD,
they are refracted, and the position of photon conversion is calculated
probabilistically from its wavelength-dependent mean free path.  If the photon converts within
the Silicon, the path of the photoelectron is followed according to a model of the electric
field in Silicon appropriate for LSST's sensors, then the electrons are
drifted to the readout.  Saturation and blooming are simulated if the number of electrons exceeds the full well
depth.  The number of electrons in each pixel is then use to build up an image of the sky.
Finally, the image is digitized through simulation of the readout process taking into account
charge transfer inefficiency for each amplifier and various sensor
defects.  An example of a simulated image is shown in Figure~\ref{ampimage}.
Figure~\ref{twelvepsf} shows simulations with different physics turned on and off.

Therefore, the DESC has the capability to construct images efficiently, with high fidelity, since the most
important complex wavelength-dependent physics is included. The
simulations can generate over half a million photons per second.  A typical
galaxy in a single LSST visit can be generated in a millisecond, and even a bright
star can be simulated in seconds.  For full fields covering the entire 10 degree$^2$
of the LSST field of view and employing all 189 camera sensors, the codes have been implemented for grid
computing and have been used to generate millions of images (tens of Terabytes) using this
method.
The codes also contain a full validation framework to evaluate the
fidelity of the physics models.  The simulations have been able to reproduce:  (i) the typical
ellipticity from atmospheric turbulence by comparing with existing telescope
data, (ii) the typical ellipticity de-correlation patterns due to the different
wedges of atmospheric turbulence, (iii) the typical astrometric jitter on small
spatial scales, (iv) the spot diagrams produced through optical designs from
alternative benchmark raytrace codes, and (v) charge diffusion patterns similar
to actual laboratory measurements.

In many cases, the fidelity in the simulator is sufficient for many scientific
purposes.  For example, the simulator could be used to estimate the approximate photometric depth of an
exposure; produce images with reasonable PSF shapes, size, and patterns that LSST might
have; and, model the photometric response and its variation across the
field.  However, the DESC will use the simulator to make
detailed quantitative statements about dark energy measurements, so the
fidelity of the models will have to have fully validated physical and
numerical details. In addition, we expect that the wide range of uses
for image simulations outlined for Weak Lensing (Section \ref{sec:workplan_weaklensing}, Tasks~\ref{task:wl:psf}, \ref{task:wl:req} and \ref{task:wl:imsim}), Large Scale Structure (Section~\ref{sec:workplan_lss}, Tasks
\ref{task:lss:known} and \ref{task:dithersys}), Supernovae (Section \ref{sec:workplan_sne}, Tasks \ref{itm:sn:endtoend} and \ref{itm:sn:imsim}), Clusters (Section~\ref{sec:workplan_clusters}, Tasks~\ref{task:cl:shearcalib} and \ref{task:cl:photoz}), and Strong
Lensing (Section~\ref{sec:workplan_sl}, Tasks~\ref{task:sl:detection} and \ref{task:sl:timedelays}) require us to extend its functionality to support a
variety of different uses.  Thus, in this collaboration we expect we will need to improve
both its fidelity further and extend its usability to meet the various
analysis challenges associated with sensitive dark energy measurements.
    In Section \ref{phoSimImprove}, we discuss detailed plans
to adapt its use for the DESC, and outline the main areas for fidelity improvement.

The main advantage of using a physics-based simulation approach outlined above is that the detailed physical models of this simulator can then be used to probe the
limitations and systematics with dark energy measurements.  To see this,
consider that there are multiple image quality characteristics that describe
an image:  the point-spread-function (PSF) size and shape, the astrometric scale,
and the photometric response.  The physics in the simulator affects each of
these in different ways.  The atmospheric turbulence, charge diffusion in the
detector, and thermal/mechanical perturbations of the optics, for example,
affect the size and shape of the point-spread-function; whereas, the
opacity in the atmosphere, the coatings on the optical surfaces, and the
photo-electric conversion affect the photometric sensitivity in a
wavelength-dependent manner.  The dark energy measurements depend on each of
these.  For example, in Chapter~\ref{sec:analysis}, we discuss how weak and strong lensing depend critically on
understanding the shape of the PSF to measure galaxy shapes accurately as well
as the photometry response needed to generate photometric redshifts of
galaxies.  Weak lensing also depends on the ability to combine different
exposures with the range of observational destails, as well as the non-ideal
astrometric scale.  Supernovae measurements, clusters, and large-scale structure
depend critically on PSF and background details that affect faint source
detection as well as the photometric response and calibration.  Large scale
structure also depends on the ability to combine exposures with different
dithering patterns, sky brightness patterns, field-dependent opacity variation
patterns, and details large-scale
PSF wings from bright stars.  Thus, the
astrophysical observables can be directly linked to the relevant atmospheric
and instrument physics in the simulator. We can then assess what improvements
in algorithms are needed to obtain the sub-percent level of precision
cosmology.  High fidelity photon-based image simulations play an essential
role in that process.

%% file: workplan/framework.tex
The ultimate goal of the Dark Energy science working groups is to
transform the data into constraints on dark energy. This over-arching
challenge will be present at all stages of the project, starting from
2012, when the working groups begin to assess the effects of
systematics on the science, all the way until 2030 when the full data
set is complete. Even at the outset, one of the most important
challenges is to construct a pipeline that transforms
simulated data into dark energy constraints. The vision is that any
working group member can generate a simulation using project tools
such as ImSim; insert a systematic effect into such a simulation; run
a pipeline to quantify the extent to which this systematic will bias
the dark energy parameters; develop an algorithm that accounts for
this systematic; and, run the pipeline again with this new algorithm to
see if the bias has been removed and by how much the dark energy
constraints have weakened (they will necessarily get weaker as
nuisance parameters that account for systematics are allowed to vary).

To turn this vision into a reality, the collaboration will develop a
unified software framework that can be used by all the science working
groups. The framework will enable scientists to easily use project
simulation tools, develop and run analysis code that will be used for
all tasks taken on by the DESC. The framework will incorporate ImSim
and the Data Management tool suite within a flexible and robust
environment. Every simulation will be tracked and stored in a common
repository or repositories. Failures or transient delays will be
monitored and aggressively attended to, with the goal of mitigating
problems. Collaboration members will also have the capability to
import external data sets (CMB, X-Ray, radio) in their
simulations. The goal is to place the power of the project simulations
in the hands of the scientists. The framework can be used in any of
three deployments: High performance machines typically found at
national labs; university and small-scale clusters; and laptops.

The framework will help literally every task listed in Section~\ref{sec:workplan_weaklensing}, as each
will use the project simulations to test systematics. Some of them
will not use the full level 3 toolkit, but even this will become
relevant for more and more of the tasks as progress is made and
metrics are required.

%% file: workplan/ComputingModel.tex
The Working Groups will require a wide range of computing resources to reach their analysis goals. A typical analysis sequence starts with significant processing on supercomputers to create cosmology simulations, followed by thousands of processing hours on mid-range computers to trace simulated photons through to the camera, rounded out by processing for source localization. The needs of the Working Groups will evolve with time from less intensive tasks in the first few years to more heavy-duty tasks in intermediate years. 
Once LSST is operating, needs will be dominated by significant reprocessing of images to satisfy DESC.

The LSST Data Management team produces semi-annual ``Data Challenges"
in which simulated data or real data from
another project are processed with the
latest LSST analysis pipeline to test specific functionalities in each Challenge.
The DESC will need processed data sets to address questions that 
are not necessarily a match to the priorities of DM.
This will require generating and processing both end-to-end simulations for short-turnaround studies
and large-scale simulations similar in size to the Data Challenges of the Project.

The Data Management team produces semi-annual LSST ``Data Challenges",
in which the LSST analysis pipeline is run on either simulated images or images from existing telescopes.
To satisfy the specific needs and address the priorities of the DESC,
we will produce data products beyond the LSST Data Challenges.

The Working Groups have identified some 30 high priority tasks to address in the next three years. There are clear patterns in the types of simulations and post-processing steps called for in their work plans. Table~\ref{workinggrouptasks} summarizes which simulations and toolchains (both standard and customized) are called for by each WG.
Clearly, ImSim is a central element of the Working Group agenda.  

\begin{table}[htdp]
\caption{Working Group Task Requirements\ ``Std DM'' refers to standard catalogs; ``Custom DM'' allows for modification or replacement of DM algorithms. ``SimpleSky'' refers to a limited set of stars or galaxies in the field; ``Custom ImSim" involves special studies of seeing conditions and other telescope- or atmosphere-related parameters.}
\centering
\begin{tabularx}{\textwidth}{l c c c c c c c } 
WG  & \# tasks & FullSky & SimpleSky & Std ImSim  & Custom ImSim & Std DM & Custom DM \\
\hline
WL      & 5         &        4     &          5          &       5           &    3                      &        5      & 4           \\
SL        &4         &        2     &        0            &       2           &   1                       &        2      &  1 \\
LSS      &3        &        2     &       0              &        2         &     0                       &      0         &   1 \\
SN        &3        &         1       &         1          &  1              &    0                      &    1           &  0 \\
Clusters &6      &          5      &     0              &       5          &   0                       &  5            & 0 \\
Theory &4 &  0  &        0               &      0             &     0                     &   0              &  0 \\
Photo-$z$  &3      &            2   &    0                &   2             &   2                         &   2           &  0  \\
\hline
\end{tabularx}
\label{workinggrouptasks}
\end{table}%

This high level summary can be fleshed out with two examples:

{\bf Weak lensing star-galaxy separation.} This task will require flexibility: ImSim runs will be needed with varying star and galaxy brightness distributions to study contamination of the PSF star sample. Dense star field simulations with limited numbers of galaxies will also be performed. In addition to running the standard DM pipeline, custom runs will be needed for specific sets of objects to determine the PSF, with resulting PSF models as outputs. Small sets of simulations that can be run locally are expected to satisfy many of these needs, with community large scale simulations (e.g., latest Data Challenge runs or shared simulations with other Working Groups, like LSS or Photo-$z$) for data analysis validation.

{\bf Photo-$z$.} The photo-$z$ group will characterize the impact of various systematics, observational characteristics, and different source properties on the accuracy of the photometric redshift, the photometric redshift error or probability distribution function, and associated galaxy spectral classification. Their analyses will be used to investigate and mitigate the following basic issues: host morphology or spectral energy distribution, internal host absorption, AGN contamination, intergalactic absorption, Galactic extinction, sky brightness, atmospheric and telescopic throughput, seeing, and stellar density. In addition, the photo-$z$ group will need simulations that allow testing of systematics induced by  tiling or stacking of individual images, including photometric uniformity and the effect of the stacking algorithm. Likewise, they will need to understand the effects of the actual dithering algorithm employed for a single pointing, and will likely want to test the stacking and tiling at different Galactic latitudes to verify that the aforementioned systematics are under control at all survey pointings.

Much of this analysis could be acomplished by having simulated images and resultant catalogs at a grid of different Galactic coordinates (both longitude and latitude). Sampling at approximately ten to fifteen degrees in both coordinates is probably sufficient (this can likely be adjusted larger at higher Galactic latitudes), although the team might have to revisit the sampling rate after looking at the data.  This leads to several hundred simulated fields in total, with some of these fields simulated under different observing conditions to quantify the effects of throughput or sky brightness

In terms of the measured catalog parameters from these different fields, interest focuses primarily on photometry and photometric errors. Ideally there will be at least two apertures (but more is generally better): a PSF-type aperture and a model aperture measured in all filters. Other parameters that might prove useful include basic shape parameters including a radius, and the object coordinates, as well as parameters for general observing conditions (e.g., airmass, sky brightness, seeing).

\subsubsection{ImSim resource usage}

The LSST Project has already performed numerous simulations, allowing us to establish an upper limit on resources needed for ImSim by extrapolation from results obtained in previous full sky visit simulations (using back-to-back 15-second exposures). The number of visits needed will be collectively estimated from a bottom-up process in conjunction with the analysis groups. Analysis computing may well turn out to be comparable to simulation computing. I/O, data access and storage issues are as important as the raw computing CPU-hour issues, given the large volumes involved.

For one visit, phoSim takes roughly 1000 CPU hours to simulate all 6048 amplifier images. Taking into account the catalog constructor, DM pipelines (and variants), as well as different configurations on the same images and custom analyses, we expect about 2000 hours of mid-range compute time per visit. The Weak Lensing group (for example) expects to need very simple sky definitions - these are far less compute-intensive and can be run on single CPUs. The total output of the sims is about 10 to 15 compressed Gbytes per visit in images; reasonable development timeframe simulations of perhaps 2500 focal planes would need on the order of 50 TB of disk space and 3 million CPU hours to produce. Note that cosmological simulations will be run on exascale machines whose cycles are obtained separately.

It is expected that the Working Groups will need a range of data products: catalogs and single images, produced with different cosmologies, sky definitions and instrument parameters, including atmosphere and telescope properties. These will demand a wide range of computing resources, from single cores for work involving a handful of stars, to roughly 100 cores for a day for a few amplifier images, on up to thousands of cores employed for weeks to months for the large-scale work.

Our resource estimates are also guided by results from recent Data Challenges. Between 2009 and 2011, the Project ran three Challenges, each requiring roughly 1000 visits; in 2012, DM tested its level 2 piplines in a Challenge involving roughly 4000 visits. These Challenges have mostly relied on opportunistic resources available at Purdue, UW, SLAC, and OSG, and computing problems have related more to simple coordination of these ad hoc resources rather than supply of CPU hours.

\subsubsection{Development Phase needs}

In the first few years, the Working Groups will be using various data products, producing them with different cosmologies and varied parameters of the image simulations. They will need flexible access to the simulation tools to easily run end-to-end simulations for short turnaround studies on up to hundreds of cores for a day, and will need much larger scale community simulations occupying thousands of cores for weeks to months, similar in size to Project Data Challenges that have taken place to date. The large scale simulations need organization and excellent data access to handle the hundreds of thousands of jobs and multi-hundred TBs of data. Very efficient work flow engines are needed to push this scale of jobs through on multiple collaboration sites as well as the GRID, with excellent fault tolerance and retry capabilities to achieve failure rates below 0.1\%. Catalog and data access tools are also essential to select and fetch the desired data. Tools (e.g., xrootd) exist already to handle PBs of local and WAN data in logical clusters. Manpower will be needed to run and keep track of produced simulations; once it is all automated with access tools in place, the manpower needed should drop to a fraction of an FTE -- 
Fhence the further need for an early start on this infrastructure.

\subsubsection{Construction Phase needs}

It is anticipated that Data Challenges would be scaled up by a factor of 2 to 4 to create 5--10k visits, requiring upwards of 10k cores and a few hundred TB of storage per year. By this time, infrastructure should be in place for the efficient management of job submission, datasets and storage. A baseline level of manpower will still be needed to handle the production runs.

\subsubsection{Production Phase needs}

During this phase, Level 3 code will be run to reprocess pixel-level data. A full reprocessing of 2 years worth of LSST data (e.g., with different algorithms) will likely take 3 months with an installation of 400--1000 TFlops, and would place an upper bound on the resources needed.

%% file: workplan/workplan.tex
\chapter[Detailed Work Plan for the Next Three Years]{Detailed Work Plan}
\label{sec:workplan}

\input{workplan/chapterintro.tex}

   \section{Weak lensing}
   \label{sec:workplan_weaklensing}
   \input{workplan/weak-lensing/workplan.tex}

   \section{Large scale structure}
   \label{sec:workplan_lss}
   \input{workplan/lss/workplan.tex}

   \section{Supernovae}
   \label{sec:workplan_sne}
\input{supernovae/supernovae_tasks.tex}

   \section{Clusters}
   \label{sec:workplan_clusters}
   \input{workplan/clusters/workplan.tex}

   \section{Strong lensing}
   \label{sec:workplan_sl}

\input{workplan/strong-lensing/workplan.tex}

   \section{Theory and joint probes}
   \label{sec:workplan_tjp}
   \input{workplan/theory/tasks.tex}

   \section{Photometric redshifts}
   \label{sec:workplan_photoz}
   \input{workplan/photoz/workplan.tex}

   \section{Cross-working group tasks}
   \label{sec:workplan_crosscutting}
   \input{workplan/crosscutting.tex}

\section[Simulations improvements]{Simulations improvements and
  development}
\input{workplan/simulations/cosmo.tex}
\input{workplan/simulations/catalog.tex}

\input{workplan/simulations/phosim.tex}

\section[Software framework]{Development of a software framework}
\label{sec:software-frameworkitasks}

\input{workplan/frameworktasks.tex}

\section[Computing model]{Development of a computing model}

\input{workplan/ComputingModelTasks.tex}

\section[Technical coordination]{Technical coordination}

\input{workplan/technical_coordination.tex}

%% file: workplan/chapterintro.tex
In this Chapter, we lay out a detailed work plan for the first three years.  
The tasks that have been identified are distilled from the more general discussions outlined in Chapter~\ref{sec:analysis}, and reflect the prioritization criteria identified in Section~\ref{sec:workplan_overview}. We use the designation ``H'' to denote ``high priority'' tasks, which are especially time urgent, and ``L''T to denote ``longer term'' tasks, which are equally important, but can be pursued on a more relaxed schedule leading up to the onset of data taking.  We have organized the analysis tasks by working group, as was done in Chapter~\ref{sec:analysis}. However, there are several tasks that are especially ``cross-cutting" and are called out separately.  The tasks related to simulations, computing infrastructure, and technical coordination with the LSST Project are organized as described in Chapter~\ref{chp:sims}.

%% file: workplan/weak-lensing/workplan.tex
\subsection{High-Priority Tasks}

The weak lensing task list was constructed based on several considerations.  Of primary importance
are those tasks that (a) might result in a change in survey plans (hardware, software, survey
strategy); or (b) involve developing tools that are of long-term value for the work that the LSST
DESC needs to accomplish (i.e., the tasks described below and, eventually, 
the validation of WL analysis methods to better
than 1\%).  However, we want to avoid excessive overlap with the work being done on many parts of
the weak lensing analysis by Stage III surveys, learning from them as much as
possible, yet focusing our own efforts on issues where LSST will be in
such a different regime that new approaches might be required.  As a result, the weak lensing
task list is rather narrow in scope, focusing on the steps of the WL analysis process that occur
relatively early on, except for the tasks related to development of tools that will facilitate all
other tasks on the list (including the development of an overall analysis framework, as described in Section~\ref{sec:software-framework}).  Our neglect of the later steps of the WL analysis process does not reflect
a belief that those are unimportant; it simply reflects a conscious choice as to where we should
spend our (limited) effort in the next 3 years.  We expect the focus of the weak lensing working
group to evolve considerably over the coming decade.

\begin{tasklist}{H}

\tasktitle{WL: Estimate importance of PSF effects of currently unknown size}

\begin{task}
\label{task:wl:psf}
\motivation{
One of the largest sources of systematic error in weak lensing measurements is improper estimation of the point-spread function (PSF).  Most aspects of this problem are fairly well
understood; however, there are several that may be of particular importance for LSST.  We must 
determine how important these effects are and whether LSST Data Management needs to change their
algorithm(s) for dealing with these issues.  
(1) We need to determine how best to deal with the fact that stars and galaxies have different
spectral energy distributions (SEDs), which means that the measured PSF from stars is not the
correct PSF to apply to the galaxies.
(2) We need to determine
how well we need to know the small-scale stochastic PSF variation and whether additional 
information can be useful to improve the PSF interpolation between stars.  
(3) We need to determine 
how best to deal with the fact that the atmospheric PSF includes a centroid shift.}
\activities{
We will quantify the error in the inferred galaxy shapes by (1) using the stellar PSF rather 
than the correct PSF for galaxy SEDs, (2) interpolating using only the PSF at the locations of the stars,
and (3) ignoring the centroid shift in the PSF.  We will then explore algorithms
for dealing more effectively with these effects by 
(1) using the photometric color information ($r-i$, $g-r$) as a 
proxy for the galaxies' SEDs, (2) incorporating additional sources of information about the 
small scale PSF variation such as the wavefront and guidance sensors, and (3) including 
the centroid shift as part of the PSF, to be interpolated along with the rest of the PSF.}
\deliverables{
We will inform LSST Data Management of the requirements on the PSF estimation routines
given what we learn about each of these effects; delivery of specific guidance on algorithms to
ameliorate these effects is an enhanced goal.}

\end{task}

\tasktitle{WL: Place requirements on Data Management algorithms that impact weak lensing analysis}

\begin{task}
\label{task:wl:req}
\motivation{
As described in Section~\ref{systematics:wl}, 
the requirements for the star-galaxy separation algorithm and the solution of the world 
coordinate system (WCS) are somewhat uncertain, so we need to provide Data Management 
with more specific
requirements for the algorithms they are developing.
In particular, for the former, we need to know how many non-stellar objects (small
galaxies, binary stars, stars with cosmic rays, etc.) may be permitted to be used for PSF
estimation.
For the latter, 
we need to define more precisely the accuracy required for the Jacobian of the WCS and 
for determining the location of the same point from one observation to another.}
\activities{
For both cases, we will start by running the current Data Management algorithms on a 
realistic ImSim run, compare the results
to the true values (stellarity or true position, respectively), 
propagate the errors into their effect on the galaxy shapes, 
and finally, determine how these errors impact the dark energy constraints.}
\deliverables{We will inform Data Management how much (if any) the current star-galaxy and WCS 
algorithms need to be improved, including useful target requirements.}
\end{task}

\tasktitle{WL: Plan ImSim simulations to test WL analysis}
\label{task:wl:imsim}
\begin{task}
\motivation{Many of the studies of systematic errors that need to be performed by the WL working group will require simulated images, where we have knowledge of the truth values.  A well-motivated, staged series of simulations will take significant time to plan and coordinate, considering the amount of computing time necessary to generate and analyze the simulations, and the fact that some capabilities that are eventually required are not currently available in ImSim (see Section~\ref{sec:phosimwp}).}
\activities{
There will be some WL analysis tests that need information or a level of fidelity that is not already present in the ImSim framework, and these should be identified as a priority for development within the ImSim group. 
We will evaluate each area of study to determine whether the ImSim team has already produced
simulations appropriate for the work, or whether further simulations are required.
These can be roughly divided into categories according to whether the required simulations are wide, or deep.  Where further
simulations are required, we will identify the simulation volume and basic requirements (e.g., any
unusual inputs). In the longer term, we will want to test not just Data Management but also
higher-level codes that use DM outputs to constrain cosmology; we will also consider what
simulations are needed, both the cosmological simulations that are the basis for galaxy catalogs and
also the phoSim settings that are needed.  
It is likely that this list can be condensed into a few specific simulation requests to pass to the ImSim team. 
We will also need to determine whether the WL analysis tests require raw images, processed images, or catalog-level data.}
\deliverables{
A list of features/inputs that need to be included in ImSim in order to enable WL systematic studies, and a list of simulations that will be necessary for WL analysis work and systematic studies. 
}
\end{task}

\tasktitle{WL: Multi-epoch data processing algorithm and optimal dithering strategy}
\begin{task}
\label{task:wl:multi}
\motivation{
As described in Section~\ref{sys:wl:multiepoch}, we need to determine how best to combine information
from the $\sim 100$ observations per bandpass for each galaxy while minimizing systematic errors.  We also need to develop an 
optimal dithering strategy such that these observations help to minimize systematic errors
in shape measurements.}
\activities{
We will continue to develop two algorithms for dealing with this, ``MultiFit'' and
``StackFit'', 
with a focus on validating them using both simulated and real survey data.
We will also carry out a suite of image simulations and determine the optimal sets of 
dithering strategies to maximally reduce PSF systematics.}
\deliverables{ We will provide LSST Data Management with performance information (related to systematic
  errors, speed, parallelization, etc.) about the two algorithms,
  with recommendations for how best to proceed with development.  We will continue to work closely
  with the development team to test the shear accuracy of the algorithms and help to improve them.
  Requirements on dithering strategies to minimize PSF-related systematics will also be delivered to
  the LSST Project.}
\end{task}

\end{tasklist}

\subsection{Longer-Term Tasks}

We also have several longer-term tasks, for which work will begin
during the next three years, but will likely continue for some time
beyond that.

\begin{tasklist}{LT}

\tasktitle{WL: Develop non-canonical WL statistics that have the potential to improve dark energy constraints}

\begin{task}
\motivation{
The canonical WL statistic for constraining dark matter is the two-point shear
correlation function.  However, there are several other WL statistics that have the 
potential to add significantly to overall dark energy constraints from the LSST
experiment.  Examples include lensing peak statistics, cosmic magnification, three-point
shear correlations, and multiple source-plane cluster tomography, among others.}
\activities{
We will investigate how these statistics might impact the total dark energy constraints for LSST 
and what might be done to help mitigate systematic errors in them, using realistic models for those systematic errors (e.g., photo-$z$ errors) in LSST.}
\deliverables{
We will produce algorithms to calculate these statistics and to reduce systematic errors 
in the measurements.  As appropriate, we will interface with LSST Data Management concerning
requirements or desirable improvements in their pipeline algorithms.}
\end{task}

\tasktitle{WL: Extend WL data analysis methods from Stage III surveys to LSST}

\begin{task}
\motivation{
For several major weak-lensing--related systematic errors (e.g.,
related to PSF correction of galaxy shapes), we anticipate major
progress to be associated with Stage III surveys in the next few
years.  It is important to assess whether new algorithms that are
being developed now and in the next few years can be used for LSST, and/or
whether the more stringent requirements on systematic errors in LSST
require these algorithms to be enhanced.
}
\activities{
As new algorithms are developed for Stage III surveys, we will test
them on simulated LSST data (ImSim), and assess the level of
systematics in that context.  In nearly all cases, LSST will be in a
different regime from Stage III surveys in terms of galaxy population, depth, and number of
exposures, so we will evaluate how those differences could affect Stage III survey algorithm performance in an LSST context.
}
\deliverables{
Recommendations for which algorithms from Stage III surveys should be
adopted, either as-is or with recommended modifications; recommendations for which steps of the analysis process might require development of
completely new algorithms.
}
\end{task}

\end{tasklist}

%% file: workplan/lss/workplan.tex
\subsection{High-Priority Tasks}
\label{sec:lss:high}

The large scale structure task list was constructed based on several considerations. 
Of primary importance are those tasks that (a) may cause a change in survey plans (for example, dithering pattern, survey strategies, hardware); (b) involve developing tools that are of long-term value for the LSST DESC (for example, developing software to reach the maximum DE constraining power via large 
scale structure); or, (c) provide stringent requirements on the telescope system  or data processing steps (DM). 
Since photometric BAO is a recently matured technique, our experience and knowledge is based largely on analysis performed on SDSS data. Therefore, there are many 
unknowns in our analysis steps, since photometric BAO has not been done in any datasets except SDSS.  We decided to encompass a large range of systematics when designing our tasks. Thus, our tasks may appear to be over-encompassing, and not 
necessarily focused on one or two systematics, but very often on a long list of systematics. This is merely a reflection of our belief that the LSST data may offer a 
significantly different challenge from SDSS data.
We expect the focus of the LSS working group to evolve considerably over the coming decade.

 \begin{tasklist}{H}

  \tasktitle{LSS: Tools to estimate, mitigate and remove key known potential systematics, more specifically sky brightness, stellar density obscuration and contamination, extinction, and seeing
}
  \label{task:lss:known}
  \begin{task}
    \motivation{In order to deliver the promised Dark Energy constraints from Baryon Acoustic Oscillations in LSST, we need to develop ways to 
not only detect but also mitigate currently known systematics in photometric BAO analysis. We will need to understand the extent of their effects and mitigate or remove them in order to extract the best constraints from photometric BAO in LSST}.
    \activities{
We will first produce near full-sky simulations with stars and galaxies input from N-body simulations and observed stellar maps; the simulations will then be imported into an LSS pipeline, which will be developed to detect, estimate and mitigate/remove the effects of stars.  
We will then 
 estimate the level of residual systematics given existing methods 
for their removal, such as cross-correlating systematic maps against tracer overdensity maps, as done in SDSS III. 
In addition to modifying existing methods used in precursor surveys for LSST, we will develop new LSST-specific methods. 
In particular, we will show two new methods as examples here: (a) using multiple epoch data in LSST to remove systematics: This has only been discussed as a possibility so far, and has not been realized for the purpose of photometric BAO;  
(b) Full pipeline reconstruction: full-sky Monte Carlo by injecting fake sources into real data, and then processing the data via the full photometric pipeline to examine the final output. We can hope to recover (in principle)  the intrinsic source densities for each position in the sky and the source (flux,color) distribution. 
We will develop a validation and verification pipeline to make sure that we recover the intrinsic object densities in sky (for galaxies, quasars). We will pay particular attention to systematics listed in Section~\ref{challenges:lss}. }
    \deliverables{Existing systematics detection and/or removal pipelines, new systematic removal codes;  Full-sky simulations of stellar contamination and obscuration, along with other key systematics, as described in Section~\ref{challenges:lss} }.
  \end{task}

  \tasktitle{LSS: Analyze image simulations of multiple contiguous LSST pointings}
  \label{task:dithersys}
  \begin{task}
    \motivation{
While there will be $\sim$ 100 visits in multiple filters, there may be print-through of residual systematics in magnitude zeropoints and PSF on dither scales.  This may affect the LSS signals in BAO, WL shear, and WL magnification. A sky area of several contiguous fields will reveal all these systematics.}
    \activities{Analyze image simulations of multiple contiguous LSST pointings to full depth, with results reduced and calibrated through the DM pipeline, using several translational and rotational dither patterns. ImSim must have full LSS weak lens shear included, the corresponding LSS galaxy distribution, as well as realistic detector systematics (e.g., crosstalk).  
All astronomical systematics listed in Section~\ref{challenges:lss} will be included.  
We will verify that the dithering pattern produces sub-dominant LSS systematics in
 WL galaxy-mass correlation, galaxy bias, and BAO signal - even though the BAO scale is similar to the LSST pointing size.  
}
    \deliverables{ An optimum observing strategy, including dither distribution and scheduler optimization using weather and DQA inputs and site monitoring requirements; analysis algorithm for maximum reduction in systematics due to dithering pattern on LSS scales.}
  \end{task}

  \tasktitle{LSS: Setting requirements on systematics}
  \label{task:lss:require}
  \begin{task}
    \motivation{In order to provide feedback on the systematic requirement to both the telescope system and Data Management, we need to quantify the expected impact of various systematics on LSS deliverables.}
    \activities{We will employ a multi-pronged approach here. 
In the first approach, we will pixelate the sky and accumulate the pixelized galaxy counts from full-sky catalog simulations as well as the mean value of the systematic in each pixel. We will transform these data counts and systematic values to over/under densities and calculate the pixel angular correlation function over a wide range of angular scales. By comparing the cross-correlation to the data auto-correlation, we can quantify the effect on angular power spectrum of specific systematics, such as stellar density, seeing, and reddening. We then propagate this into an LSS cosmological parameter estimation code to estimate effects on cosmological parameters directly (see Task~\ref{task:lss:software} below). In the second approach, we will develop parameterized models, incorporate them into a DE forecast tool and quantify their effects on DE constraints}.
    \deliverables{From the first approach: Pixelized angular correlation function and power-spectrum codes, pixelized systematic maps, quantification of angular scales where systematics impact angular-correlation measurements and the BAO signals; from the second approach: Parameterized models of residual systematics, a DE forecast tool which includes model of systematics. From both approaches: we will quantify the requirements on these systematics.}
\end{task}

\end{tasklist}

\subsection{Longer-Term Tasks}
\label{sec:lss:low}

 \begin{tasklist}{LT}
  
  \tasktitle{LSS: Scalable optimal LSS analysis software development}
  \label{task:lss:software}
  \begin{task}
     \motivation{The motivation is two-fold. First, as known systematics are studied and new ones are uncovered, it is essential that mitigation strategies and requirements be developed in the context of the expected ultimate impact on cosmological parameters. Second, with an efficient, standardized, and fully functional pipeline will enable the full LSST DESC LSS collaboration to participate in the BAO analysis efficiently. }
     \activities{We will develop an LSS software pipeline that propagates a photometric catalog through to estimates of cosmological parameters. This activity will include assembling (and possibly upgrading) existing open-source tools and constructing new tools as necessary. The pipeline will be tested on both simulations and precursor survey catalogs. The development schedule will need to be closely coordinated with the activities of Task~\ref{task:lss:known} and Task~\ref{task:lss:unknown}, in order to facilitate their progress and avoid unnecessary duplication of effort. In particular, the development of a functional end-to-end pipeline that is useable for other tasks will be the first priority, rather than improvements to specific components. The LSS pipeline will conform to the interface standards to be developed under general Task~\ref{task:framework:requirements}, adapting existing codes as necessary. A significant undertaking of this task would include producing a scalable optimal quadratic estimator angular power spectrum code that will allow us to make the best statistical measurement of LSST angular power spectrum of any sources. An optimal quadratic estimator can improve our signal-to-noise by a factor of 2 for large scales and a factor of 50\% at the BAO scale when compared to the 
usual FKP power-spectrum methodology. There are so far only 2 existing quadratic estimator codes for LSS, but neither is scalable to meet LSST requirements.}
     \deliverables{A portable, open-source pipeline with documented installation and running instructions, and reproducible benchmarks on simulated and precursor survey data, to be made available to the entire DESC collaboration.}
  \end{task}

\end{tasklist}

%% file: supernovae/supernovae_tasks.tex
\label{sec:sn:tasks}

\subsection{High-Priority Tasks}
\label{sec:sn:priority}
We have prioritized three pressing high-priority tasks specific to supernovae.

\begin{tasklist}{H}

 \tasktitle{SN: Optimize analysis methods and quantify the error budget of photometric redshift and classification of SNe and their host galaxies}
 \label{itm:sn:photoz}

\begin{task}
\motivation{LSST is faced with a  ``new'' source of uncertainty as most
objects will not be observed spectroscopically to determine the redshift and classification.  We 
currently have no motivated estimate of the full science reach of the LSST SN
program due to a lack of ability to fully predict the power of the photometric
sample.}
\activities{Develop and refine algorithms for photometric classification
and redshift determination for supernovae discovered in LSST.  Develop
quantitative predictions of cosmological systematic uncertainties from these
methods based both on models of supernova properties as well as comparison
with data from current and near-future data sets.}
\deliverables{Algorithms for photometric-only analysis with
corresponding tools that calculate Hubble Diagram biases and covariance
matrices due to photometric classification and/or redshift determination. 
These algorithms shall be implemented as modules that plug into the framework
established by Task~\ref{itm:sn:endtoend}.}
\end{task}

\tasktitle{SN: Design an end-to-end analysis pipeline that starts with survey properties and finishes with cosmology projections}
\label{itm:sn:endtoend}
\begin{task}
\motivation{We will continuously need to project the impact of the LSST survey on measuring dark energy parameters.  
Although we have monolithic tools and some pieces of code
to address these questions now, we anticipate that the analysis will evolve between now and the commissioning of LSST.
We need a flexible, modular analysis pipeline that can support this evolution.  
Code used for projections can also be used for the real data analysis.
Writing good software requires careful and thoughtful design; it is never too early to begin this process.}
\activities{Identify and collect current existing codes.  Identify major gaps to fill in interfaces and usability.  Design prototype modular system.  Provide these to DESC supernova group for review and testing.
Establish standard formats and analysis in communication with Stage III projects.
Coordinate with those involved in Tasks~\ref{itm:sn:photoz}, \ref{itm:sn:imsim} and \ref{itm:sn:distance} to define standards and include those new modules when available.}
\deliverables{A reviewed design of the analysis pipeline with a prototype implementation. }
\end{task}

\tasktitle{SN: Supernova realizer for simulations}
\label{itm:sn:imsim}
\begin{task}
\motivation{A lesson learned from DES: realistic SNe must be designed into project simulations.  
The supernova population is an initial input needed for all of the other supernova tasks, and is therefore urgently needed.} 
\activities{Implement realistic supernova populations into LSST ImSim.  This will start with basic models and will grow to include more complicated models of host galaxy and supernova properties and covariance.  Collaborate with the ImSim team to test and integrate this code into the LSST ImSim.}
\deliverables{Code that realizes supernovae and furnishes their intrinsic SED at specified observing epochs.}
\end{task}

\end{tasklist}

\subsection{Longer-Term Tasks}
\label{sec:sn:longterm}

We fundamentally need to understand supernova better as distance indicators in the cosmos to fully exploit the potential of LSST to measure dark energy.  This will require sustained effort and different investigations over the next decade.

\begin{tasklist}{LT}

\tasktitle{SN: Develop theoretical/numerical/empirical SN models to better describe or improve the distance indicator}
\label{itm:sn:distance}
\begin{task}
\motivation{Improved supernova models can reduce statistical uncertainty and quantify systematic uncertainties.
They are therefore important in determining the error budget and projecting LSST science reach.}
\activities{LSST scientists must remain active in cutting-edge projects that lead into LSST. These include both observing and theory/simulation programs that
allow us to identify methods of utilizing supernovae from optical surveys with the least biased and smallest variance distance indicators.
}
\deliverables{Code for SN models that can work within the analysis pipeline; written reports of application of improved models to LSST SN cosmology.}
\end{task}

\end{tasklist}

%% file: workplan/clusters/workplan.tex
\subsection{High-Priority Tasks}
\label{sec:cl:high}

\begin{tasklist}{H}
\tasktitle{Cl: Optimized methods for absolute cluster mass calibration}
\label{task:cl:masscalib}
\begin{task}
\motivation{Accurate absolute calibration of the key mass-observable
scaling relations using weak lensing techniques is critical to the
extraction of robust cosmological constraints. The shear signals of
background galaxies viewed near the centers of clusters are larger but
more sensitive to systematics originating from projection effects (due
to, for example, triaxiality, large scale structure) and
mis-centering. Determining the optimal radial ranges and shear
profiles to fit, and quantifying the expected accuracy of mass
calibration, will impact on the entire analysis strategy.
}
\activities{The task will employ cosmological simulations spanning the
mass and redshift ranges of interest, with realistic galaxy
populations superimposed (to estimate mis-centering effects).
Ray-traced shear patterns for the simulated clusters will be analysed
to quantify the accuracy and precision of the mass calibration
obtainable, as well as the scatter (size and shape) about the mean, as
a function of radial filter and fitting profile.
}
\deliverables{The deliverables will be tabulated mass calibration
precisions and accuracies, and measurements of the scatter (size and
shape) about the mean, as a function of cluster mass and redshift, for
the various algorithms employed.
}
\end{task}

\tasktitle{Cl: Extending shear calibration programs into the cluster regime}
\label{task:cl:shearcalib}
\begin{task}
\motivation{The weak lensing shears associated with galaxy clusters
exceed those of typical large scale structure. Shear measurement
algorithms must be calibrated in the cluster regime. The innermost
regions of clusters also exhibit higher-order lensing effects that
will impact on shear measurements, if unaccounted for.
}
\activities{The task will develop STEP-like simulations spanning the
shear regime of clusters and enabling blind tests of the
shear measurement algorithms in this regime.  The LSST image
simulation codes will also be extended to incorporate higher order
lensing distortions such as flexion and multiple imaging that will be detected near cluster centers.
}
\deliverables{This task will be deliver STEP-like simulations spanning
the shear regime of clusters and enabling blind tests of the shear
measurement algorithms.  This will enable robust quantification of
shear biases (and therefore mass biases) as a function of algorithm,
shear and other parameters of interest.
}
\end{task}

\tasktitle{Cl: The impact of photometric redshift uncertainties on cluster
mass calibration}
\label{task:cl:photoz}
\begin{task}
\motivation{Photometric redshifts are important both for determining
cluster redshifts and for estimating the redshift distributions of
background galaxies employed in weak lensing mass calibration.
However, the cluster environment is not typical and the photo-$z$
calibration requirements for clusters may differ from the field.
}
\activities{The activities of this cross-cutting task (to be carried
out in collaboration with the photometric redshifts group) are: (1) to
quantify the expected bias in cluster mass calibration as a function
of mass and redshift due to various photometric redshift errors; (2) to
determine in particular the impact of faint cluster members on these
results; and, (3) to identify any external data (such as NIR photometry)
necessary to obtain cluster mass calibration to an accuracy of 2\% or
better across the mass and redshift range of interest.
}
\deliverables{Predictions for the impact of
photo-$z$ uncertainties on cluster mass calibration across the mass and
redshift range of interest; establishment of requirements for
external data sets to meet the mass calibration goals.
}
\end{task}

\end{tasklist}

\subsection{Longer-Term Tasks}
\label{sec:cl:longterm}

\begin{tasklist}{LT}

\tasktitle{Cl: Optimizing magnification-based cluster mass calibration}
\label{task:cl:magnification}
\begin{task}
\motivation{Magnification information can be used to provide
shear-independent mass calibration, with a different sensitivity to
systematic uncertainties. Although the statistical precision of
magnification measurements may be lower than those of shear, the
potential of this technique for mass calibration with LSST has yet to
be explored.
}
\activities{The task will use ImSim simulations to assess the
precision and accuracy of magnification-based cluster mass
calibration, as a function of mass and redshift. It will explore the
impact of systematic uncertainties, including photometric calibration
uncertainties, for various measurement strategies.
}
\deliverables{The deliverables include a robust assessment of the
precision and accuracy of mass calibration achievable with
magnification-based measurements, and the improvement in overall mass
calibration obtained when combined with shear measurements.
}
\end{task}

\end{tasklist}

%% file: workplan/strong-lensing/workplan.tex
\subsection{High-Priority Tasks}
\label{sec:sl:high}

\begin{tasklist}{H}

   \tasktitle{SL: Lens external mass distribution characterisation}
   \label{task:sl:environment}

   \begin{task}
     \motivation{Estimation of the external convergence estimation is known to
     be the  dominant source of uncertainty in time delay cosmography for
     overdense lines of sight. The residual systematic errors arising from the
     choice of mass assignment recipe, calibration simulation and so on could
     dominate the systematic error budget in all LSST lens samples.
     This key systematic must be understood as soon as possible in order
     to allow mitigation plans to be developed. Such schemes could involve
     long lead-time projects: developing
     more sophisticated halo model and photometric analysis codes, processing
     more detailed calibration cosmological simulations, and investigating
     extensive corollary observations.} 
     \activities{Use ray-traced cosmological simulations to
     investigate the methodology of lightcone mass reconstruction given
     LSST-grade photometric object catalogs and provide
     an optimal estimate of the prior PDF  ${\rm Pr}(\kappa_{\rm ext})$ (or
     other related  nuisance parameters), and test its accuracy in
     LSST-sized mock lens samples. This analysis will be needed during
     Stage III, but at lower required performance.}
     \deliverables{Code to generate the above prior PDFs; quantification of
     the systematic error floor dictated by line of sight mass structure.}
   \end{task}

   \tasktitle{SL: Automated lens candidate detection in the LSST catalogs and images}
   \label{task:sl:detection}

   \begin{task}
     \motivation{Automated lens classification by database search followed
     by image modeling is 
     the key process that will define strong lens yield.
     What purity, completeness and rejection rate can be obtained with LSST?
     How does this depend on model PSF quality and the 
     deblender performance? Byproducts from such an analysis include image lightcurves
     and covariance matrices, which are the key inputs to time delay
     inference.
     The sensitivity of the lens sample size to the performance of the level 2 LSST Data Management software
	suggests that, to optimize the strong lensing DE science
     output, we need to play a supporting role in its development. 
     Image-based automatic classification will be implemented via the DM 
     level 3 API, and could require a significant amount of computing
     resources to run. Early prototyping will allow this risk to be evaluated
     and reduced, and will provide feedback to the LSST Project on the level 3
     interface. } 
     \activities{Use ImSim (DC and custom) and Stage III (PS1, DES, HSC)
     catalogs and images to 
     (a) test the level 2 deblending software, 
     (b) develop algorithms for  fully-automated catalog-based lens candidate
     selection, 
     (c) develop algorithms for fully-automated image-based lens 
     classification and measurement, and
     (d) implement these algorithms using the level 3 API.}
     \deliverables{A set of requirements on the source de-blender level 2
     software, in the form of mock datasets for unit tests, and a set of
     write requirements on the level 3 API; new algorithms and their
     prototype level 3 implementations, with their performance assessed.}
   \end{task}

   \tasktitle{SL: Time delay estimation}
   \label{task:sl:timedelays}

   \begin{task}
     \motivation{Time delay uncertainties (statistical and systematic) need to
     be a few percent or lower for each lens in the cosmographic ensemble for
     them not to dominate the error budget. It is not known what fraction of
     LSST time delay lenses will meet this criteria.
     The basic feasibility of high-precision cosmography with time delay
     lenses is determined by the number of well-measured systems: 
     this needs to be determined as soon as possible in order to identify the
     required improvements in the observing strategy of analysis software.} 
     \activities{Use simulated lightcurves in 6 filters, sampled at realistic
     OpSim-generated cadence, to probe the available time delay precision and
     accuracy, using both ``standard'' and newly-developed algorithms.
     Realistic microlensing and intrinsic variability must both be included,
     as must realistic observing conditions.}
     \deliverables{Analysis code for inferring time delays from LSST data,
     and a pipeline for computing
     time delay uncertainty metrics given OpSim inputs, to allow requirements
     on the observing strategy to be explored.}
   \end{task}

\end{tasklist}

\subsection{Longer-Term Tasks}
\label{sec:sl:low}

\begin{tasklist}{LT}

\tasktitle{SL: Explore multiple source plane cosmography as a competitive DE probe}
\label{task:sl:jackpots}
\begin{task}
  \motivation{Preliminary studies suggest that multiple source plane
  compound strong lens systems could provide valuable additional information 
  about dark energy, but detailed forecasts for DE from plausible (but as yet
  poorly understood) samples of galaxy- and cluster-scale systems
  need to be developed to assess the impact of systematic errors associated
  with the mass modeling.
  This analysis needs to be done soon, to enable planning for a high-precision analysis to begin if it is justified, but depends on the joint
  analysis pipeline developed in Task~\ref{task:thjp:de_cap}.} 
  \activities{Simulate (a) realistic compound lens mass distributions,
  including line-of-sight structures down to subhalo mass scales, and (b)
  realistic LSST imaging data; model them, in order to
  estimate (a) the useability and (b) the detectability of multiple source
  plane lenses for dark energy science.}
  \deliverables{A mock catalog of compound lenses, and a simple pipeline
  plug-in to test its cosmographic information content.}
\end{task}

\end{tasklist}

%% file: workplan/theory/tasks.tex
\subsection{High-Priority Tasks}
\label{sec:thjp:high}

To fulfill the mission of the DESC, we must build a comprehensive 
analysis pipeline that extracts as much DE information as possible 
from LSST with enhancement of external data. Working toward 
this goal, we plan a set of tasks that will set up the architecture of 
the analysis pipeline and deliver many of the building blocks. 

\begin{tasklist}{H}

\tasktitle{TJP: Dark energy analysis pipeline}
\label{task:thjp:de_cap}
\begin{task}
\motivation{
To design the best dark energy experiment with LSST, we must be able
to quantify the impacts of the LSST data model and our control of systematic
uncertainties. We also need to parameterize, and account for, the many different effects that dark energy can
have on LSST survey data. The effort will be responsive
to evolution in the survey and developments in dark energy theory, as they occur.}
\activities{Characterization of the LSST data model, including instrumental
and site-based atmospheric uncertainties and the most up-to-date survey
information, in collaboration with LSST personnel; develop a
comprehensive list and parameterizations for key astrophysical systematics; develop likelihood techniques; build a suite of software tools to integrate systematics, dark energy
theory/modified gravity models to predict all observables for LSST, external
datasets and cross-correlations.}
\deliverables{Cosmological codes that (1) take the most current 
LSST data model and systematic uncertainties as input, (2) incorporate 
various calibration methods for different systematics, (3) jointly 
analyze multiple LSST DE probes with properly determined cross 
correlations, tailored likelihood techniques and covariance matrices, and (4) forecast constraints for 
a range of dark energy models of interest.}
\end{task}

\tasktitle{TJP: Exploring LSST DE science capability -- galaxy modeling}
\label{task:thjp:gal_mod}
\begin{task}
\motivation{ Improve and validate predictions for the galaxy 
bias and intrinsic alignments using numerical simulations and precursor 
surveys}
\activities{ Perform large-volume, baryonic cosmological simulations to develop and test 
parameterized phenomenological models of baryonic effects. Develop analysis algorithms that reduce the sensitivity of DE 
probes to uncertainties in the galaxy bias and intrinsic alignments. Tie in with the large number of high-resolution
N-body simulations for representative DE models to further reduce the
uncertainty of matter power spectrum on nonlinear scales.}
\deliverables{Large scale baryonic simulations and phenomenological models for parameterizing baryonic effects on galaxy clustering statistics.}
\end{task}

\tasktitle{TJP: Understanding statistics over very large scales}
\label{task:thjp:ls}
\begin{task}
\motivation{Unprecedented LSST survey volume enables DE 
studies over very large scales, though one must disentangle real DE 
signal from observational and other theoretical effects.}
\activities{ %
(1) Analyze how the clustering on large scale depend on the dark 
energy/modified gravity parameters.
(2) Develop an accurate theoretical model for analyzing the clustering 
on very large scales, including wide-angle and GR corrections.
(3) Identify optimal statistics to extract information from clustering 
analyses on those scales.
(4) Investigate degeneracies and dependencies of the improved model.
(5) Develop simulations that incorporate those very large scale effects.
(6) Use simulations to improve our understanding of large-scale 
clustering signatures and systematics for the range of plausible dark 
energy models of interest.
}
\deliverables{%
(1) Theoretical understanding and precise modeling of clustering on very 
large scales. (2) Codes for analyzing clustering on large scales that 
include a realistic description of the geometry of the system, 
wide-angle and general relativistic corrections. (3) A simulation code 
and simulated data that incorporate effects on very large scales.
}
\end{task}

\tasktitle{TJP: Exploring LSST DE science capability -- photometric redshifts}
\label{task:thjp:photo-z}
\begin{task}
\motivation{Photo-$z$ systematics impact all LSST DE probes and must be 
well understood and modeled for DE analyses.}
\activities{In collaboration with the Photo-$z$ WG (\autoref{sec:pz:high}), 
develop a realistic model for the photo-$z$ error distribution and its 
uncertainties that should capture features such as catastrophic errors and 
correlations between photo-$z$ errors at different redshifts; assess the 
impact of different features in the photo-$z$ error distribution on individual LSST 
DE probes, as well as the combined probes; 
explore methods to mitigate their adverse impact; 
incorporate results of this task and  of cross-correlation techniques for
self-calibration in Task~\ref{task:thjp:de_cap}.}
\deliverables{A realistic model of the photo-$z$ error distribution and
integration into LSST DE forecast tools.}
\end{task}

\end{tasklist}

%% file: workplan/photoz/workplan.tex
\subsection{High-Priority Tasks}
\label{sec:pz:high}

 \begin{tasklist}{H}
 
  \tasktitle{Phz: Calibration strategies}
  \label{task:photoz:calibration}
  \begin{task}
    \motivation{Calibration of photo-$z$ algorithms requires a ``truth'' set of secure redshifts for a representative subsample of galaxies.  Incompleteness in training data will lead to biases in LSST redshift estimates, which will propagate directly into the dark energy constraints if not accounted for otherwise (see also Task~\ref{task:pz:xcorr}).  Obtaining training sets must begin soon due to the large investments of telescope time required.
   }
    \activities{We will develop a detailed plan for targeted spectroscopic observations sufficient to meet LSST calibration/training set requirements, and reach out to potential partners. We will also explore obtaining very deep multi-wavelength imaging that enables very accurate photo-$z$ estimates of faint galaxies for which spectroscopic measurements are difficult.  We will investigate synergies with upcoming large space based missions (EUCLID, WFIRST) that could be mutually beneficial.}
    \deliverables{A comprehensive photo-$z$ calibration plan; the first proposals to fulfill that plan; results from negotiations with the EUCLID/WFIRST teams. See also Task~\ref{task:photoz:tools}.}
  \end{task}

  \tasktitle{Phz: Produce realistic tools to test photo-$z$ strategies and impact on science requirements}
  \label{task:photoz:tools}
  \begin{task}
    \motivation{Lacking LSST-scale data sets, we must rely on simulations to predict photo-$z$ performance, but those currently available have insufficient fidelity.    
 We need simulations with realistic template and training set incompleteness to accurately set LSST science requirements~\citep{SRD} that are driven by photo-$z$ errors, as well as to optimize methods for rejecting objects with problematic photo-$z$ determinations.}
    \activities{We will develop a detailed framework for simulating LSST photo-$z$ performance, which can be used to test the impact of template or training set incompleteness.  We will explore methods for identifying and removing objects with problematic photo-$z$'s, investigating tradeoffs between photo-$z$ performance and sample size, and investigate approaches that optimize dark energy errors.}
    \deliverables{Realistic simulation code and outputs; improved algorithms to identify problematic areas of parameter space; updated Science Requirements.}
  \end{task}
 
  \tasktitle{Phz: Testing cross-correlation techniques}
  \label{task:pz:xcorr}
  \begin{task}
   \motivation{Cross-correlation methods can provide accurate photo-$z$ calibrations while avoiding the problem of incompleteness in deep spectroscopic samples, but tests to date have not been in the high-precision regime required for LSST, making this a potential risk that must be explored}.
    \activities{We will test this method by dividing the BOSS spectroscopic sample into subsets, differing in observed color or magnitude (and hence in $z$ distribution and LSS bias), and attempting to reconstruct the true $z$ distribution of one sample using only position information and redshifts from the other sample.  We will use both this dataset and realistically-complex mock catalogs to explore methods of dealing with bias evolution and to test the impact of correlations between bias and errors}.
    \deliverables{Comparison of actual to predicted reconstruction errors; assessment of residual reconstruction errors from bias evolution and from covariance of photo-$z$ errors and galaxy properties.}
\end{task}

\end{tasklist}

\subsection{Longer-Term Tasks}
\label{sec:pz:secondary}

\begin{tasklist}{LT}
  \tasktitle{Phz: Optimal methods of storing and using $p(z)$ information}
  \label{task:pz:pofz}
  \begin{task}
    \motivation{Dark energy inference will be more accurate if we use full redshift PDF information ($p(z)$) for each object; exploration of optimal methods to determine and store this are needed soon as this can affect LSST database design.  Significant progress is expected from Stage III surveys.}.
    \activities{We will focus on activities that need to happen soon as they impact data management requirements, and will not occur as part of Stage III; e.g., determining the most compact representation of the redshift PDF for an object that yields bias-free results, whether to store multiple $p(z)$ estimates from different algorithms, and how to combine different $p(z)$ estimates.}.
    \deliverables{Storage requirements for LSST data management.}
      \end{task}

\end{tasklist}

%% file: workplan/crosscutting.tex
\subsection{High-Priority Tasks}
\label{sec:cross:high}

\begin{tasklist}{H}

\tasktitle{CWG: Metrics for automated data quality assessment}
\begin{task}
\label{task:cross:metric}
\motivation{The galaxy shear extraction algorithm (and system hardware) must be capable of
  delivering the level of galaxy shear systematics residual defined in the Science Requirements for individual and
  dithered exposures.  LSST system performance at this level must be continuously validated during
  the survey.  We
  thus need to develop diagnostics for important systematics (also including photometric and
  astrometric performance), creating metrics that can be
  incorporated into the Automated Data Quality Assessment pipeline to be run in real-time.  This is
  important to determine early, since the computational and algorithmic difficulties in running such code as observations are being carried
  out can be significant.}
\activities{Data quality diagnostics of particular relevance to dark energy will be developed.  In
  the next few years this will require fully realistic simulations of PSF systematics based on
  measured subsystem performance, weather statistics, and dither scenarios. A prioritized list of
  per-exposure and auxiliary data products will be made, based on their ability to monitor and
  diagnose data quality issues during the survey. New, currently unplanned, diagnostic data products
  need to be defined early.
}
\deliverables{A set of well defined metrics will be developed and specified for incorporation into the LSST ADQA pipeline, along with an algorithm for interpreting the results.  A deliverable for each metric is validation via full simulation and tests on precursor data.}
\end{task}

  \tasktitle{CWG: Full sky simulations of galaxy density and color systematics using the LSST Operations Simulator}
  \label{task:lss:fullsky}  
  \label{task:cross:fullsky}
  \begin{task}
    \motivation{As described in Section~\ref{challenges:lss}, 
systematics 
cause correlated variations in galaxy density and color accuracy across the largest angular scales and hence must be studied collectively in 
full-sky LSST 
simulations.  
Coupled variations in survey depth, photometric calibration, and galaxy colors will affect LSS, cluster-finding, cluster ``self-calibration'', cosmic magnification, and photo-$z$ accuracy.}
    \activities{We will work with the LSST Project to use outputs from the Operations Simulator (OpSim), along with Calibration Simulation (CalSim).  The meta-data output by OpSim and CalSim trace additional systematics, including sky brightness, photometric calibration, throughput (atmospheric and telescope), and seeing (PSF).  Analytical formulae will be used to estimate the effect on survey depth, photometric zeropoint, and color errors from these systematics.  Scaling relations observed in stage III surveys will be used to propagate these quantities into full-sky maps of photo-$z$ accuracy and artificial fluctuations in galaxy density.}
    \deliverables{For a variety of dither patterns, we will estimate the large-scale angular correlation function and power spectrum of the observed galaxy density and the achieved photo-$z$ accuracy.   We will determine the optimal dither pattern for minimizing residual power on the BAO scale and maximizing the performance for clusters, magnification and photo-$z$.}
  \end{task}

  \tasktitle{CWG: Develop tools to detect, mitigate and remove unknown observational and astrophysical systematics in both LSS and WL to satisfy the DE requirement of LSST} 
  \label{task:lss:unknown}  %
  \label{task:cross:unknown}
  \begin{task}
     \motivation{In order to achieve the required dark energy constraints from both weak lensing and large scale structure, we need to develop tools to identify
unknown observational and astrophysical systematics, and hopefully develop methods to detect and mitigate/remove unknown systematics}.
     \activities{
This task will be carried out in a multi-pronged approach. We will test several existing methods to detect unknown systematics. Such current methods include cross-correlating different redshift slices in a photometric survey and compare the expected signals (due to overlap in galaxy redshift distribution) against observed signals.  We need to develop not only the existing method, but new methods. An example of possible new method includes inputing observational estimates in how each known systematics affect the observation into ``clean simulated observations'' and compare it with observations; any residual systematics can show up 
in the comparison.  Another approach will be using stellar observations to detect various effects such as inadequate ghost-pupil removal in the imaging, to zero point descrepancies between the pointings. We will develop a validation and verification pipeline to make sure that we recover the intrinsic extragalactic object densities in sky. The systematics detection pipeline will be geared
towards LSS, however any found unknown systematics will have effects on a wide variety of DE probes other than LSS. Future development can include a systematic detection pipeline that have 
emphasis on a larger variety of DE probes}. 
     \deliverables{Existing systematics detection pipelines, new systematic detection pipelines}. 
  \end{task}

  \tasktitle{CWG: Deblending for weak lensing and cluster cosmology}
  \begin{task}
\label{task:cross:deblend}
     \motivation{
As described in Section~\ref{sys:wl:deblending}, the high degree of galaxy overlap in LSST stacks 
requires new deblending algorithms and will likely introduce some level of bias to the resulting 
shape measurements. Potential biases need to be studied and quantified, resulting in an initial 
set of deblender requirements for the data management (DM) group, who are scheduled to begin 
work on deblending algorithms in 2013.
In particular, galaxy clusters are among the densest source environments
that LSST will encounter and pose some of the severest challenges to
its deblending algorithm. Limitations in the deblending algorithm can
impact the cosmological constraints from clusters --  e.g., by inducing
density-dependent errors in shear measurements, photometric
calibration, and possible misidentification of the central galaxy.
}
     \activities{
We will characterize the level of blending expected in LSST, using realistic ImSim runs,
in terms of the distance, size, and magnitude of objects falling into the typical aperture 
used by shape measurement algorithms.  We will determine to what level these interloping objects
bias the shape measurements of galaxies and thus the level to which they need to be 
correctly removed.
We will use the ImSim software to generate
realistic simulated cluster images (based on deep multi-band HST
images of known clusters of galaxies) spanning a wide range of
richness and redshifts.  These will be used to determine the biases in
shape and photometric measurements -- and therefore mass calibration --
induced by the deblending algorithm, as a function of magnitude and
overdensity (radius).
}
     \deliverables{
We will provide LSST DM with specifications of the expected level of overlaps, and performance 
targets for either masking or weighting overlap pixels as a function of the relevant parameters
of the blended objects.
We will quantify the performance of the deblending
algorithm in the high galaxy density environments of clusters,
enabling calibration of its deficiencies. A list of specific ``error
modes'' will be provided, identifying the limitations of the deblending
algorithm that have the greatest impact on cluster cosmology and the
highest priority for improvement.
}
  \end{task}

\end{tasklist}

\subsection{Longer-Term Tasks}
\label{sec:cross:secondary}

\begin{tasklist}{LT}

\tasktitle{CWG: Enhancing LSST DE science with external CMB lensing maps}
\label{task:cross:ex_lens_data} 
\label{task:thjp:ex_lens_data} %
\begin{task}
\motivation{External CMB lensing datasets can be very useful in determining galaxy bias, 
and in mitigating LSST galaxy lensing systematics. 
We need to find an optimal way to analyze all the datasets and know how well they perform in 
combination.
}
\activities{ 
Explore the utility to LSST cosmological parameter estimation and systematic mitigation of 
high-precision CMB lensing potential maps reconstructed from 
next-generation CMB polarization maps covering nearly the entire LSST 
survey footprint. 
Work will be done with a Fisher matrix analysis in a 
staged progression of complexity of the description of systematic error 
sources, eventually to be combined with larger LSST simulation 
efforts. Statistical quantities that will be studied are the various two-point functions 
possible with the collection of lensing maps (one CMB + multiple LSST 
maps) and galaxy-count maps.
}

\deliverables{
(1) Tools to explore cross-correlation of CMB lensing and LSST galaxy lensing maps. 
(2) Assessment of impact of CMB-galaxy cross-correlations on reducing uncertainties in astrophysical  and observational systematics and improvements on dark energy constraints.
}
\end{task}

\tasktitle{CWG: Enhancing LSST DE science with external galaxy-cluster datasets}
\label{task:cross:ex_clus_data}
\label{task:thjp:ex_clus_data} %
\begin{task}
\motivation{External cluster datasets can be very useful to help  provide important complementary cluster mass estimates when analyzed jointly 
with LSST data.
We need to find an optimal way to analyze all the datasets and know how well they perform.
}
\activities{
Investigate opportunities  to improve cluster scaling relations and cosmological constraints from combined analyses of LSST cluster lensing and CMB SZ and X-ray cluster data, with particular focus on fields overlapping with the LSST survey footprint. }
\deliverables{
Assessment of improved cluster cosmology constraints and cluster-mass scaling relations from combining CMB SZ, X-ray and LSST cluster surveys.
}
\end{task}

\tasktitle{CWG: Enhancing LSST DE science with external spectroscopic galaxy data}
\label{task:cross:ex_spec_data}
\label{task:thjp:ex_spec_data} %
\begin{task}
\motivation{External spectroscopic data can be very useful when analyzed jointly 
with LSST data. We need to find an optimal way to analyze all the 
datasets and know how well they perform.}
\activities{ %
(1) Cross-correlation with spectroscopic surveys:
Develop algorithms to measure clustering statistics jointly between 
lensing, photometric, and spectroscopic galaxy data. (2)~Develop the theoretical framework 
to interpret these measurements for dark energy and modified gravity theories. (3)~Develop tools and models that enable us to push as far as possible into the nonlinear regime,  including the behavior of deviations from the simple, separable power spectrum, with well-understood limitations. 
Work in collaboration with the simulation effort to model redshift-space clustering in simulations.
(4) Work with the cross-cutting photo-$z$ effort to use spectroscopic galaxy data to study how imaging data can be used not 
only to estimate photo-$z$'s but to identify groups of photometric galaxies with a ``controllable'' distribution of biases.
}
\deliverables{%
(1) Cross-correlation with spectroscopic surveys:
rudimentary code to make multi-tracer clustering measurements on 
overlapping mock datasets. (2)~Theoretical fitting code implementing our current 
best understanding of how to deal with the onset of nonlinearity and 
other complications in the various tracers. A by-product of this will be 
more realistic, comprehensive projections of constraining power. 
(3)~Forecasted constraints on prospective modified-gravity/dark-energy constraints from a 
combination of relativistic and non-relativistic tracers.
}
\end{task}

\tasktitle{CWG: Enhancing LSST DE science with external supernova datasets}
\label{itm:sn:external} %
\label{task:cross:ex_sn_data}
\begin{task}
\motivation{
External datasets such as spectroscopy and NIR imaging can be used to improve supernova distance measurements and cosmological analyses, increasing the power of LSST SNe.  We need new tools to identify the optimal uses of outside resources and advanced planning to acquire them.}
\activities{
Extend the end-to-end analysis tool from Task~\ref{itm:sn:endtoend} to include additional surveys or information.
Model additional information such as spectra of supernovae or host galaxies and how that will affect the systematics error budget for distance determination. 
Investigate opportunities to coordinate with complementary surveys such as Euclid to develop and test joint strategies. }
\deliverables{ Code that takes as input characteristics of LSST and external-survey observations and outputs the resulting figure of merit.  Assessment of improved supernova cosmology constraints as a function of amount of external spectroscopy and NIR imaging.  A written report on findings.}
\end{task}

\end{tasklist}

%% file: workplan/simulations/cosmo.tex
\subsection{Cosmological Simulation Tasks}
\label{task:cosmosim}

\subsubsection{High-Priority Tasks}
\label{sec:cosmosim:high}

\begin{tasklist}{H}

\tasktitle{CoSim: Simulations for mock catalog generation}

\begin{task}
  \motivation{The discussion in Section~\ref{sec:analysis} makes it clear that
    sophisticated and detailed mock catalogs will play a crucial role
    for investigating possible systematics, survey strategies, and
    development of analysis tools during all three phases of the LSST
    project -- Development, Construction, and Operation. The accuracy
    of the mock catalogs will have to be improved throughout this time
    and continuous validation against observational data will be very
    important. The resolution and coverage of the underlying
    cosmological simulations has to be therefore improved as well, and
    over the years, simulations with ever-larger volumes and higher
    mass resolution will be carried out. In addition, the model space
    has to be increased to accommodate investigations of different
    dark energy and modified gravity scenarios as outlined in
    Section~\ref{sec:theory}.}

  \activities{During the first three years, we will work closely with
    the analysis groups and the ImSim team to generate a first set of
    mock catalogs suitable for some of the tasks outlined in
    Chapter~\ref{sec:workplan}.  We will explore different ways of
    generating these mock catalogs and start developing accurate
    validation procedures. We will develop a detailed simulation plan
    with analysis groups to map out their needs with respect to
    accuracy and coverage. }

  \deliverables{First set of detailed mock catalogs that will be used
    as input to the ImSim pipeline and by the analysis teams; comprehensive plan for necessary simulation activities for the
    next phase of the project. These deliverables have a direct impact
    on tasks from all the working groups, in particular, high priority
    tasks for WL (Task~\ref{task:wl:imsim}), %
    clusters~(Task~\ref{task:cl:masscalib}), strong
    lensing~(Task~\ref{task:sl:environment} and lower priority
    Task~\ref{task:sl:jackpots}), and theory~(Task~\ref{task:thjp:de_cap} --
    Task~\ref{task:thjp:ls}).}
\end{task}

\tasktitle{CoSim: Data analysis and prediction tools}
\label{task:cosim:pred}
\begin{task}
  \motivation{The LSST DESC is aiming to derive constraints on dark
    energy and modified gravity with unprecedented accuracy. As
    emphasized by the different analysis groups in
    Chapter~\ref{sec:analysis}, simulations will play an important role in
    achieving this ambitious goal. In particular, high precision
    predictions in the nonlinear regime of structure formation will be
    crucial. A few examples from Chapter~\ref{sec:analysis} include:
    cosmological parameter estimation from weak lensing with two-point
    statistics (Section~\ref{analysis:wl}), the 3-d power spectrum from LSS
    measurements (Section~\ref{analysis:lss}), prediction of the
    distribution of massive halos as a function of mass and redshift
    (Section~\ref{analysis:clusters}), and the integration of different
    cosmological probes (Section~\ref{challenges:theory}). As outlined in
    Section~\ref{sec:cosim}, we have recently developed a framework to build
    emulators for observables spanning a range of parameters from a
    relatively limited set of simulations. Such emulators and their
    improved descendants will be necessary LSST DESC prediction
    tools. The emulator strategy will be extended to provide estimates
    for covariance matrices. In particular, for combining different
    probes, large volume simulations and many realizations of these
    are needed. The brute force approach (simply generating thousands
    of large simulations) is very undesirable. More intelligent strategies
    for tackling this problem are therefore needed and will be
    investigated as part of this program. }

  \activities{Much as for the mock catalog generation, the first
    step will be to develop a detailed plan together with
    the different analysis groups that covers all desired prediction
    capabilities, the requirements on their accuracy, and the
    cosmological model space (including dark energy/modified gravity
    models) to be covered. For some of the desired prediction
    capabilities, existing tools for extracting them from the
    simulations have to be improved or newly developed. Again, close
    collaboration between the analysis and simulation groups will be
    essential. Some of the new emulators developed for Stage III
    experiments will be useful for testing LSST DESC analysis
    pipelines. We will build upon our experience with those tools to
    refine the prediction capabilities as needed. We will generate a
    first set of simulations that can be used to test out different
    ideas for covariance estimation and will allow us to derive better
    estimates of the actual computing costs for different
    approaches. The simulations and methods will be improved over
    time. Following the approach for the prediction tools, some of
    this work will build upon our experience with Stage III
    experiments. While this task addresses longer term
    priorities, the work for it will need considerable lead time, as
    substantial development effort is needed. }

  \deliverables{A first set of emulators for different observables
    will be delivered along with a detailed work plan covering the
    needs of the different working groups. The developments will
    address the reqirements for high priority 
    Task~\ref{task:lss:require} for WL and Task~\ref{task:thjp:de_cap} for
    theory. We will also deliver an initial set of simulations that
    can be used to test and develop new methods for covariance
    estimations. In the long term, this task will impact all of the
    dark energy probes, in the current list it will explicitly
    addresses Task~\ref{task:thjp:de_cap} in the theory section.}

\end{task}

\tasktitle{CoSim: Astrophysical systematics}

\begin{task}
\motivation{Understanding, quantifying, and mitigating astrophysical
  systematics on the observables of interest will be an enormous
  challenge for the simulation working group. In particular, focus
  topics include: understanding the theoretical systematics issues in
  measurements of various two-point (and higher) statistics,
  cross-correlations between probes, and cluster masses. Without
  substantial progress in this area, it will be very difficult to probe
  deeper into the nonlinear regime of structure formation and obtain
  the measurement accuracy on dark energy parameters aimed at by the
  LSST DESC. }

\activities{This task will require improvements in our modeling and
  simulation capabilities to accurately capture the baryonic physics
  that influence the observables of interest. Together with the
  analysis groups, we will develop a plan on how to address these
  uncertainties, map out at what level of accuracy they have to be
  understood, and create a first test simulation campaign including
  different astrophysical effects. This will be particularly important
  for weak lensing and clusters. Since it will be impossible to
  simulate all of these effects from first principles, new 
  approaches have to be developed to allow us to model some of them in
  post-processing.}

\deliverables{Establish an initial program for estimating
  astrophysical uncertainties in individual probes and their
  correlations; based on the results of this program, design a set of
  simulations to investigate the possible range of
  outcomes; understand what observations can be used to mitigate the
  theoretical systematics. The findings will directly feedback into
  the development of the prediction tools (Task~\ref{task:cosim:pred}) and
  be of importance for clusters (Task~\ref{task:cl:masscalib}) and theory
  (Task~\ref{task:thjp:de_cap}).}
\end{task}

\end{tasklist} 

%% file: workplan/simulations/catalog.tex
\subsection{Catalog Improvements}
\label{sec:workplan:catalog}

Accurate and validated realizations of the Universe are at the heart
of the analyses described in Section~\ref{sec:analysis}. For simulated
catalogs, the properties of extra-Galactic sources, Galactic
foregrounds, reddening and extinction, and uncertainties due
to the LSST system and its observing strategies must all be modeled. The
fidelity of these models must be coupled to the timeline of the studies of 
systematic and statistical limitations on the measurement of dark
energy. Herein, we describe the high-priority developments and tasks that
must be undertaken to support the work described in
Section~\ref{sec:analysis} and Section~\ref{sec:workplan}.

\subsubsection{High-Priority Tasks}
\label{sec:catalog:high}

\begin{tasklist}{H}

\tasktitle{CatSim: Mock catalogs that trace astrophysical properties that
  impact the DESC}

\begin{task}

  \motivation{A significant number of analysis tasks and use cases
    described in Section~\ref{sec:analysis} require catalogs that reproduce
    the observed properties of galactic and extragalactic
    sources. These properties include: accurate representations of
    the distribution of sizes, colors, magnitudes, and redshifts of
    extragalactic sources (with appropriate clustering statistics);
    the distributions of stellar sources with realistic colors and
    number densities; accurate realizations of foreground
    contamination from extinction and reddening, as well as from the
    variability of Galactic and extra-Galactic sources.  Existing
    catalogs described in Section~\ref{sec:mock} need to be improved to
    better represent the size and color distributions of galaxies, and
    extended to include supernovae, and gravitational lensing.}

  \activities{The initial component to this work is the definition of a
    detailed timeline for the inclusion of additional astrophysical
    components within the mock catalogs. Based on this roadmap,
    catalogs derived from semi-analytic models, N-body and
    hydrodynamic simulations, and empirical halo models will be
    developed. These catalogs will include SNe; a general model for
    variability (for strong lensing, supernovae, and transient and
    variable foregrounds); improved representations of galaxy spectral
    energy distributions; weak gravitational lensing tied to the mass
    distributions within the underlying N-body simulations; and, the
    implementation of models for strong gravitational lensing. To
    accomodate a range of different models (optimized for the tasks
    described in Section~\ref{sec:workplan}), a framework will be require to
    enable the easy ingestion of different catalogs into the databases
    that serve the mock catalogs. This work will need to be undertaken in
    collaboration with the analysis groups.}

  \deliverables{A series of mock catalogs with progressively enhanced
    fidelity (matched to the needs of the analysis teams) will be
    delivered together with databases that can serve these data to the
    analysis teams and to other simulation tools. 
This work will
   support the high priority tasks: Weak Lensing (Tasks~\ref{task:wl:req} and \ref{task:wl:multi}), LSS
   (Tasks~\ref{task:lss:known} and \ref{task:dithersys}), Supernovae (Task~\ref{itm:sn:imsim}), Clusters (Task~\ref{task:cl:masscalib}), Strong lensing (Tasks~\ref{task:sl:environment}
  and \ref{task:sl:timedelays}), Theory (Task~\ref{task:thjp:photo-z}), Photometric redshifts (Task~\ref{task:photoz:tools}), Crosscutting
   (Tasks~\ref{task:lss:fullsky} and \ref{task:cross:unknown}), and Technical (Task \ref{tech:calib}).
}
\end{task}

\tasktitle{CatSim: Validation of mock catalogs}

\begin{task}
  \motivation{Mock catalogs used as inputs to the image simulations
    need to be validated against observational data sets. The fidelity
    of this validation is dependent on the requirements of individual
    analysis tasks, but will include requirements on the sizes, shapes,
    redshifts and colors of galaxies as well as the photometric
    properties of stars and the impact of the interstellar
    medium. Validation of mock catalogs is a complex process, since the
    underlying observational data contain inherent systematics and
    biases (e.g., the efficiency of measuring a redshift for a galaxy
    is dependent on its color, redshift and magnitude). Validation of
    the input and mock catalogs will enable these data sets to be
    utilized in studies of systematics and the ability to constrain
    cosmological parameters.}

  \activities{To undertake a detailed validation of catalog data will
    require the compilation of large ground and space-based
    observational data sets. For each data set, the inherent selection
    effects will need to be derived or measured and the source
    properties corrected for these effects.  The observed source
    properties will be compared with the mock catalogs and outputs of
    semi-analytic and empirical models in order to characterize the
    capabilities and limitations of the catalogs (and to determine
    what corrections are require to mitigate these limitations).}

  \deliverables{A data set will be delivered, consisting of the
    observational properties of stars, galaxies and transients
    sources, compiled from existing ground and space based
    observations; the source properties will also be characterized in
    terms of their distribution functions (as a function of redshift,
    magnitude, etc); the comparison and validation procedures will be
    automated to enable them to be applied to the multiple mock
    catalogs generated by the DESC. 
This work will
   support the high priority tasks: Weak Lensing (Tasks~\ref{task:wl:req} and \ref{task:wl:multi}), LSS
   (Tasks~\ref{task:lss:known} and \ref{task:dithersys}), Supernovae (Task~\ref{itm:sn:imsim}), Clusters (Task~\ref{task:cl:masscalib}), Strong lensing (Tasks~\ref{task:sl:environment}
  and \ref{task:sl:timedelays}), Theory (Task~\ref{task:thjp:photo-z}), Photometric redshifts (Task~\ref{task:photoz:tools}), Crosscutting
   (Tasks~\ref{task:lss:fullsky} and \ref{task:cross:unknown}), and Technical (Task~\ref{tech:calib}).
}
 
\end{task}

\tasktitle{CatSim: The implementation of errors within the source catalogs}
\begin{task}

  \motivation{Understanding the constraints on cosmology from LSS, SNe
    and weak lensing will require catalogs of representative subsets
    of LSST data.  Characterizing the impact of systematics on
    these constraints requires the generation of the uncertainties associated
    with the measurements. These uncertainties will depend on the
    astrophysical properties of the sources, the nature of the
    observations and the nature of the LSST system itself. Of
    particular importance will be the impact of non-Gaussian tails
    within these distributions).}

  \activities{Multiple runs of the image simulator will be generated
    for a broad range of astronomical and observing conditions in
    order to characterize the uncertainties on the photometric,
    astrometric and shape properties. These distributions will be
    propagated to the catalog framework to enable the generation of
    large-scale catalogs with appropriate error distributions. }

  \deliverables{Distribution functions that characterize uncertainties
    in the astrometric, photometric, and shape performance of the LSST
    system will be delivered through this task. The distributions will
    be characterized as a function of survey properties (e.g., airmass,
    flexure in the instrument, signal-to-noise, etc), with an emphasis
    on non-Gaussian properties.  This work will support weak lensing
    (Task~\ref{task:wl:req}), LSS (Tasks~\ref{task:lss:known} and
    \ref{task:dithersys}), Clusters (Task~\ref{task:cl:masscalib}), Strong
    lensing (Tasks~\ref{task:sl:environment} and \ref{task:sl:timedelays}),
    Theory (Task~\ref{task:thjp:photo-z}), Photometric redshifts
    (Task~\ref{task:photoz:tools}), Crosscutting (Tasks~\ref{task:lss:fullsky}
    and \ref{task:cross:unknown}), and Technical (Task~\ref{tech:calib}). 
    }
\end{task}

\end{tasklist}

%% file: workplan/simulations/phosim.tex
\subsection{Photon Simulator Improvements \label{phoSimImprove} }
\label{sec:phosimwp}

As described in Section~\ref{phosimdescription}, the photon simulator was designed to help determine if the LSST hardware and data
management software achieve nominal performance goals, but in many cases this
is not necessarily sufficient for high precision dark energy measurements.  In
addition, its detailed use among many users having a wide variety of
scientific goals was not the initial focus.   Therefore, to achieve the
analysis objectives described previously, we must (1) improve the
usability of the code for both individual and large-scale grid computing applications,
(2) improve the physics fidelity for detailed dark energy measurement
studies, (3) validate the simulator for dark energy measurement use, and (4)
connect dark energy systematics to simulation physics.

\subsubsection{High-Priority Tasks}
\label{sec:phosim:high}

\begin{tasklist}{H}

\tasktitle{PhoSim: Photon simulation usability improvements}

\begin{task}

\motivation{Support for the large number of users of the photon simulator in the
  collaboration and the wide variety of use cases involved entails a significant 
  development effort. We anticipate that
  individuals will run the code on their laptops, as well as 
  needing large-scale runs on grid computing platforms.
  The simulator is already capable of running on a wide variety of grid
computing systems, such as the Open Science Grid via CONDOR scripts.
However, the previous work all centered around using the simulator in single,
monolithic and approximately annual data challenges.  Since the development of the
analysis efforts in the DESC requires rapid prototyping and simulations in a
variety of different contexts, we will need to facilitate the simulator's use by many
users in a grid computing environment.  This task is necessary to begin the
wide range of analysis tasks using the simulator.}

\activities{In order to improve the photon simulator for individual use, we plan to
streamline the installation and remove other hinderances to widespread deployment.
We will maintain a released source code respository in the LSST DESC, and
pre-compile binaries for various platforms.  We also plan to develop a subset
of power users that will become proficient at using the code in a complex
manner and be aware of major development changes and limitations.  We will
develop a series of cookbooks for its detailed use.  We will also have all
DESC users obtain access to the base Universe catalog, as well as make it simple
to ingest user-defined catalogs.  As we build up the user community for the
photon simulator, we expect to reap benefits in functionality (as users 
provide ideas for enhancements) and robustness (as users identify
limitations and errors in the implementation).  We expect the simulator
will improve significantly from this active user involvement.
We expect a large variety of simulations because the scale of the simulation and 
the photon simulator and astrophysical catalog physics 
will be different for many applications.  Therefore, we will install and maintain release versions 
of the simulator on one or more grid computing submission sites.  We will build a user request
simulation system that interfaces with these grid computing sites.  We expect
significant challenges in developing an efficient monitoring system and 
fully connecting the input and outputs of catalog, photon simulator, and
analysis codes.  We also plan to maintain an organized and
user-friendly storage system for the simulation and analysis outputs.}

\deliverables{New simulator versions every three months, with code that is more useable for both grid computing and
  individual applications; complete simulator documentation.}

\end{task}

\tasktitle{PhoSim: Photon simulation fidelity improvements}

\begin{task}
\motivation{In order to improve the fidelity of the code, we will pursue an ambitious
development plan.  There are several major physics components that need
improvement that we know may affect dark energy science.  In general, all dark
energy science tasks require updates to the relevant physics, which affects
both the photometric response (simulated by whether photons make it to the detector) and
the point-spread-function size, shape, and astrometric scale (simulated by the particular path of the photon).  The ambitious tasks
described for Weak Lensing (Tasks~\ref{task:wl:psf}, \ref{task:wl:req},
and \ref{task:wl:imsim}), Large Scale Structure (Tasks~\ref{task:lss:known} and
\ref{task:dithersys}), 
Supernovae (Task~\ref{itm:sn:imsim}),
Clusters (Tasks~\ref{task:cl:masscalib} and \ref{task:cl:shearcalib}), 
Strong Lensing (Task~\ref{task:sl:detection}), Crosscutting
(Tasks~\ref{task:cross:metric} and \ref{task:cross:deblend}) and Technical
(Tasks~\ref{task:tech:wave} and \ref{task:tech:model})
require us to improve and assess simulation fidelity for each
case.  We need to begin this task to be able to start with detailed analysis
tasks covering DE systematics.
As discussed in Section~\ref{phosimdescription}, the main advantage of
  using a physics-based simulation is that the detailed physical models of the
atmosphere, telescope, and camera can be connected to the astrophysical dark
energy measurements.  Thus, we expect an active synergy between analysis and
simulations, in which the particular limitation of an algorithm might be
connected to a particular physical aspect of the simulations.  In some cases,
this could allow us to build a better algorithm that exploits the particular
physical form of the effect; in other cases, we might be able to remove or
weight particular exposures for variable effects; in yet other cases, we would
simply reach the systematic floor because the effect cannot be removed.
}

\activities{There are several areas of fidelity of the simulator that we
  already know can be improved for higher-fidelity simulations.  These are
  particularly well-suited for the DESC, where the exact details of the PSF and
  photometric response matter for dark energy science.  One area is the perturbations to the
optical elements that are currently set by tolerances and estimates, but not based
on a physical model that predicts the perturbations from thermal drifts and
the change in gravity vectors from the telescope motion.  In addition, the
control system feedback loop attempts to compensate for these perturbations.
Much of the modeling for this has already been developed by the camera,
telescope, and wavefront sensing teams, but we have not collected this into a
complete physical model.  Another area for improvement is the physics of the
multilayers, which can be properly formulated through an EM calculation of the
transmission/reflection through the layers.  Sensor defects are another large
area for improvement that will become easier to model accurately when further testing of
prototype sensors in the camera team is complete.  A particularly important aspect
of the camera physics is whether there is any anisotropic charge diffusion due
to non-parallel field lines in the 100 micron thick LSST sensors.
Additional LSST atmospheric site data will also become available to improve or
validate the atmospheric model.  Finally, the implementation of the
astrophysical catalogs needs improvement in the coupling of galaxy positions
to gravitational shear and cosmic magnification parameters.  
Many other
improvements will result from studying the specific analysis tasks and
assessing the appropriate level of fidelity required.
Regardless of the type of physics improvement needed, the process of
implementing the improvement is clear.  First, the need for improvement is
identified from one of two possible sources:  (1) analysis of dark energy
algorithms that indicate a deficiency in the simulator or impose a stricter fidelity
requirements, and (2) improvement in the understanding of the LSST site, camera,
and telescope.   Second, the physics model is developed to add to the LSST
instrument model -- a large collection of documentation, experimental codes, and
studies using engineering codes.  The instrument model then attempts to capture
all the relevant physics for LSST, whether or not those elements are critical to put into
the photon simulator.  Third, the required data for fidelity improvement is
placed in the LSST-specific data repository and the code is updated if
necessary to provide the greater fidelity.  The code enhancements must then 
adhere to the strict requirements of the photon simulator for modularity,
software standards, physics consistency with other parts of the code, and
documentation for the photon simulator.  Often, the code enhancement is a
relatively simple task compared to the earlier design work.
After upgrading the fidelity, for each algorithmic task we will either turn on or off
  particular physical effects to gauge their relative importance, or we will
  reach dark energy measurement systematics limits that we will attempt to
  mitigate by a deeper physical understanding of the measurement process.
  We expect that the great multiplicity of simulations will lead to several
  breakthroughs in dark energy science techniques, and there will be several
  cross-cutting innovations developed by using these high-fidelity simulations.}

\deliverables{Release of a new version photon simulator code every three
  months with improved fidelity; a matrix connecting irreducible physical effects to key dark
  energy systematics.}

\end{task}

\tasktitle{PhoSim: Photon simulation validation}

\begin{task}
\motivation{Another major aspect of the improvement of the photon simulation is its
ongoing validation.  Validation is a complex task, involving careful study
of the output of the simulator in a specific configuration.  This is essential
to be able to justify the conclusions of the analysis work, and make progress
in understand dark energy systematics.}

\activities{All validation
requires a comparison source, either some measurements from real data, an alternative code that can
calculate a similar quantity, or a design expectation based on a detailed study.  Direct validation
with data or other codes is a significant element of the simulator improvement work.  In particular, we have experience
with existing codes developed for use in the adaptive optics
community and codes for active optics wavefront estimation and feedback
loops.  In many cases, a specific test simulation with as many alternative
codes as possible will help validate the simulations.  We also expect to
simulate either existing telescopes (we have already been testing the simulator for
the Subaru telescope) as well as future telescopes where joint analyses are
expected, such as Euclid.}

\deliverables{We will run validation simulations and collect either alternative code
  simulations or real data.  The outputs will be compared with metrics
  using a uniform analysis techniques and documented.}

\end{task}

In summary, the work on the photon simulations will focus on (1) facilitating
the use of the high-fidelity simulation tools for both individual use and
deployment on large-scale grid computing, (2) improving the detailed physics in
the simulator according to an aggressive development plan adding another level
of physics detail, and (3) validating the physics models quantitatively in the
simulator by comparisons with existing data.  These improvements will be used
to build a thorough scientific understanding of the detailed connection between the physics of LSST, the astrophysics of dark energy probes, and the best methods of making dark energy measurements.   Dark energy science with LSST will not be successful without high fidelity simulations, and therefore we will pursue very detailed simulations connecting dark energy probes with the relevant atmosphere and instrument physics.

\end{tasklist}

%% file: workplan/frameworktasks.tex
\subsection{High-Priority Tasks}
\label{sec:framework:high}

\begin{tasklist}{H}

\tasktitle{SW: End-to-end simulator capable of running at small scales}
\begin{task}
\motivation{Every user in the collaboration should be able to exploit the powerful simulation tools built by the project in order to test a wide variety of systematics. A framework that makes importing and using these tools as easy as possible for even the casual user will greatly enhance the science reach of the collaboration.}
\activities{The first activity is to streamline the installation and remove other hinderances for widespread use on the scale of tens of CCD's. The released source code will be maintained in repositories available to all members of the DESC; binaries for various platforms will be pre-compiled. A subset of power users will be identified; these will become proficient at using the code in a complex manner and be aware of major development changes and limitations. A simple set of procedures and scripts will be developed that will aid in job submission. The framework will also provide data handling tools that will help 
organize results and temporary files. Aggressive provenance tracking will not only spot failures but also follow up on them to reduce transient delays.}
\deliverables{The small scale simulator will be delivered as a core part of the software analysis framework.}
\end{task}

\tasktitle{SW: Requirements for software framework}
\label{task:framework:requirements}
\begin{task}
\motivation{The software framework will be developed to address the needs of all science working groups. However, scientists are not necessarily well-versed in framework development, so a collaboration of computer professionals and scientists will work together to generate requirements for the framework. As the needs of scientists evolve, changes in requirements will be managed to address these needs.}
\activities{Develop the requirements for the software framework based on the use patterns of the analysis group. Every WG will develop use cases in conjunction with computing professionals. The collaboration will then define a list of packages to be included in the framework, a set of dependencies, and guidance for usage by the scientists. A series of workshops will be run in order to ensure the face-to-face time needed for this collaborative work.}
\deliverables{A set of requirements for the software framework to be used by the DESC.}
\end{task}

\tasktitle{SW: Level 3 software framework}
\label{task:framework:level3}
\begin{task}
\motivation{A Level~3 software framework will enable the science WG's to assess the significance of the systematics under consideration. The collaboration will then be able to rank systematics based on quantitative apples-to-apples comparisons of their effects on dark energy constraints and to make educated choices about which algorithms to use to address every systematic effect considered.}
\activities{During the development phase of the
project, these tools will look very different than the final versions
that will be used during the production phase. However, even at this
early stage, the Working Groups will need a uniform, robust way to
quantify the effects of systematics and the algorithms used to
mitigate them.  The challenge is to build this toolkit in close coordination with the WG's, starting with a relatively simple, yet robust, set of modules, and gradually incrementing, improving upon these, keeping careful track of the different versions. Analysts will have access
to all code developed by members of the collaboration and to previous
systematics tests.}
\deliverables{Over the next 3 years, the Level~3 toolkit will start with: a suite of two-point function estimators;
simple analytic covariance matrices; a theory code such as camb; and a
Fisher estimator of bias and errors. These will gradually be extended.}
\end{task}

\tasktitle{SW: Repository for code and data}
\begin{task}
\motivation{The collaboration will want access to a tested set of routines that is tracked over time and tested. This set will help every collaboration member do her science without re-inventing the wheel and further ensure that differences in results obtained by different groups can be easily tracked down.}
\activities{The repository will be established so that it is easily accessible to everyone in the collaboration, and the various WG's will be prompted to add analysis code. Data from simulations will likely be housed in data stores, but the framework will link these so users can import both and operate on any data set with any set of code. External data sets, such as WMAP, Planck, DES, will also be easy to import when conducting analyses.}
\deliverables{The deliverable is a documented repository with easily categorized modules for analysis, linked to other stores and repositories that contain external data sets and code packages.}
\end{task}

\end{tasklist}

%% file: workplan/ComputingModelTasks.tex
\subsection{High-Priority Tasks}
\label{sec:compmodel:high}

 \begin{tasklist}{H}

\tasktitle{CM: Defining and updating computing models}
\begin{task}
\label{task:comp:define}
 \motivation{ There will be a ramp of computing needs ranging from the relatively modest needs for the early studies through to significant resources needed for Level 3 reprocessings of large chunks of observatory data. We need to understand the roadmap of needs over a 10+ year period to ensure the necessary resources can be obtained and funded.  }
 \activities{ We will collaborate with the working groups, via workshops and other discussions, to build a model of needs vs time. This will require understanding their activities as well as tracking usage to assist in realistic projections.}
 \deliverables{ We will provide projections of significant cycle and storage needs spanning a running 10 year period. These roadmaps would be refreshed annually.}
 \end{task}
 
\tasktitle{CM: Providing computing resources to the working groups}
\begin{task}
 \motivation{As described in Task \ref{task:comp:define}, a profile of computing resources will be needed over the collaboration's life. In addition to planning for the resources levels, we will need to provision the collaboration for mid-range cycles and storage at each phase in time. Workflow engines and data catalogs and access tools will be needed. Additionally the usual operational resources needed by a collaboration will be required: code repositories, automated builds and quality testing.  }
 \activities{ A period of understanding working group needs will be needed to start. We will need to address both the scale of computing required along with the ways analyses are done. This will allow us to craft the workflow engines and access tools, as well as the development tools, to suit the needed usage patterns. }
 \deliverables{ We will provide early versions of workflow engines and data catalogs, as well as code development environments in the first year. These will need to be iterated as we develop the software framework and understand the analysis patterns. }
 \end{task}

\tasktitle{CM: Providing storage and infrastructure for catalogs}
\begin{task}
 \motivation{ Realistic mock-catalogs and ImSim simulations will require non-trivial storage. It is important to provision for multi-TB databases as well as hundreds of TB of data files. Large databases require significant tuning for efficient data ingest and to maintain query performance, while large collections of files must be distributed appropriately with capable servers, filesystems and networking to survive significant load, minimizing failure rates.}
 \activities{ Reasonable scale catalog databases and file collections exist for the LSST project. We will explore these to develop our own tools and techniques to support them.}
 \deliverables{We will deliver well-managed and tuned databases and file systems within the first year or two. We will then need to keep up with increasing volumes of data to ensure that the solutions scale.}
 \end{task}
\end{tasklist}

\subsection{Longer-Term Tasks}
\label{sec:compmodel:longterm}

\begin{tasklist}{LT} 
\tasktitle{CM: Providing resources for N-body simulations}
\begin{task}
 \motivation{We anticipate very large n-body simulations runs, requiring significant compute cycles and storage. We need to understand the time profile of such simulations to project resource needs, as well as working with the supercomputing centers for efficient access to the produced data. We will also need to understand what entities will be charged with obtaining the cycles.}
 \activities{ We will work with the cosmology simulations experts, both in the collaboration and in external large cosmology collaborations to understand the scales of cycles needed as well as under what auspices they may be obtained.}
 \deliverables{ We will create a timeline of resources needed for cosmology simulations and identify the entities who will be charged with delivering the resources for the collaboration.}
 \end{task}

\end{tasklist}

%% file: workplan/technical_coordination.tex
The technical coordination tasks lie at the interface between the DESC science objectives and the LSST Project's technical and engineering
team. The success of the Dark Energy programs with LSST depend on a combination of a deep knowledge of the 
apparatus and software, and a clarity of scientific focus. Each of the priority tasks listed below span the dark energy science goals and the 
technical aspects of the LSST project, and for diverse reasons they require progress in the near term. The coordination of these
tasks and the facilitation of the interaction with the LSST project team will be undertaken by C.~Stubbs (DESC Technical Coordinator)
and J.~A.~Tyson.

\subsection{High-Priority Tasks}
\label{sec:tech:high}

\begin{tasklist}{H}

\tasktitle{TC: Wavefront sensing and pointspread function characterization}

\begin{task}
\label{task:tech:wave}
\motivation{The corner rafts of the LSST focal plane each contain a wavefront sensor, which obtains simultaneous images on either side of the
telescope's focal surface. These sensors are used to measure the shape of the arriving wavefront at these positions. This information is 
important for two reasons:  First, we use it to make an ongoing set of quasi-static corrections to the optical train, especially the shape of the primary
mirror.  Second, the out-of-focus images provide a means to measure the in-focus point spread function on the focal plane. The determination of the 
point spread function and its variation across the field is essential for the weak lensing analysis, and can also facilitate the determination of the 
optimal frame subtraction kernel.}

\activities{Working with the Project, we will study the implementation of the out-of-focus wavefront sensors, the choice of their spacing, the development and validation of wavefront characterization algorithms, 
and the incorporation of this information into the downstream weak lensing shape analysis and frame subtraction photometry (for SN measurements).  This work will involve collaboration between the technical experts on these subsystems working on the Project, and representatives from each of the analysis working groups who can assess the scientific impacts of various design choices.}

\deliverables{Informal feedback to the Project, plus detailed scientific assessments of itemized design choices.}

\end{task}

\tasktitle{TC: Assessment of the approach toward photometric calibration}

\begin{task}
\label{tech:calib}
\motivation{Achieving precise flux measurements is essential to all aspects of LSST dark energy science. At present, band-to-band photometric 
normalization uncertainties dominate the systematic error budget for Type Ia supernova cosmology. Moreover, obtaining precise 
photometric redshift measurements requires a consistent and precise definition of flux measurements across the entire survey area.}

\activities{The LSST project is in the final design stage for both the overall photometric calibration philosophy and its implementation in both hardware
and software. The choices include how to best measure the focal-plane-position--dependent instrumental sensitivity function $S(\lambda,x,y)$ 
for the various LSST passbands, how to best determine the optical transmission function of the atmosphere $A(\lambda)$, and how to 
determine and correct for Galactic extinction $G(\lambda,\alpha,\delta)$  at right ascension $\alpha$ and declination $\delta$.  The DESC team will bring both experience and insight to bear on the assessment of this calibration strategy, and help inform
the design choices. DESC scientists will be fully engaged in the assessment and refinement of the LSST calibration plan, and with 
testing, validating, and implementing the LSST calibration approach using a combination of ground-based and space-based
observations.}

\deliverables{Informal feedback to the Project, plus detailed assessment of expected residual photometric calibration errors, and their impact on dark energy science.}

\end{task}

\tasktitle{TC: Instrument model development}

\begin{task}
\label{task:tech:model}
\motivation{The ``instrument model'' is a predictive tool, or collection of predictive tools, that can be used to quantitatively estimate the response of the LSST system to monochromatic radiation at an arbitrary wavelength emanating from an arbitrary position on the sky with respect to the optical axis.  It will comprise our complete understanding of all physical effects within the camera and telescope that can contribute to the imaging and throughput performance of the system.  An effort will be made to embed as much of the instrument model as possible into the photon simulator, however it may prove more convenient and/or feasible to also utilize other computational approaches.  The instrument model will evolve as we acquire additional calibration and characterization data on the as-built components.  Clearly, our understanding of subtle systematics affecting dark energy analysis will be critically dependent on the fidelity of this instrument model.}

\activities{The requirements for dark energy analyses present the most stringent requirements we currently know of for our knowledge of the system behavior.  Therefore, the DESC will play a driving role in working with the project to define a set of measurement and analysis requirements for component level testing.  Experts from the hardware teams will work with representatives from the analysis working groups to make these determinations.  Continuing assessments will be made of the extent to which modifications can and should be made to the photon simulator to reflect our evolving knowledge of the system.}

\deliverables{Informal feedback to the Project, along with detailed scientific assessment of impacts associated with identified physical effects.}

\end{task}

\tasktitle{TC: Cadence and operational simulation, and optimization of main and deep-drilling surveys}

\begin{task}

\motivation{The decisions of where to point the LSST telescope, in what filter to expose in, how long to expose for, and and how frequently to return to a given field, are
the knobs that will be adjusted during system operations. We need to understand, well in advance, how to best make these choices in order to 
maximize the dark energy science impact of LSST.  Since the extracted constraints on the properties of the dark energy depend upon both the 
image archive and the data analysis algorithms, the DESC team will play a vital role in learning how to optimize operation of the LSST system.}

\activities{The DESC will assess observing strategies that balance the tensions between the operating modes that are deemed optimal for different dark energy probes. The supernova
technique favors repeated observations of a few fields, while the weak lensing approach favors wide field coverage. We will explore and understand these
tradeoffs and strive to understand the operational approach that will best use LSST for the understanding of the dark energy.}

\deliverables{A cadence toolkit for optimizing dark energy science that can be delivered to the Project for assessment against comparable toolkits delivered by collaborations working on other science topics.}

\end{task}

\end{tasklist}